\documentclass[twocolumn]{aastex631}
\usepackage{graphicx, amsmath, bbm, mathtools, capt-of} 
\usepackage[para]{threeparttablex}

\begin{document}

\title{New Evidence for a Flux-independent Spectral Index of Sgr~A* in the Near-infrared}

\author[0000-0002-2603-6031]{Hadrien Paugnat}
\author[0000-0001-9554-6062]{Tuan Do}
\author[0000-0002-2836-117X]{Abhimat K. Gautam}
\author[0000-0002-7476-2521]{Gregory D. Martinez}
\author[0000-0003-3230-5055]{Andrea M. Ghez}
\author[0000-0001-5972-663X]{Shoko Sakai}
\author[0000-0003-4081-1839]{Grant C. Weldon}
\author[0000-0003-2874-1196]{Matthew W. Hosek, Jr.}
\author[0009-0004-0026-7757]{Zoë Haggard}
\affiliation{Department of Physics and Astronomy, UCLA, Los Angeles, CA 90095-1547, USA}
\author[0000-0003-2400-7322]{Kelly Kosmo O'Neil}
\affiliation{Department of Physics and Astronomy, UCLA, Los Angeles, CA 90095-1547, USA}
\affiliation{Physics Department, University of Nevada, Reno, Nevada 89557, USA}
\author{Eric E. Becklin}
\affiliation{Department of Physics and Astronomy, UCLA, Los Angeles, CA 90095-1547, USA}
\author[0000-0003-2618-797X]{Gunther Witzel}
\affiliation{Max Planck Institute for Radio Astronomy, Auf dem Hügel 69, D-53121 Bonn (Endenich), Germany}
\author[0000-0001-9611-0009]{Jessica R. Lu}
\affiliation{Department of Astronomy, University of California, Berkeley, CA 94720-3411, USA}
\author{Keith Matthews}
\affiliation{Division of Physics, Mathematics, and Astronomy, California Institute of Technology, MC 301-17, Pasadena, California 91125, USA}

\begin{abstract}
In this work, we measure the spectral index of Sagittarius~A* (Sgr~A*) between the $H$ (1.6 $\micron$) and $K^\prime$ (2.2 $\micron$) broadband filters in the near-infrared (NIR), sampling over a factor $\sim 40$ in brightness, the largest range  probed to date by a factor $\sim 3$.  
Sgr~A*-NIR is highly variable, and studying the spectral index $\alpha$ (with $F_\nu \propto \nu^{\alpha}$) is essential to determine the underlying emission mechanism. 
For example, variations in $\alpha$ with flux may arise from shifts in the synchrotron cutoff frequency, changes in the distribution of electrons, or multiple concurrent emission mechanisms.
We investigate potential variations of $\alpha_{H-K^\prime}$ with flux by analyzing 7 epochs (2005 to 2022) of Keck Observatory imaging observations from the Galactic Center Orbits Initiative (GCOI).  
We remove the flux contribution of known sources confused with SgrA*-NIR, which can significantly impact color at faint flux levels. We interpolate between the interleaved $H$ and $K^\prime$ observations using Multi-Output Gaussian Processes. 
We introduce a flexible empirical model to quantify $\alpha$ variations and probe different scenarios.
The observations are best fit by an $\alpha_{H-K^\prime} = - 0.50 \pm 0.08 _{\rm stat}  \pm 0.17_{\rm sys}$ that is constant from $\sim 1$ mJy to $\sim 40$ mJy (dereddened 2 $\micron$ flux). 
We find no evidence for a flux-dependence of Sgr~A*’s intrinsic spectral index. In particular, we rule out a model explaining NIR variability purely by shifts in the synchrotron cutoff frequency. We also constrain the presence of redder, quiescent emission from the black hole, concluding that the dereddened 2 $\micron$ flux contribution must be $\leq 0.3$~mJy at 95\% confidence level. 
\end{abstract}


\section{Introduction} \label{sec:intro}

The Galactic center (GC) of the Milky Way hosts a compact source, Sagittarius~A* (Sgr~A*), first detected in the radio \citep{Balick_Brown1974} and now confidently associated to a supermassive black hole (SMBH) with a mass $\approx 4 \times 10^6 M_\odot$ \citep[e.g.,][]{Schodel2003_massBH,Ghez2003, Ghez2008, Gillessen2009}.
Observations have been realized across the electromagnetic spectrum, revealing the existence of a rapidly varying counterpart at X-ray \citep{Baganoff2001} and near-infrared (NIR) wavelengths \citep{Genzel2003, Ghez2004}. This observed variability is most likely powered by physical processes in the accretion flow surrounding the black hole, but the exact mechanisms are still not completely understood.

Sgr~A* offers a particularly compelling window into the complex puzzle of SMBH accretion. First, it can be observed at high angular resolutions due to the relative proximity of the GC \citep[$d\approx 8 $ kpc, e.g.][]{GR2019,GRAVITY2019_distGC}. Second, Sgr~A* is remarkably faint, radiating at $\lesssim 10^{-8}$ less than the Eddington luminosity \citep[e.g.,][]{Baganoff2003, Genzel2010}. Thus, it probes low accretion rate regimes more characteristic of the low-luminosity active galactic nuclei (AGN) population found in the nearby universe \citep[e.g.,][]{Ho2008, Contini2011, Eckart2018}, whereas accretion mechanisms are understood more thoroughly in the context of very luminous AGN. In fact, Sgr~A* is faint even by low-luminosity AGN standards, and its study has motivated the development of sophisticated models for radiatively inefficient/advection-dominated accretion flows \citep[RIAF/ADAF, e.g. ][]{Narayan1995, Yuan2003,Yuan2014}. 

These models can be constructed using semianalytic prescriptions, but also numerical simulations, which have become increasingly refined, incorporating general relativistic (GR) effects, intricate magnetic field interactions, and plasma turbulence into the dynamics of accretion. State-of-the-art general relativistic magnetohydrodynamic (GRMHD) simulations of the accretion flow, paired with GR ray-tracing codes, are now able to match reasonably well some of the observed properties of Sgr~A*'s emission \citep[e.g.,][]{Anantua2020, EHT2022, Ressler2023}. These simulations are mostly driven by observational constraints in the radio domain \citep[the Event Horizon Telescope results in particular,][]{EHT2022_obs} - at mm/sub-mm wavelengths, the emission is believed to predominantly arise as synchrotron from a population of electrons with a thermal energy distribution \citep[e.g.,][]{Genzel2010, EHT2022}.

Connecting GRMHD predictions to other wavelength regimes, however, is not straightforward. Although the NIR emission is similarly thought to be dominated by optically thin synchrotron radiation from a population of relativistic electrons near the event horizon ($\sim 10$ Schwarzschild radii) \citep[e.g.,][or \cite{Genzel2010} for a review]{Eckart2006,Eckart2009,Trippe2007}, it cannot be directly related to the radio. The NIR emission displays a strong variability, often explained by a fluctuating contribution from a distinct, non-thermal population of electrons \citep[e.g.,][]{Yuan2003, Yuan2004,Chael2018,Davelaar2018}. It remains unclear what processes govern the evolution (injection through particle acceleration, escape, heating, cooling) of this population, and whether the flares arise from outflows or orbiting hotspots in the disk \citep[e.g.,][]{Ripperda2020,Porth2021, Ball2021}. For instance, recent efforts suggest that some of the observed NIR properties \citep[e.g., realistic lightcurves, ][]{Dexter2020, Chatterjee2021, Grigorian2024} can be reproduced with magnetic reconnection in GRMHD simulations, but a consensus model for the NIR emission is still lacking.

To constrain the existing models and, in turn, better understand the complex physics in the vicinity of the SMBH, measurements in the NIR in unexplored regimes (e.g., of the flux distribution or the spectral index) are needed, all the more since Sgr~A*'s variability can be best probed at these wavelengths. Some fluctuations are observed in the X-ray and radio domains as well, but X-rays mostly probe strong flares that punctuate an otherwise quiescent emission state \citep[e.g.,][]{Baganoff2001,Baganoff2003,Porquet2008,Ponti2017}. Radio (sub-mm in particular) wavelengths capture the bulk of Sgr~A*’s steady radiation, and show relatively small flux variations ($\sim10-30\%$) around the mean value \citep[e.g.,][or \cite{Genzel2010} for a review]{Mauerhan2005, Maquart2006, Subroweit2017}. Sgr~A*-NIR, on the other hand, is continuously variable, and observable as a distinctive source at nearly all times using ground-based telecopes with adaptive optics. This allows to probe the variability at many timescales and flux levels \citep[over a factor of $\sim 500$,][]{Weldon2023}.

For this reason, a wealth of studies report observations of Sgr~A* in the near-infrared, with a lot of focus on the flux distribution and timing properties \citep[e.g.,][]{Do2009_variability,Dodds-Edden2011_2states,Witzel2018,Chen2019, Gravity2020_twostate, Weldon2023} or on correlation with X-ray flares and/or sub-mm emission \citep[e.g.,][]{Eckart2004, Eckart2012,Yusef-Zadeh2012,Ponti2017, Witzel2021, GRAVITY2021_spectral_index}.

For NIR-only observations, while investigating the flux distribution and timing helps to constrain Sgr~A*'s flux states, a more complete physical picture requires information about the variations in the spectral energy distribution. For example, probing the flux-dependence of the spectral index can be important to differentiate between models of the variability. Determining the NIR color/spectral index variations of Sgr~A* is, however, more challenging than single-band photometry, since this often faint source has to be observed within its very short variability timescale \citep[$\sim 1-10$ min,][]{Witzel2018, Do2019flare} in at least two wavelengths. 

Amongst the existing measurements, there is a general agreement in both photometric \citep{Ghez2005_color, Hornstein2007, Bremer2011, Trap2011, Witzel2014, GRAVITY2021_spectral_index} and spectroscopic \citep{Krabbe2006, Gillessen2006, Ponti2017} studies over a value $\alpha \sim -0.6$ (where $F_\nu \propto \nu^{\alpha}$) - typical for optically thin synchrotron radiation - when Sgr~A*-NIR is a bright state. There are, however, two questions left open in the literature about the NIR spectral index of Sgr~A*, that we attempt to address in this paper.

First, Sgr~A* showed heightened NIR activity in 2019, completely challenging previous statistical models of its variability \citep{Gravity2020_twostate, Murchikova2021_flare, Weldon2023}. In particular, 2009 May 13 displayed an unprecedented flare in the near-infrared, reaching twice the level of any historical measurement, with a drop in brightness from $m_{K^\prime}\sim 13$ to $m_{K^\prime}\sim 17$ (ie. a factor $\sim 50$ in flux) over a $\sim 2$ hr timescale \citep{Do2019flare}. The origin of this episode is still unknown - some proposed explanations include a magnetic reconnection event involving unusually strong magnetic fields \citep[e.g.,][]{Gutierrez2020, Dexter2020}, or a temporary increase in the accretion rate  \citep[e.g. due to the delayed infall of material from a dusty G-object, tidally stripped  during its pericenter passage,][]{Murchikova2021_flare}. A NIR spectral index measurement has yet to be reported for this epoch. Since it explores a new flux range, this would allow to test whether the usual value for Sgr~A*-NIR flares remains valid, even at very bright flux levels.

Second, fainter states of Sgr~A*-NIR spur much more debate. Some studies report measurements consistent with a constant spectral index value over large flux ranges \citep{Hornstein2007, Trap2011, Witzel2014}. Other works find that Sgr~A*-NIR becomes redder at low flux densities, with $\alpha \lesssim -2$ \citep{Krabbe2006, Gillessen2006, Bremer2011, Ponti2017, GRAVITY2021_spectral_index}. At least two explanations for this possible flux dependence have been advanced. \cite{Witzel2018} proposed that shifts in the synchrotron cutoff frequency create the NIR variability, which would cause the spectral index to vary linearly with magnitude (ie. logarithmically with flux). Alternatively, based on its flux distribution, it has been argued that Sgr~A*-NIR is best described by a model with two components: a low-level, quiescent emission state, supplemented by another mechanism generating flares \citep{Dodds-Edden2011_2states, Gravity2020_twostate}. If the two processes exhibit different spectral indices, color variations are to be expected as flux densities transition from one state to the other.

In this work, we present new measurements of the NIR spectral index of Sgr~A*, from both previous and new observations obtained with the Keck Observatory in the $H$ ($1.6 \mu m$) and $K^\prime$ ($2.2 \mu m$) broadband filters. They cover the largest dynamic range ever explored for a study of this type: observed flux in $K^\prime$ $\approx 0.1-4$ mJy, or equivalently  dereddened flux in $K_s$ $\approx 1-42$ mJy  \citep[using $A_{K_s}=2.46$,][]{Schodel2010_extinction}. Our dataset also includes the brightest NIR measurements to date thanks to the inclusion of 2019 May 13; and bright detections reduce systematic uncertainties such as confusion with stellar sources that may occur when Sgr~A*-NIR is faint. 

Furthermore, we propose a new method to interpolate correlated lightcurves, based on Multi-Output Gaussian Process (MOGP) regression - a recent development in applied mathematics that is still scarcely employed for astrophysical applications \citep{Alvarez_MOGP_2, GP_astro_review}. 

This paper is organized as follows. 
Section~\ref{sec:obs+red} presents our observations and data reduction method. Section~\ref{sec:SpecIndexMeas} explains how we obtained spectral index measurements, giving details regarding extinction correction, confusion correction, MOGP interpolation, and motivating the magnitude cuts for the final datasets. In section~\ref{sec: model}, we introduce an empirical model encompassing the three pictures discussed above, allowing to infer the intrinsic spectral index of Sgr~A*-NIR from noisy measurements. Section~\ref{sec:results} describes the results of this analysis, first for a very robust subset composed of bright observations only, then for our complete dataset. We discuss our conclusions in perspective with the literature in section~\ref{sec: discussion}, before summarizing in section~\ref{sec: summary}.

\vspace*{\baselineskip}
\section{Observations and data reduction} 
\label{sec:obs+red}

\subsection{Observations}
\label{subsec:obs}

Our data consists of high-resolution images from the Galactic Center Orbits Initiative (GCOI; PI: A. Ghez), obtained using the Laser-Guide Star Adaptive Optics system \citep{LGS_AO} on the NIR imager NIRC2 (PI: Keith Matthews) at the 10 m W. M. Keck II telescope. These observations were performed with one of two broadband filters - either $K^\prime$ ($\lambda_0 = 2.124 \mu$m, $\Delta \lambda = 0.351 \mu$m) or $H$ ($\lambda_0 = 1.633 \mu$m, $\Delta \lambda = 0.296 \mu$m) - over seven epochs: one in 2005, one in 2019, and five in 2022. In particular, the $H$-band observations in 2022 are newly reported in this work. We draw attention to the fact that this dataset includes the brightest NIR flare ever recorded for Sgr~A* (2019 May 13), part of its period of enhanced activity in 2019. Studying color variations is especially interesting at that time, since the physical mechanism responsible for this episode is still debated (see section~\ref{sec:intro}).

For 2005 July 31, observations cycled through $H$-band (3 coadds with exposures of $t_{int} = 7.4$ s each), $K^\prime$-band (10 coadds, $t_{int} = 2.8$ s each), and a third filter not used in this work ($L^\prime$, $30$ s integrations) over a total time of 113 minutes \citep{Hornstein2007}. The filter changed after each frame, ie. every $\sim 30$ s, so the time between $H$ and $K^{\prime}$ observations varied between $30$ s and $1$ min. Images for this night were not dithered, but Sgr~A* was held fixed in an area of the detector free from bad pixels.

The other six epochs had a different setup, switching only between H and $K^\prime$, with multiple dithered frames taken before a change of filter. For 2019 May 13 and the two epochs in May 2022, observations alternated between six frames in each band, meaning that the time to the closest observation in the other band ranged from $\sim 30$ s to $ \sim 1.5$ min. For the three epochs in August 2022, only three dithered images were taken before switching filters. This resulted in a temporal separation between H and $K^\prime$ of $\sim 30$ s to $\sim 3$ min. We kept the same coadds and integration time in $K^\prime$ as 2005 July 31. In $H$, however, the number of coadds was increased to 4 (with the same $t_{int}$) in order to improve the signal-to-noise ratio (SNR). The various observational setups are summarized in Table~\ref{ObsChar}.

\begin{ThreePartTable}
\begin{deluxetable*}{CCCCCCCCC}[hbtp]
    \tablehead{\colhead{}  & \colhead{Total observation} & \colhead{Dithered} & \colhead{Filter switch}  & \colhead{Temporal separation} & \colhead{} & \colhead{Integration} &  \colhead{Number of} & \colhead{} \\[-5pt]
    \colhead{Date (UT)}  & \colhead{time (min)} & \colhead{(yes/no)} & \colhead{every x frames} & \colhead{between bands (s)} & \colhead{Band}  & \colhead{time (s)} &  \colhead{co-adds}  & \colhead{Reference}}
    \startdata
        \text{2005 July 31}  & 113 & \text{no} & 1 & 30-60 & K^\prime & 2.8 & 10 & \tnote{a}\\
         &  &  &  &  & H & 7.4 & 3  & \tnote{a}\\
    \hline
      \text{2019 May 13}  & 213 & \text{yes} & 6 & 30-180 & K^\prime & 2.8 & 10 & \tnote{b}\\
         &  &  &  &  & H & 7.4 & 4  & \tnote{b}\\
    \hline
       \text{2022 May 21}  & 105 & \text{yes} & 6 & 30-180 & K^\prime & 2.8 & 10 & \tnote{c}\\
         &  &  &  &  & H & 7.4 & 4  & \tnote{d}\\
    \hline
       \text{2022 May 25} & 134 & \text{yes} & 3 & 30-90 & K^\prime & 2.8 & 10 & \tnote{c}\\
         &  &  &  &  & H & 7.4 & 4  & \tnote{d}\\
    \hline
       \text{2022 Aug 16} & 101 & \text{yes} & 3 & 30-90 & K^\prime & 2.8 & 10 & \tnote{c}\\
         &  &  &  &  & H & 7.4 & 4  & \tnote{d}\\
    \hline
       \text{2022 Aug 19} & 114 & \text{yes} & 3 & 30-90 & K^\prime & 2.8 & 10 & \tnote{c}\\
         &  &  &  &  & H & 7.4 & 4  & \tnote{d}\\
    \hline
       \text{2022 Aug 20} & 116 & \text{yes} & 3 & 30-90 & K^\prime & 2.8 & 10 & \tnote{c}\\
         &  &  &  &  & H & 7.4 & 4  & \tnote{d}\\
    \enddata
    \label{ObsChar}
\begin{tablenotes}[flushright, para]
  \item[a]{ \cite{Hornstein2007}}
  \item[b]{ \cite{Do2019flare}}
  \item[c]{ \cite{Weldon2023}}
  \item[d]{ This work.}
  \end{tablenotes}
\caption{Description of the observational setups for the epochs used in this work.}
\vspace*{-\baselineskip}
\end{deluxetable*}
\end{ThreePartTable}
\vspace*{-2\baselineskip}

Following standard image reduction methods, each image was flat-fielded, sky-substracted, corrected for bad pixels, cosmic rays and optical distortions \cite[e.g., ][]{Yelda2010_distortion} using the Keck AO Imaging (\texttt{KAI}) data reduction pipeline \citep{KAI}.

Our group employs two main formats for the imaging data: individual frames, and composite images. The composite images are produced by combining  individual frames within a single night, and applying a weighting scheme based on each image's quality - quantified by the Strehl ratio \citep{Gautam2019}. Since Sgr~A* can be variable on very short timescales ($\sim 1-10$ min) in the NIR \citep{Witzel2018, Do2019flare}, we focused on individual frames to maximize the amount of information obtained from a lightcurve. Still, positional and photometric data extracted from the composite images proved useful for the analysis of the individual frames (see below).

\subsection{Point Source Detection \& Photometry}
\label{subsec:data_reduction}

To identify point sources in images (both individual and composite), then extract their astrometric and photometric data, we used the Point Spread Function (PSF) fitting software \texttt{AIROPA} \citep{AIROPA}, based on the code \texttt{StarFinder} \citep{Starfinder_code}, an IDL package designed for crowded stellar fields. Stellar crowding is extremely high in the central arcsecond region, making measurements of Sgr~A*-NIR especially challenging. To maximize the number of frames with a detection of Sgr~A*-NIR, we employed a modification of \texttt{StarFinder} introduced by \cite{Hornstein2007}, applied by \cite{Weldon2023}, and described in Appendix~\ref{Appendix: SF_force}. This method makes use of prior knowledge on the location of Sgr~A*, determined within $<1$ mas thanks to an IR astrometric reference frame - constructed from SiO masers with extremely accurate radio positions \citep{Sakai2019_refframe}.

\begin{deluxetable*}{CCCCCCCC}[hbtp]
    \tablehead{\colhead{} & \colhead{} & \colhead{$N_{\rm obs}$} & \colhead{ $N_{\rm obs}$ after }& \colhead{$N_{\rm obs}$ after}  & \colhead{$N_{\rm obs}$ after} & \colhead{$N_{\rm obs}$ with} & \colhead{$N_{\rm obs}$ with} \\[-5pt]
    \colhead{Date (UT)} & \colhead{Band}  & \colhead{before cuts} &  \colhead{Strehl ratio cuts} & \colhead{confusion correction} & \colhead{interpolation filters} & \colhead{$m_{K^\prime} \leq 16.5$} & \colhead{$m_{K^\prime} \leq 17.2$} }
    \startdata
        \text{2005 July 31}  &  K^\prime & 32  & 32 & 30 & 15 & 8 & 13\\
         &  H & 32  & 29 & 16   & 16  & 9 & 14 \\
    \hline
      \text{2019 May 13}  &  K^\prime & 97  & 90 & 90 & 10 & 9 & 10 \\
        &  H & 70  & 52 & 34 & 10 & 9 & 10 \\
    \hline
       \text{2022 May 21} &  K^\prime & 51  & 51 & 49 & 10 & 6 & 8\\
       &  H & 46  & 45 & 29 & 10 & 6 & 8\\
    \hline
       \text{2022 May 25} &  K^\prime  & 65  & 65 & 53 & 6 & 0 & 2\\
       &  H & 61  & 46 & 25 & 7  & 0 & 2\\
    \hline
       \text{2022 Aug 16} &  K^\prime  & 42  & 39 & 39 & 5 & 1 & 3\\
       &  H & 37  & 10 & 9 & 5 & 0 & 2\\
    \hline
       \text{2022 Aug 19} &  K^\prime & 49  & 48 & 48 &22 & 12 & 22\\
       &  H & 44  & 39 & 39 &23 & 13 & 23\\
    \hline
       \text{2022 Aug 20} &  K^\prime  & 56  & 52 & 52 & 12 & 6 & 12\\
       &  H & 45  & 25 & 25 & 14 & 8 & 14\\
    \hline \hline
       \text{Total} &  \text{Both}  &  -  & -  & -  & 165 & 87 & 143\\
    \enddata
    \caption{Number of measurements for each epoch, before and after the various data quality cut that are applied in this study (Strehl ratio cut: see section~\ref{subsec:data_reduction} and Appendix~\ref{Appendix: Strehlcut}; confusion correction cut: see section~\ref{subsec:confusion correction} and Appendix~\ref{Appendix: conf_corr+err}; interpolation filters: see section~\ref{subsec: GP interp} and Appendix~\ref{Appendix: GP tests}; magnitude cut: see section~\ref{subsec: bright end results}). We note that the interpolation filter are re-applied after the magnitude cut.}
    \label{NbObs}
    \vspace*{-\baselineskip}
\end{deluxetable*}
\vspace*{-2\baselineskip}

For this work, we employed the version of \texttt{AIROPA} from \cite{GR2019, Do2019flare}; and photometric calibration was performed with the same calibrator stars (IRS16NW, S3-22, S1-17, S1-34, S4-3, S1-1, S1-21, S3-370, S3-88, S3-36, and S2-63) and reference flux measurements as those from \cite{Weldon2023}. We remark that this version of \texttt{AIROPA} is different than the one employed by \cite{Weldon2023}, since we find that it separates close-by sources more effectively, improving photometry around Sgr~A*. We include more details in Appendix~\ref{Appendix: Single_vs_legacy}.


 
Since poor AO performance can lead to unreliable photometry, we applied an image quality cut for our dataset. The Strehl ratio $S$ was estimated for each frame using \texttt{KAI} \citep{KAI}, then we removed frames $S<0.2$ in $K^\prime$, and  $S<0.175$ in $H$ (these values are motivated in Appendix~\ref{Appendix: Strehlcut}). Typically, this removes $0-30\%$ of the frames per night (Table~\ref{NbObs}), and more frames in $H$ than in $K^\prime$. The worst epoch (2022 Aug 16, in $H$-band) has $\sim 70\%$ of the frames removed because of this cut. 

Finally, to estimate the photometric uncertainties, we applied the same procedure as described in \cite{Do2009_variability}. We considered all non-variable stars within some radius of Sgr~A*, and fitted a power-law relation to the the rms uncertainty in flux as a function of flux:
\begin{equation}
     \frac{ \sigma(F_{\rm band})}{1 \ \rm mJy} = C_{\rm band} \times \left( \frac{F_{\rm band}}{1 \ \rm mJy}\right)^{\beta_{\rm band}}
    \label{Eq:flux_err}
\end{equation}
with a scaling $C_{\rm band}$ and a power-law exponent $\beta_{\rm band}$.

The enclosing radius was chosen to be either 0.5\arcsec, 1\arcsec, or 2\arcsec, depending on the night and band, to ensure that the number of stars was sufficient to obtain a fit that did not vary when including additional stars. To compute the uncertainty on the flux measurement of Sgr~A*-NIR in each frame, equation~(\ref{Eq:flux_err}) was used with the values of $C_{\rm band}$ and $\beta_{\rm band}$ fitted for that night. Table~\ref{Noisefitvalues} reports the values of $C_{\rm band},\beta_{\rm band}$ found for each night.

\begin{deluxetable}{CCCCC}[hbtp]
    \tablehead{ \colhead{Epoch}  &\colhead{$\log_{10}C_{K^\prime}$} & \colhead{$\beta_{K^\prime}$} & \colhead{$\log_{10}C_H$} & \colhead{$\beta_H$}}
    \startdata
        \text{2005 July 31} & -1.71 & 0.52 & -1.70 & 0.64 \\
      \text{2019 May 13} & -1.54 & 0.51 & -1.67 & 0.61\\
       \text{2022 May 21} & -1.53 & 0.54 & -1.65 & 0.61\\
       \text{2022 May 25} & -1.57 & 0.48 & -1.56 & 0.65 \\
       \text{2022 Aug 16} & -1.39 &  0.53 & -1.50 & 0.57\\
       \text{2022 Aug 19} & -1.55 & 0.49 & -1.69 & 0.53 \\
       \text{2022 Aug 20} & -1.62 & 0.45 & -1.57 & 0.64 \\
       \hline
       \text{Mean values} & -1.56  & 0.50 & -1.62  & 0.61 \\
       \text{Standard deviation} & \phm{-} 0.09 & 0.03 & \phm{-} 0.07 & 0.04
    \enddata
    \caption{Normalization $C_{\rm band}$ and power-law index  $\beta_{\rm band}$ of the the noise law (\ref{Eq:flux_err}), fitted for each night and band. }
    \label{Noisefitvalues}
    \vspace*{-\baselineskip}
\end{deluxetable}
\vspace*{-\baselineskip}

\begin{figure*}[!t]
\includegraphics[width=.99\linewidth]{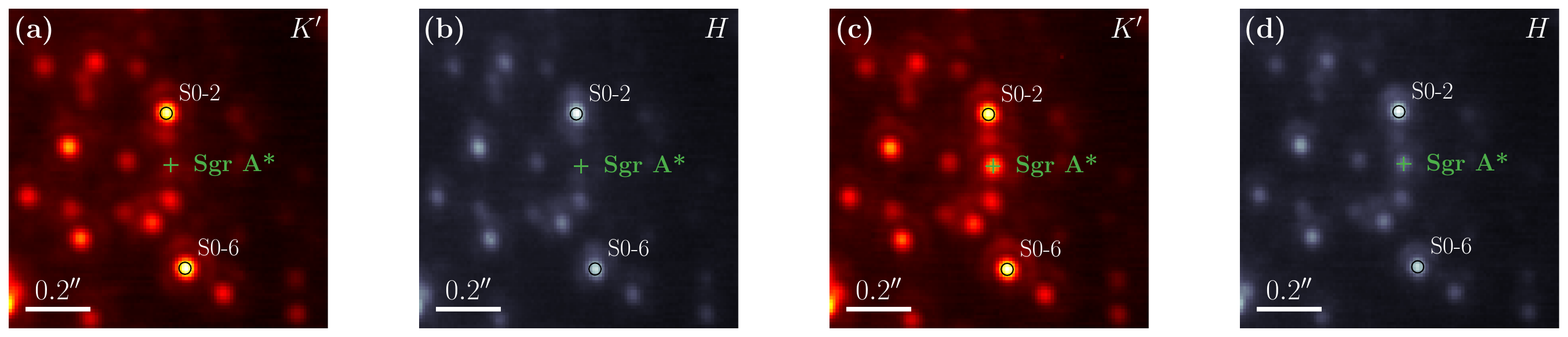} \\
\vspace{4mm}
\includegraphics[width=.99\linewidth]{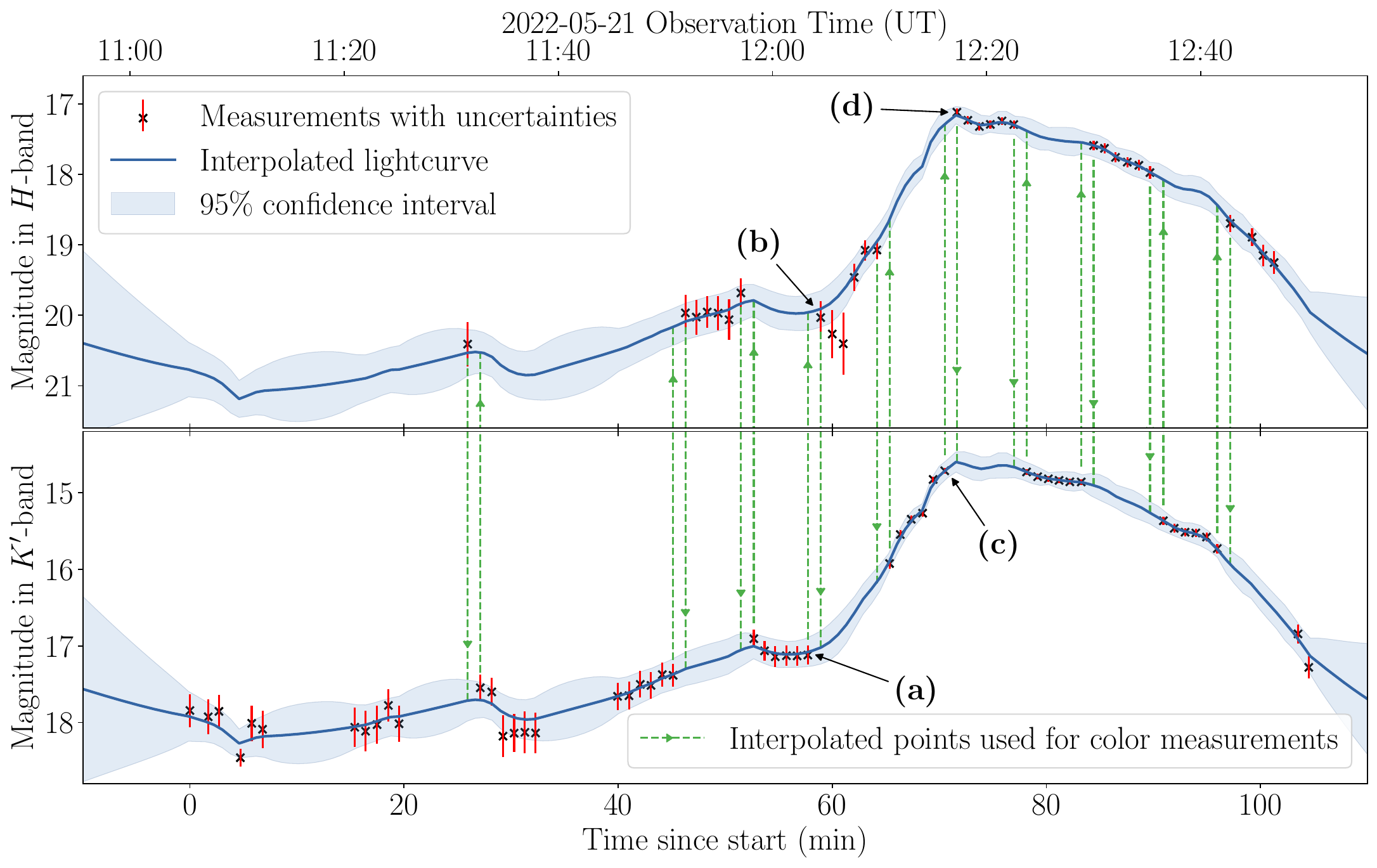} \\
\caption{Top row: example images in $K^\prime$ (red) and $H$ (blue)  from 2022 May 21, zoomed in on a $1\arcsec \times 1\arcsec$ box around Sgr~A*, showcasing its large variations in brightness. The bright nearby stars S0-2 and S0-6 ($m_{K^\prime} \sim 14$, $m_H \sim 16$) are labeled for reference. Bottom panel: Sgr~A*-NIR lightcurves in $H$ and $K^\prime$-band for the same epoch, interpolated using the method described in section~\ref{subsec: GP interp}. The green arrows show which points in the lightcurve are selected by the interpolation filters to serve for the color/spectral index measurements. The black arrows indicate the location of the four images in the panels above.} 
\label{fig:lightcurve}
\end{figure*}

Our measurements have photometric uncertainties on flux ranging from 1\% to 11\% in $K^\prime$, and from 3\% to 18\% in $H$, depending on the brightness of Sgr~A*-NIR. Figure~\ref{fig:lightcurve} displays two example frames centered on Sgr~A* for each band, as well as the lightcurves for the 2022 May 21. This night has good AO performance (S $\sim0.35$) and shows strong variation in the Sgr~A*-NIR flux.  


\vspace*{\baselineskip}

\section{Spectral index measurements}
\label{sec:SpecIndexMeas}

Throughout this work, we use the convention $F_\nu \propto \nu ^\alpha$ (where $F_\nu$ is the flux density) for the implicit definition of the spectral index $ \alpha$.


In this section, we detail the procedure used to derive spectral index measurements from photometric data, explaining how we correct for extinction (section~\ref{subsec:extinction}), for confusion (section~\ref{subsec:confusion correction}), and how we interpolate the lightcurves (section~\ref{subsec: GP interp}). We specify the amplitude of the corrections in Table~\ref{OoMcorrection}, along with the uncertainty introduced by each step.
We then motivate magnitude cuts for the final samples in section~\ref{subsec: mag cuts}.

\begin{deluxetable*}{CCC}[hbtp]
    \tablehead{ \colhead{Name of step}  & \colhead{Amplitude of correction (value or range)} & \colhead{Added uncertainty (value or range)} }
    \startdata
    \text{Extinction correction} & \Delta \alpha = 7.25 & \sigma_{\rm add}(\alpha) = 0.17 \text{ (systematic)} \\
    \hline
    \text{Confusion correction} & \Delta m_{K^\prime} \in [0.07, 0.67] & \sigma_{\rm add}(m_{K^\prime}) \in [0.01, 0.20] \\
    \text{(without 2019 May 13)} & \Delta m_H \in [0.08, 1.06] & \sigma_{\rm add}(m_H) \in [0.02, 0.34] \\
      & \Delta \alpha \in [0.01, 1.88] & \sigma_{\rm add}(\alpha) \in [0.01, 0.80] \\
    \hline
    \text{Interpolation} & \text{N/A} & \sigma_{\rm pred}(m_{K^\prime}) \in [0.04, 0.17]\\
     &  & \sigma_{\rm pred}(m_H) \in [0.05, 0.16]\\
    &  & \sigma_{\rm add}(\alpha) \in [0.14, 0.60] \\
    \enddata
    \label{OoMcorrection}
    \caption{Summary of the amount of correction (if any) and the uncertainty introduced by each step of the spectral index measurement estimation. Typical photometric uncertainties on magnitude (before all the correction steps) range from 0.02 to 0.10 mag in $K^\prime$-band, and from 0.04 to 0.15 mag in $H$-band. For this table, we only considered points that end up in the extended dataset ie. that survive the various cuts (see Table~\ref{NbObs} and section~\ref{subsec: mag cuts}). Comparing the spectral index $\alpha$ before and after confusion correction is only possible after interpolation, so the ranges for $\Delta \alpha$ and $\sigma_{add}(\alpha)$ reported in the "confusion correction" row are only estimates. Appendix~\ref{Appendix: uncertainty_comparison} provides more detail on the uncertainties added by confusion correction and interpolation. }
    \vspace*{-2\baselineskip}
\end{deluxetable*}

\newpage
\subsection{Extinction correction}
\label{subsec:extinction}

The Galactic center is characterized by strong, highly spatially variable interstellar extinction - so much so that the stellar population only becomes visible in the near-infrared ($A_V \gtrsim 30$ mag, $A_K \gtrsim 2.5$ mag) \citep{Nishiyama2008_extinction, Schodel2010_extinction}. Significant reddening is therefore expected in our observations, an effect which needs to be corrected for when measuring Sgr~A*-NIR's intrinsic color. The extinction curve for the GC in the NIR is usually approximated as a power law ($A_\lambda \propto \lambda^{-\beta}$), however the values for the normalization $A_0$ and power-law index $\beta$, along with possible deviations from the power-law form, are still debated \citep{Fritz2011_extinction, Hosek2018_extinction, Nogueras-Lara2019_extinction}. These uncertainties are limiting for studies concerned with precise color measurements, like this work. Even small uncertainties on $A_0$ and $\beta$ ($\sigma_{A_0} \approx 0.1$ and $\sigma_{\beta} \approx 0.06$) can lead to a significant spread in the value for the color excess needed to retrieve the true $H-K^\prime$ color of an object ($\sim 0.3$ mag, see Appendix~\ref{Appendix: Extinction}).

Fortunately, the GC is a crowded field, meaning that Sgr~A* is observed simultaneously with nearby stars of known spectral type, from which the color excess can be estimated. We dereddened the apparent color of Sgr~A*-NIR using the color of the star S0-2, (a bright, non-variable star, less than 0.2\arcsec away from Sgr~A*), avoiding the systematic uncertainties coming from the extinction law. S0-2 is spectroscopically identified as an early-type (B0-2V) star \citep{Paumard2006_S2_spectral_type, Do2009_spectroscopy}, so it has an intrinsic color $(H-K^\prime)_{\rm int}^{S0-2} = -0.08 \pm 0.03$ \citep{Ducati2001, Do2013_intrinsic_colors}.
In addition, S0-2 is photometrically stable ($<5\%$ scatter on flux values within a night and between nights), with mean magnitudes $\langle m_{K^\prime} \rangle = 14.10 \pm 0.03$, $\langle m_{H} \rangle = 16.11 \pm 0.03$ \citep{Gautam2024}. Then $\langle H-K^\prime \rangle_{\rm det}  = 2.01 \pm 0.04$ for S0-2, and using its intrinsic color, the mean color excess is $2.09 \pm 0.05$. We accounted for the difference in intrinsic spectral index between S0-2 and Sgr~A*-NIR's expected range by applying a second-order correction $\Delta E \equiv E(H-K^\prime)_* - E(H-K^\prime)_{\rm SgrA}  \approx 0.02 $ (see Appendix~\ref{Appendix: Extinction}). The final value for the mean color excess is therefore $\langle E(H-K^\prime)\rangle = 2.07 \pm 0.05$. We calculate the extinction-corrected $H-K^\prime$ spectral index of Sgr A*-NIR using:
\begin{equation}
    \alpha_{H-K^\prime}^{\rm est} = \frac{-0.4[(H-K^\prime)_{\rm est} - \langle E(H-K^\prime) \rangle] + \log_{10}\left(\frac{f_{0,H}}{f_{0, K^\prime}}\right)}{\log_{10}(\lambda_{K^\prime}/\lambda_H)}
    \label{eq:SpecIndex}
\end{equation} 
\vspace{0.5pt}

where $(H-K^\prime)_{\rm est}$ is the measured color of Sgr~A*-NIR, $f_{0,H} = 1050$ Jy, $f_{0, K^\prime}=686$ Jy are the zero-point flux densities in $H$ and $K^\prime$ \citep{Zero_pt_flux_densities}\footnote{The value originally quoted in \cite{Zero_pt_flux_densities} for the zero-point flux in H-band ($f_{0,H} = 1040$ Jy) was modified to $f_{0,H} = 1050$ Jy in an erratum \citep{Zero_point_erratum}. This difference only has an impact $|\delta \alpha| \lesssim 0.04$ on the spectral index,  ie. within our final uncertainties.}, and the central wavelengths are used for $\lambda_{\rm band}$. 

We note that choosing a different value for the mean color excess amounts to shifting all spectral index measurements by the same amount. In section~\ref{sec: model}, where we discuss an empirical model to investigate spectral index variations, we therefore fix the value of $\langle E(H-K^\prime) \rangle$. We propagate the uncertainty on $\langle E(H-K^\prime) \rangle$ as a separate systematic error term for the final spectral index value.
We also attempted to compute the color excess in each individual frame rather than just using the mean value, in order to account for slight variations in the photometry. This did not significantly impact our results.



\subsection{Confusion correction}
\label{subsec:confusion correction}

Because the Galactic center is a crowded field, the extended PSFs of individual sources frequently overlap, which can bias their respective photometry. We consider two objects to be confused if their angular separation is lower than 60 mas ($\approx 6$ pixels on the NIRC2 detector); in which case both will be fitted as a single, combined source by \texttt{StarFinder} \citep{Weldon2023}. In consequence, if there are sources within 60 mas of Sgr~A*, they will contribute to the flux measured at the location of Sgr~A*-NIR. 
Since stars are expected to have different intrinsic colors than Sgr~A*-NIR, it is necessary in that case to correct for the additional flux in the two bands, especially when Sgr~A*-NIR gets faint.

With the exception of 2019 May 13, Sgr~A*-NIR is confused with known stars in all epochs of our dataset: with S0-104 for 2005 July 31, with S0-38 for the 2022 nights\footnote{In 2022, S0-102 is also close to the confusion limit (separation $\sim 60$ mas), but \texttt{StarFinder} consistently detects a source near its expected location, with its expected magnitude. Therefore, we considered that Sgr~A*-NIR was not confused with S0-102 in our 2022 epochs.}. Based on two decades of AO observations of the GC, we determined the expected $K^\prime$ magnitudes of these stars, calculating the mean and standard deviation over all epochs for which they are detected and not confused (see Table~\ref{ConfMags}).

\begin{deluxetable}{CCC}[h]
    \tablehead{&\colhead{S0-104} & \colhead{S0-38} }
    \startdata
        \text{Number of epochs used}  & 7  & 36  \\
       \text{Magnitude in } K^\prime  & 16.75 \pm 0.08 & 17.04 \pm 0.12 \\
       \text{Magnitude in } H  & 18.85 \pm 0.14 & 19.24 \pm 0.14 
    \enddata
    \caption{Mean and standard deviation for the magnitudes of stars confused with Sgr~A*-NIR in our dataset. Values in $K^\prime$ are determined from long-term monitoring data of the Galactic center \citep{Gautam2024} ; then $H$ magnitudes are computed using the mean color excess measured from S0-2 and assuming an intrinsic color.}
    \label{ConfMags}
    \vspace*{-\baselineskip}
\end{deluxetable}
\vspace*{-2\baselineskip}


Due to less frequent observations, fewer detections of S0-38 and S0-104 were available in $H$-band, and unfortunately, these stars were confused with another source in the corresponding epochs. For that reason, we computed the expected $H$ magnitude from the measurement in $K^\prime$, using an assumption on the intrinsic color of the star and the mean color excess measured on S0-2 (see section~\ref{subsec:extinction}). S0-38 is identified as a late-type star \citep{Gillessen2009_S38_spectral_type}, so we assumed $(H-K^\prime)_{\rm int} = 0.1 \pm 0.05$ for the intrinsic color. The spectral type of S0-104 is unknown, so we took $(H-K^\prime)_{\rm int} = 0.0 \pm 0.1$, which is approximately the combined mean for early and late-type stars observed in this region \citep{Do2013_intrinsic_colors}. The errorbars reported in Table~\ref{ConfMags} take these assumptions into account, and were propagated through our process for confusion correction.

\begin{figure*}[ht!]
\plotone{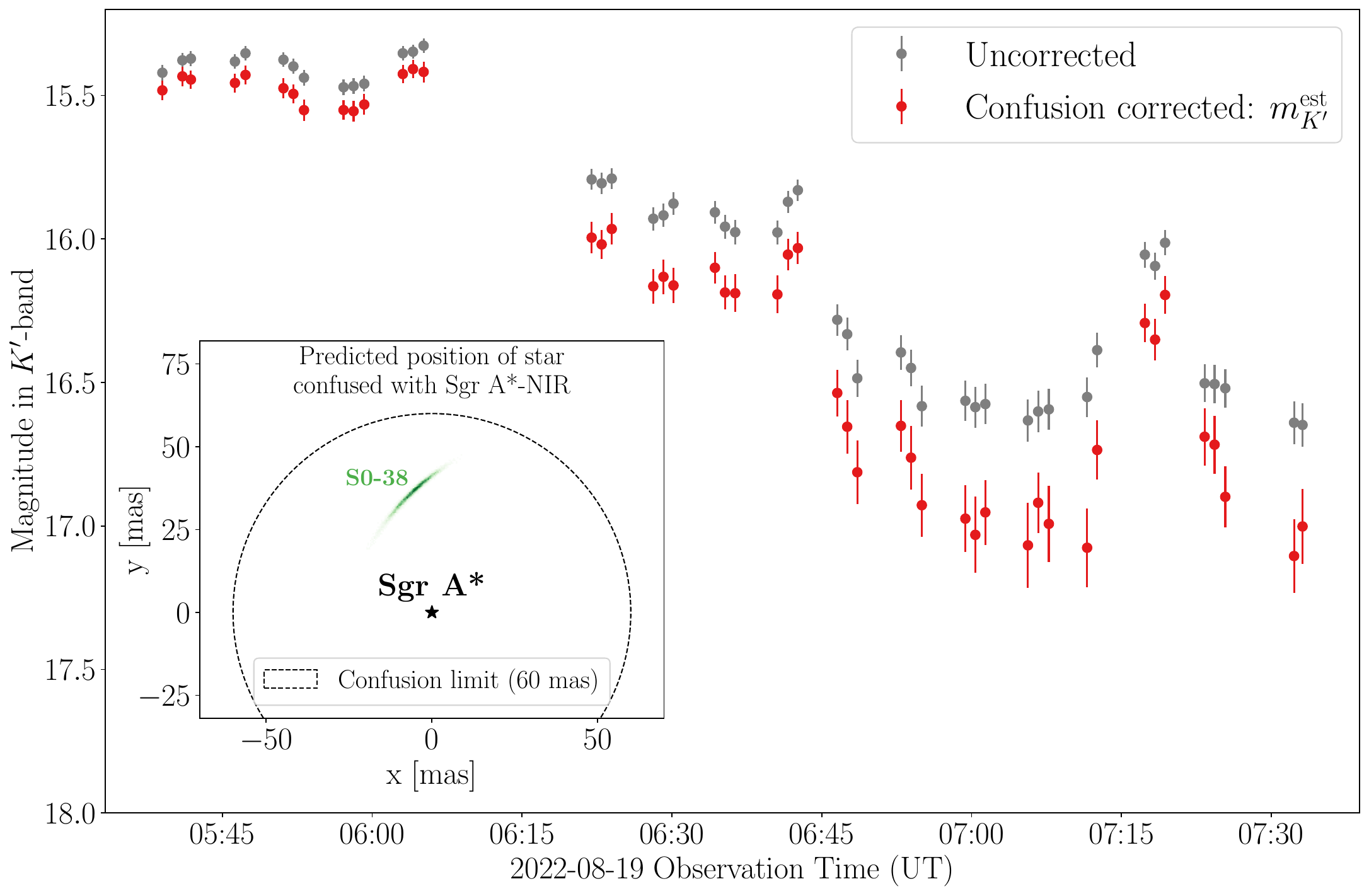}
\caption{Comparison of the uncorrected (grey) and confusion corrected (red) lightcurves for 2022 Aug 19. The inset shows the posterior on the predicted position (relative to Sgr~A*) of the star S0-38 confused with Sgr~A*-NIR, inferred from orbital fits to long-term GC astrometric data (see Appendix~\ref{Appendix: starplanting_sim}).} 
\label{fig:corrected_lightcurve}
\end{figure*}

To derive this photometric correction for confusion, we used star-planting simulations, building on the approach described in \cite{Weldon2023}. For each frame, many synthetic images were created by adding: (a) the background and stellar field inferred by \texttt{StarFinder} during its initial run, (b) Sgr~A*-NIR, injected at the known position of Sgr~A*, with various ‘‘planted’’ magnitudes, (c) the known star confused with Sgr~A*-NIR, randomly sampling its magnitude and position (making use of the values in Table~\ref{ConfMags}, and posteriors from orbital fits, respectively). For each frame and each ‘‘planted’’ magnitude of Sgr~A*-NIR, $18$ samples were realized. \texttt{StarFinder} was then run again on these images to obtain  ‘‘recovered’’ magnitudes for Sgr~A*. This resulted in a relation, for each frame, between the median recovered magnitude and the planted magnitude, as well as uncertainties on this relation.
Using this relation, we removed the flux contribution from the confusing sources and obtained corrected lightcurves with updated uncertainties.  Figure~\ref{fig:corrected_lightcurve} shows an example of such a corrected lightcurve. We note that it important to derive the relationship for each individual frames, because changes in the PSF can have a strong impact on the flux contribution from nearby sources. 
We find that, for some frames where the PSF quality is poor and Sgr~A*-NIR is faint, the correction can be unreliable, possibly making the flux of Sgr~A*-NIR negative. We removed these data points from the sample (see Table~\ref{NbObs}). Appendix~\ref{Appendix: conf_corr_starplanting} presents additional details about the star-planting simulations, confusion correction, and derivation of uncertainties. 

For 2019 May 13, no known sources are expected within the confusion limit of Sgr~A*. Regardless, we applied a ‘‘null’’ confusion correction, in order to ensure that all epochs were going through the same analysis steps. In the absence of a confusing source, this amounts to checking if a point source injected with the reported magnitude at the location of Sgr~A* is still detected by \texttt{Starfinder}. As Table~\ref{NbObs} shows, 18 $H$-band points (all having $m_H \sim 20$) did not fulfill this criterion and thus were removed from the sample.

\begin{figure}[ht!]
\centering  
\hspace{-0.5cm}
\includegraphics[width=1.04\linewidth]{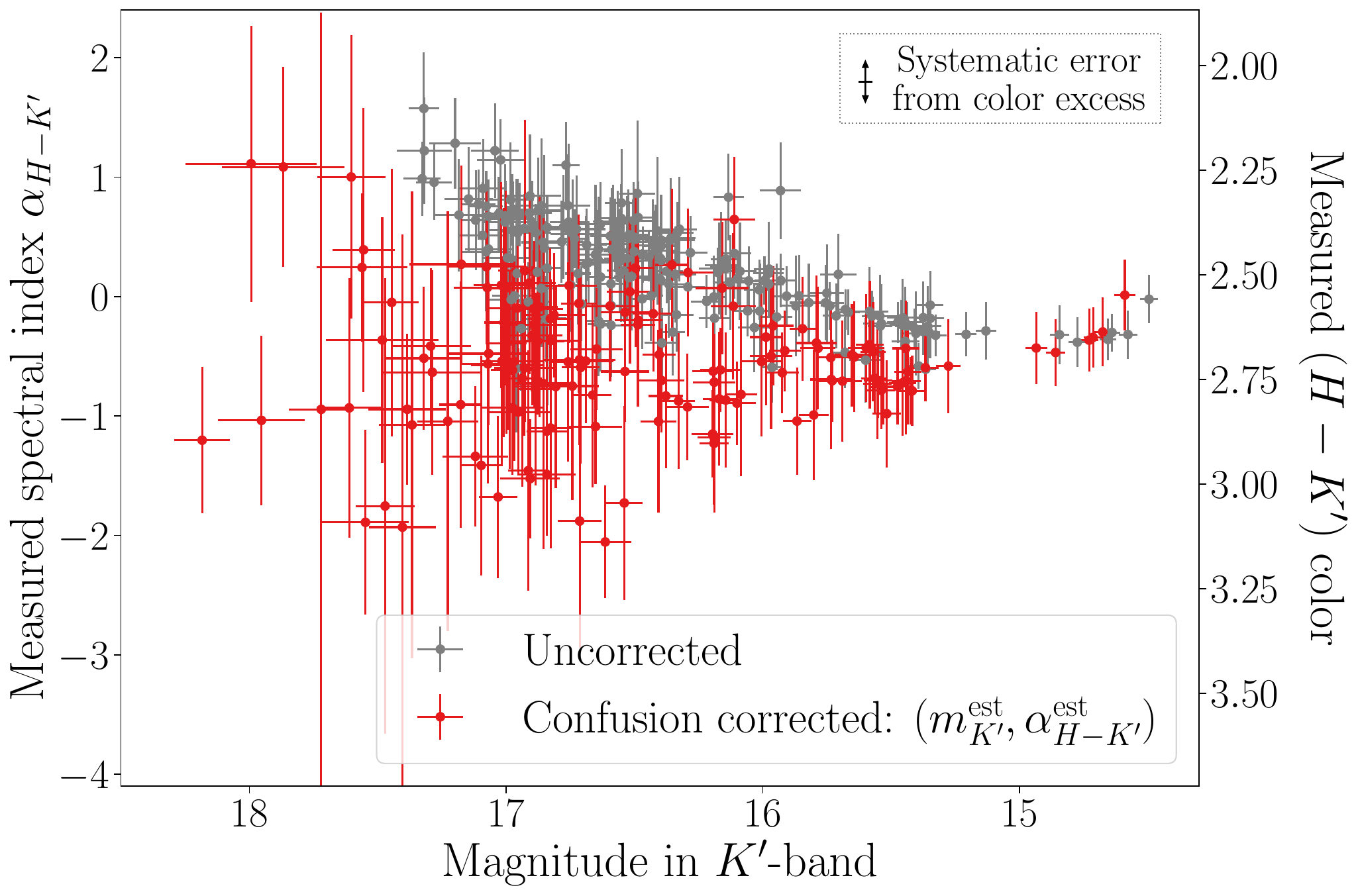}%
\caption{Comparison of the confusion corrected (red) and uncorrected (grey) spectral index measurements of Sgr~A*-NIR for the epochs when it is known to be confused (ie. all except 2019 May 13). Uncorrected spectral index values are systematically bluer and brighter, moving towards stellar values ($\alpha_* \sim 2$) when Sgr~A*-NIR becomes faint. This trend disappears after applying confusion correction. NB: here, the plotted spectral index uncertainties do not include the systematic error from the color excess, which is shown separately for reference.}
\label{fig:corrected_alpha}
\end{figure}

Despite its large computational cost, this correction step is justified  given the substantial color bias caused by stellar confusion. Stars have a spectral index $\alpha_* \sim 2$ that is very different from Sgr~A*-NIR, meaning that they contaminate the flux differently in the two bands. As a result, Sgr~A*-NIR's uncorrected spectral index is systematically bluer than its actual value. This effect becomes increasingly important as Sgr~A*-NIR becomes faint, creating an apparent trend (bluer when fainter) which does not reflect any intrinsic physical change in the accretion flow's emission. The confusion correction method presented here efficiently removes this bias (see Figure \ref{fig:corrected_alpha}).
In our dataset, the stars confused with Sgr~A* are relatively faint (see Table~\ref{ConfMags}), but confusion has a discernable impact on the measured spectral index even for bright states (e.g., $|\Delta \alpha| \sim 0.3$ for $m_{K^\prime} \sim 15.5$). The subscript or superscript ‘‘$\rm est$’’ is used throughout this paper to emphasize that the final ‘‘measured’’ quantities (e.g. magnitudes, color) are actually estimated after this confusion correction step (and, when relevant, after interpolation - see below).

\vspace{-2\baselineskip}
\subsection{Gaussian process interpolation}
\label{subsec: GP interp}

In theory, to accurately characterize the variations in $H-K^\prime$ color/spectral index, observations in both bands have to be taken on a timescale much smaller than the variability timescale. This is not feasible with Sgr~A* given its high-frequency NIR variability: large changes in brightness occur on timescales $\sim 1-10$ min \citep{Do2019flare}. Instead, we obtained simultaneous magnitude estimates by interpolating the lightcurves in $H$ and $K^\prime$.

We employed Gaussian Process (GP) regression, a non-parametric supervised learning method that makes probabilistic predictions of interpolated values. The main benefit is that this method naturally estimates the uncertainties on predicted values. 
The most important components of GPs are kernels (also called covariance functions), which specify the covariance between pairs of points and fully describe the GP. Kernels usually have hyperparameters that are fitted during the regression.

In this work, we used a combination of three common kernels: a Radial Basis Function (squared-exponential) kernel for long-term trends, an exponential kernel for short-term variations, and a white noise kernel in order to account for scatter and measurement uncertainties. Additional details on the mathematical framework of GP interpolation, as well as the description of kernels employed in this work, are included in Appendix~\ref{Appendix: GP math}.

Naïvely, one would perform a GP interpolation on the lightcurve in $K^\prime$-band (which is a real-valued function of the time variable), run a similar procedure in $H$-band ; then obtain the spectral index from equation~(\ref{eq:SpecIndex}), where the magnitudes in both bands would be computed for the same time. This approach, however, has a major shortcoming: it does not incorporate any correlation between the two bands. We expect indeed that when Sgr~A*-NIR gets brighter in one of the bands, it also gets brighter in the other (with the relative increase linked to the spectral index) since emission increases overall. This information is not exploited by two independent single-output GPs, leading to significant scatter in the inferred spectral index.


For this reason, instead of interpolating the two lightcurves separately, we used Multi-Output Gaussian Processes (MOGP) - more specifically the linear model of coregionalization (LMC) - to do a joint interpolation. The broad idea is to generalize GP interpolation from real to vector-valued functions - now relying on matrix-valued kernels, with off-diagonal terms describing the correlation between the different outputs (see \cite{Alvarez_MOGP} for an extensive review). In the astrophysics literature, GP regression can be found in a variety of contexts (see \cite{GP_astro_review} for a review), but MOGPs are not yet widespread despite promising applications in exoplanetary science \citep{MOGP_exoplanets, MOGP_exoplanets_2} or other disciplines \citep[e.g., robotics or neuroscience, ][]{MOGP_robotics, MOGP_neuroscience}. The mathematical foundations of this method are also briefly introduced in Appendix~\ref{Appendix: MOGP math}.


We emphasize that this technique does not prescribe the amount of correlation between the two outputs - which, in our setup, would be equivalent to choosing a value for the spectral index implicitly. Instead, the covariance between the two bands is allowed to vary as a function of the interpolation variable (time here) ; and is optimized during the MOGP regression \citep{Alvarez_MOGP_2}.

We also highlight that, in this work, interpolation was performed in magnitude space and not in flux space. 
First, the spectral index is a linear function of the magnitudes (see equation~(\ref{eq:SpecIndex})). Therefore, the coregionalization (linear by definition) will more naturally describe the correlation between the outputs in magnitude space.
Second, the uncertainties are more accurately represented as fractional values on flux (ie. absolute values on magnitude) than by noise independent of flux. Interpolation with a white noise kernel in magnitude space does not capture the exact dependence from equation~(\ref{Eq:flux_err}), but it is more accurate than doing the same in flux space.

We performed our MOGP regression with the Python package \texttt{GPy} \citep{gpy2014}, which already provides an implementation for the LMC and many kernels (see Appendices~\ref{Appendix: GP math} and \ref{Appendix: MOGP math}). Figure~\ref{fig:lightcurve} presents an example of two lightcurves (in $H$ and $K^\prime$, during the same night) being jointly interpolated. We remark that each epoch was interpolated independently.

To assess the performance of our MOGP method, we re-interpolated the lightcurves many times, dropping points one at a time. Then, we compared the predicted value to the actual measurement, and checked that the leave-one-out errors were distributed as expected. This test, detailed in Appendix~\ref{Appendix: GP tests}, indicates that the MOGP makes magnitudes predictions that are consistent with the observed values (within uncertainties). 


Still, to mitigate any potential biases, we applied a  ‘‘close-point filter’’ after interpolating, ie., for our final analysis of spectral index, we only kept points for which: (a) there was an observation in one of the bands, and (b) there was an observation in the other band within $\Delta t = 90 s$. We also discarded the first and last group of points in each epoch, so that interpolation was only used in-between actual measurements. The number of points left after applying these criteria are presented for each night in Table~\ref{NbObs}.

Typical uncertainties on magnitude from the interpolation are in the range [0.04, 0.17] mag, compared to [0.02, 0.15] mag for the original photometric errors (see Table~\ref{OoMcorrection}). The uncertainties on spectral index added by interpolation, also reported in Table~\ref{OoMcorrection}, are compared in more detail to the uncertainties added by confusion correction in  Appendix~\ref{Appendix: uncertainty_comparison}: for bright points (resp. faint points), interpolation (resp. confusion correction) is the dominant source of added uncertainty.

\subsection{Bright subset and extended dataset}
\label{subsec: mag cuts}

\begin{figure*}[ht!]
\centering
\includegraphics[width=0.88\linewidth]{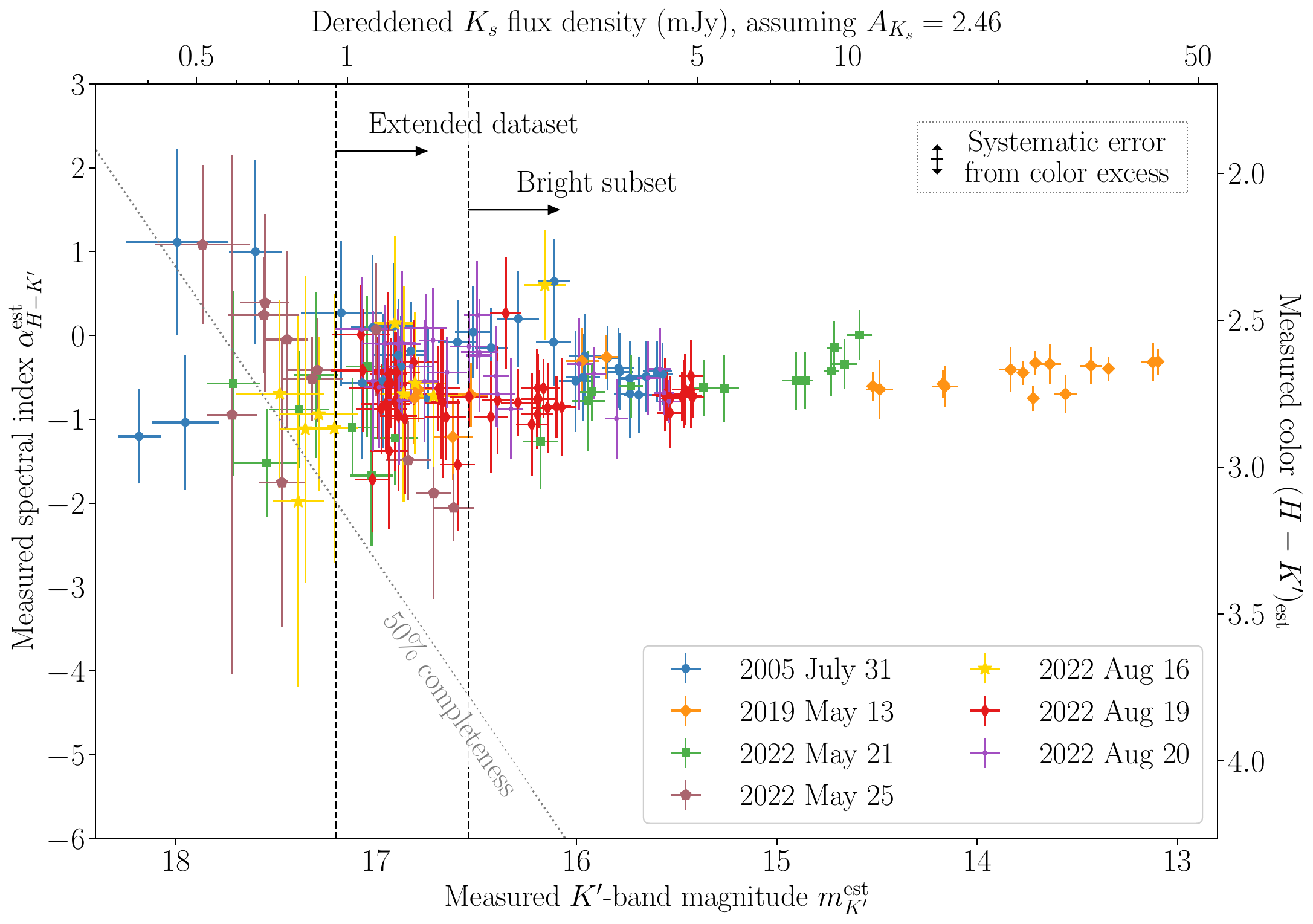} 
\caption{Measurements of the $H-K^\prime$ spectral index/color of Sgr~A*-NIR plotted against the measured magnitude in $K^\prime$. The Strehl ratio cuts, confusion correction, and interpolation filters have been applied. The dashed black lines show the location of the magnitude cuts for the bright subset and the extended dataset. The dotted grey line corresponds to $m_H^{\rm est}=20.3$, which is our estimate for the 50\% completeness limit. The plotted uncertainties on $\alpha$ do not include the systematic error from the color excess, which is shown separately for reference. The data points plotted here are also available as a machine-readable table.} 
\label{fig:alpha_vs_mag}
\end{figure*}

After the interpolation, we considered two different samples: (1) a bright subset, formed by points with $m_{K^\prime}^{\rm est}< 16.5$, and (2) and extended dataset, defined by $m_{K^\prime}^{\rm est}< 17.2$. 
More specifically, we performed the MOGP regression on the full lightcurves, then discarded points fainter than the chosen threshold before applying the interpolation filters (see section~\ref{subsec: GP interp}). The number of points surviving this procedure for each night are presented in Table~\ref{NbObs}. Figure~\ref{fig:alpha_vs_mag} shows all the spectral index measurements as a function of magnitude, along with the location of the magnitude cuts, indicating which data points belong to the two samples. These cuts were chosen to ensure completeness, as well as to minimize the impact of systematic uncertainties.

First, because confusion happens for 6 out of the 7 epochs, completeness is largely determined by our ability to correct for this effect. 
We found in the star-planting simulations that there was a frame-dependent limiting intrinsic magnitude $m_{\rm lim}$ for Sgr A*-NIR, fainter than which confusion correction became impossible (see Appendix~\ref{Appendix: conf_corr+err}). Confusion correction is more difficult in $H$, since Sgr~A*-NIR is fainter (relative to the stars) than in $K^\prime$, so our spectral index measurements are limited by $H$-band completeness. The median value of limiting magnitude for Sgr A*-NIR was $\tilde{H}_{\rm lim}\approx 20.3$ (corrected), meaning that confusion correction was possible up to this value for at least 50 \% of the frames. Therefore, we adopted a 50 \% completeness limit at $m_H^{\rm est}=20.3$ (see Figure~\ref{fig:alpha_vs_mag}).

To compare with the literature, however, we need to examine spectral index variation within a $K$-band flux range. Thus, we chose cuts in $K^\prime$, but making sure that observations were not limited by $H$-band completeness. The first cut at $m_{K^\prime}^{\rm est} = 16.5$ is conservative: it ensures that, even if Sgr~A*-NIR is a very red source ($\alpha_s \approx -4$, or equivalently, $(H-K^\prime)_{\rm est} \approx 3.7$), it would be detected in both bands. In addition, any results obtained with this subset are very robust, because using only bright points mitigates known sources of systematic uncertainties (fainter detections are kept during interpolation, but their effect is small due to the interpolation filters, see section~\ref{subsec: GP interp}). 

First, when Sgr~A*-NIR is bright, the flux removed by confusion correction (outlined in section~\ref{subsec:confusion correction}) is smaller. Since the flux contribution from the stars confused with Sgr~A*-NIR varies with the epoch and PSF (ie. is unique for each frame), we quantify this using upper limits. For points with $m_{K^\prime}^{\rm est} \leq 16.5$, we estimated that confused sources account for $\leq 40 \%$ of the detected $K^\prime$ flux (ie. $\Delta m_{K^\prime} \leq 0.55$). Furthermore, with the mean color of Sgr~A*-NIR $(H-K^\prime)_{\rm est} \sim 2.7$, $m_{K^\prime}^{\rm est} \sim 16.5$ corresponds to $m_H^{\rm est} \sim 19.2$ ; and if $m_H^{\rm est} \leq 19.2$, confused sources account for $\leq 50 \%$ of the detected flux in $H$-band (ie. $\Delta m_H \leq 0.75$).

Second, we expect less photometric uncertainty when Sgr~A* is bright -  because \texttt{StarFinder} is more accurate (in astrometry and photometry), and because confusion has a relatively smaller impact. This results in reduced scatter for the measured spectral indices. By mapping the final photometric uncertainty (ie., after confusion correction)  against the measured magnitude of Sgr~A*-NIR, we estimated that for the bright subset, $\lesssim 10 \%$ (resp. $\lesssim 20 \%$) uncertainty on flux on average could be achieved in $K^\prime$ (resp. $H$).

The observed flux distribution of Sgr~A*-NIR peaks around $m_{K^\prime} \sim 17$ \citep{Weldon2023}, so the cut at $m_{K^\prime}^{\rm est} = 16.5$ has the disadvantage of removing almost 50\% of the available measurements (see Table~\ref{NbObs}). In order to increase this fraction, we applied another less conservative magnitude cut at $m_{K^\prime}^{\rm est} = 17.2$, removing less than 15\% of the points after interpolation. Because of the limit at $m_H^{\rm est}=20.3$, observations at $m_{K^\prime}^{\rm est} = 17.2$ would be complete only for colors $(H-K^\prime)_{\rm est}\leq 3.1$ (or equivalently, $\alpha_s \geq -2$). This extended dataset is unlikely to suffer from selection effects, however, since among all points with $m_{K^\prime}^{\rm est} < 17.2$,  there are none with a color $(H-K^\prime)_{\rm est}\geq 3.1$ (see Figure~\ref{fig:alpha_vs_mag}). This point is discussed more rigorously in section~\ref{subsec: extended dataset result}, building on the results obtained from the bright subset.

\section{Models}
\label{sec: model}

This section introduces the empirical model used to characterize spectral index variations with brightness. This model can account for three different physical pictures specifying Sgr~A*-NIR's spectral index  (section~\ref{subsec: instrinsic alpha}). In addition, it includes the effect of noise and the potential presence of a background (section~\ref{subsec: alpha model}). We also review the Bayesian inference framework (section~\ref{subsec:Bayesian_framework}) employed to constrain the model parameters.

\subsection{Intrinsic spectral index models}
\label{subsec: instrinsic alpha}

The variability of Sgr~A* across the electromagnetic spectrum is a puzzle, especially in the near-infrared. The NIR flux distribution, spectral index variations or absence thereof, and relation with other wavelength regimes (e.g. X-ray and radio) should be explainable within a consistent physical picture - on which there is currently no consensus. Studies of the spectral index variations are usually driven by observations rather than derived from a physical model, but even the empirical descriptions are still debated. In this section, we discuss three such scenarios, each having different implications for the physics of Sgr~A*'s emission. 




\subsubsection{Constant spectral index model}
\label{subsubsec:ConstAlpha}

In the simplest description, Sgr~A*-NIR's spectral index does not vary with brightness. Some studies have found no significant fluctuations or correlation with flux \citep{Trap2011, Witzel2014}. For example, one of the first investigations of this question \citep{Hornstein2007} found that the spectral index is independent of the flux level and wavelength in the near-infrared, with a value of $\alpha_s = -0.63 \pm 0.16$ (weighted average over three color pairs: $H-K^\prime$, $K^\prime-L^\prime$, $L^\prime-M_s$). The value for $H-K^\prime$ specifically is also reported in this work: $\alpha_s = -0.88 \pm 0.33$.


This first picture implies interesting constraints for the physical process responsible of Sgr~A*'s emission: that process needs to generate large NIR intensity changes while keeping the spectral index constant. Assuming synchrotron emission, this would suggest that that the mechanism driving the observed flares leaves the shape of the electron energy distribution unchanged, altering only the normalization (ie. the total number of electrons in the relevant energy range for NIR emission).

These claims of a constant spectral index are disputed by other investigations, which find a strong dependence of the spectral index on the intensity of the NIR emission \citep{Gillessen2006, Bremer2011, Ponti2017}. Sections~\ref{subsubsec:Expcutoff} and~\ref{subsubsec:TwoState} present two descriptions that would allow such variations. 

\subsubsection{Exponential cutoff model}
\label{subsubsec:Expcutoff}

In order to jointly explain the observed $K$-band and $M$-band (4.5 $\mu$m) flux distributions, \cite{Witzel2018} proposed a model where the spectral index $\alpha_s$ depends linearly on $\log( F_{K})$. More specifically, they argue that this dependence naturally arises when assuming a log-normal flux distribution in the two bands, then matching the cumulative distribution functions (ie., making the lowest $n$\% in K-band correspond with the lowest $n$\% in M-band for all $n$).  The dependence could reconcile studies that find spectral index values around $\alpha_s \approx -0.6$ for bright Sgr~A*-NIR phases \citep{Ghez2005_color,  Krabbe2006, Hornstein2007, Witzel2014}, and those claiming $\alpha_s \sim -3$ for periods of lower emission \citep{Eisenhauer2005, Gillessen2006, Bremer2011, GRAVITY2021_spectral_index}.

The linear dependence on log-flux is also motivated by the following physical picture. The synchrotron spectrum is usually expressed (in the optically thin regime) as a power-law with an exponential cutoff above some frequency $\nu_0$, corresponding to a similar shape for the distribution of electron energies (or equivalently, of Lorentz factors $\gamma$). For typical electron energy cutoffs and magnetic field strengths \citep[$\gamma_0 \sim 10^3$, $B \sim 40$ G,][]{Dodds-Eden2009}, the cutoff frequency is close to the NIR range ($\nu_0 \sim \frac{3qB}{4\pi m_e c} \gamma_0^2 \sim 10^{14}$ Hz). Thus, Sgr~A*'s NIR variability could be caused by a process that shifts the cutoff energy of the electron distribution (and not the normalization ie. the total number of electrons). This would explain the changes in NIR flux density and, at the same time, make the spectral index vary with flux. If the NIR-variability is due primarily to this effect, we would observe a linear dependence of the spectral index on the log-flux \citep{Witzel2018}.

This scenario, and the one discussed in section~\ref{subsubsec:ConstAlpha}, can both be described by a parameterizing Sgr~A*-NIR's intrinsic spectral index as a linear function of magnitude:
\begin{multline}
    \alpha_s({H-K^\prime}) = -0.4 \xi \cdot (m_{K^\prime,\rm SgrA}^{\rm obs} - m_{K^\prime}^0) + \eta \\ \text{ with } m_{K^\prime}^0 \equiv 15.8
    \label{eq:linear param}
\end{multline}
where $m_{K^\prime,\rm SgrA}^{\rm obs}$ is the (reddened) magnitude of Sgr~A*-NIR alone (ie. after the background contribution and noise have been removed, see section~\ref{subsec: alpha model}) ; and we choose the value of $m_{K^\prime}^0$ (origin of magnitudes) close to the weighted average magnitude in our sample, in order to make the interpretation of $\xi$ and $\eta$ easier. We note that in this work, $\alpha, \xi$ and $\eta$ have been defined differently compared to \cite{Witzel2018}

The claim of a constant spectral index value \citep{Hornstein2007} corresponds to $\xi=0$ and in that case, $\eta$ is the constant value. Conversely, the exponential cutoff model of \cite{Witzel2018} predicts

\begin{equation}
\begin{split}
    \xi & = \frac{\sqrt{\lambda_{K^\prime}/\lambda_H} - 1 }{\log_{10}(\lambda_{K^\prime}/\lambda_H)} \\
    \eta & = - \tilde{\alpha} \left[ 1+ \xi \log_{10}(\nu_{K^\prime}/\tilde{\nu}) \right] - \xi\log_{10} \left( \frac{F_{\tilde{\nu}}}{f_{0, K^\prime}}\right) \\
    & \qquad + 0.4 \xi (A_{K^\prime} - m_{K^\prime}^0)
\end{split}
\label{Eq:exp_cutoff_pred}
\end{equation}

Quantities bearing a tilde in equation~(\ref{Eq:exp_cutoff_pred}) represent values in the synchrotron power-law regime, ie. at a frequency $\tilde{\nu} \ll \nu_0$, with $\nu_0$ the cutoff frequency. One can use, for example,  measurements of the sub-millimeter emission of Sgr~A*: observations at $\tilde{\nu}= 868$ GHz with ALMA yield $ F_{\tilde{\nu}} \sim 2 \text{ mJy}$, and $ \tilde{\alpha} \approx -0.3$ \citep{Submm_obs}. Assuming a value for extinction of $A_{K^\prime}\approx 2.5$ \citep{Schodel2010_extinction, Fritz2011_extinction}, we can estimate $\eta \sim -2$; but the calculation uses several quantities with loose constraints, thus this expected value is not very precise. 
The prediction for $\xi$, on the other hand, holds much less uncertainty: with central wavelength values for $H$ and $K^\prime$, we find $\xi \approx 1.23$ for the \cite{Witzel2018} model. Henceforth, we will refer to any model with $\xi = 1.23$ as a ‘‘exponential cutoff model’’ regardless of the value of $\eta$.

\subsubsection{Two-state model}
\label{subsubsec:TwoState}

Variations with flux of the observed spectral index can result from  another scenario, where the emission from Sgr~A*-NIR is physically different when it is in quiescence, compared to when it is bright. 
\cite{Dodds-Edden2011_2states} proposed two states: (1) a quiescent state defined by a continuously present, variable component with a log-normal flux density distribution; (2) a flaring state responsible for the high-flux tail of the flux distribution. They claim that the transition between the two states starts at $F_{K_s}^{\rm dereddened} \sim 5$~mJy (value de-reddened with $A_{K_s} = 2.5$), with a median flux in the quiescent state of $F_{K_s}^{\rm dereddened} \approx 1.1$ mJy. The conversion to observed $K^\prime$ magnitudes for Sgr~A*-NIR can be performed using $F_{K_s}^{\rm obs} = 1.09 F_{K^\prime}^{\rm obs} $ \citep{Chen2019}, such that:


\begin{equation}
\begin{split}
    F_{K_s}^{\rm dereddened} &= 1.09 F_{K^\prime}^{\rm obs} 10^{0.4 A_{K_s}}\\
    &=1.09 f_{0, K^\prime} 10^{-0.4(m_{K^\prime}-A_{K_s})}
\end{split}
\label{Eq:Ks_Kp_conversion}
\end{equation}

yielding $m_{K^\prime}\sim 15.5$ and $m_{K^\prime}\sim 17$, respectively. A more recent analysis of the flux distribution of Sgr~A* within this model \citep{Gravity2020_twostate} fits a log-normal distribution to the quiescent state, with an expected log-flux value $\mu_{ln} = -0.21 \pm 0.23$ ie. a median flux $F_{K}^{\rm dereddened} = 0.81^{+0.21}_{-0.17}$ mJy (assuming $A_{K_s} = 2.43$). This corresponds to $m_{K^\prime} = 17.65 \pm 0.25$, using equation~(\ref{Eq:Ks_Kp_conversion}) and $F_K^{\rm obs} = 1.29 F_{K_s}^{\rm obs} $ \citep{Chen2019} to convert to $K^\prime$.

This third scenario could naturally explain a difference in spectral index between bright and faint flux levels (as claimed by some studies): if the spectral index of the flaring state is different from the spectral index of the quiescent state, one expects a transition between those two regimes depending on the brightness of Sgr~A*-NIR. 

Such a two-state description cannot be expressed with the intrinsic spectral index law from equation~(\ref{Eq:exp_cutoff_pred}), but it will be incorporated using other parameters in the full empirical model (see section~\ref{subsec: alpha model}).

\subsection{Empirical model}
\label{subsec: alpha model}

\begin{figure*}[p]
\vspace{\baselineskip}
\begin{center}
    \includegraphics[width=\linewidth]{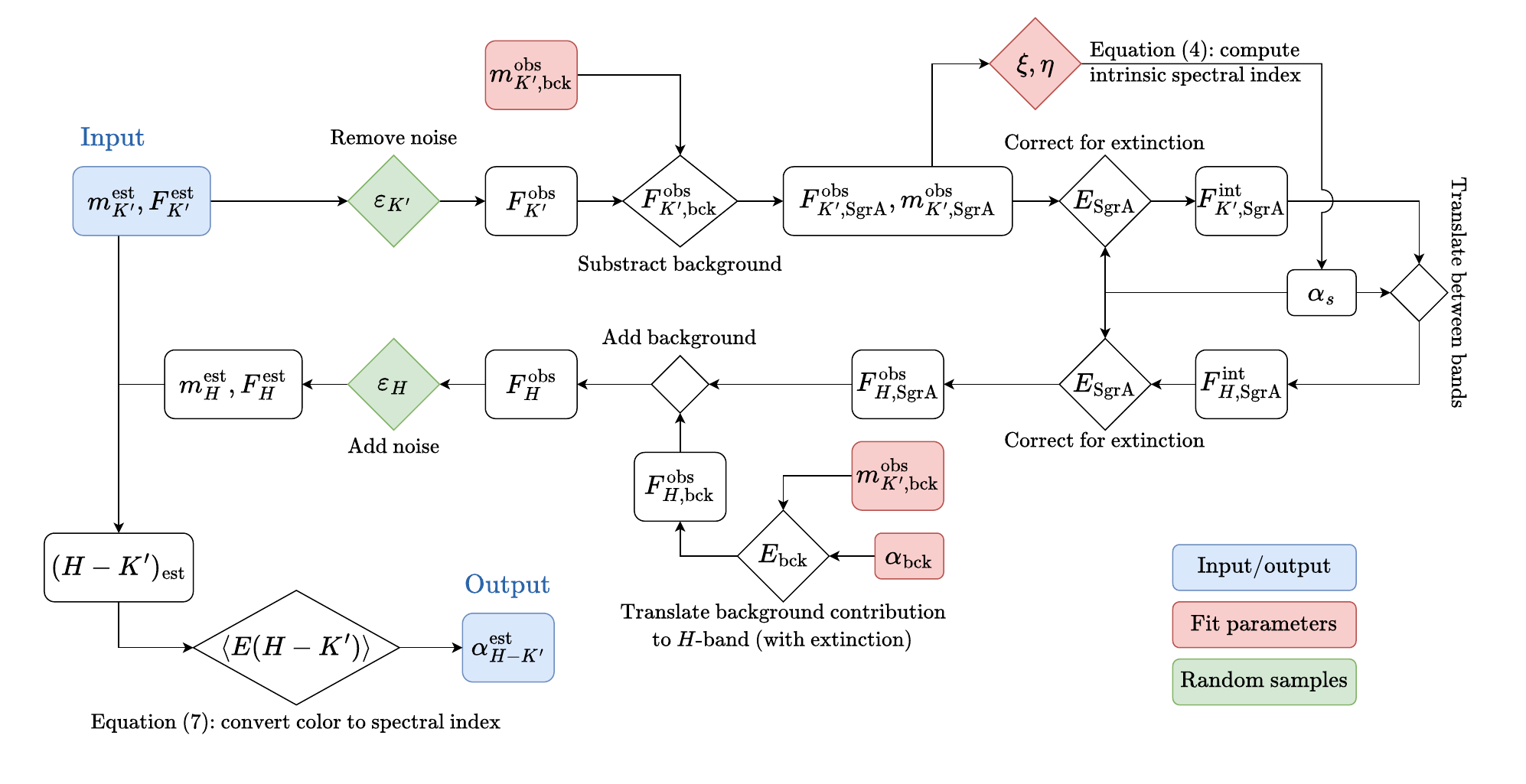} 
\end{center}
\vspace*{-0.5\baselineskip}
\begin{align}
\tag{\ref{eq:linear param}}
\alpha_s &= -0.4 \xi \cdot (m_{K^\prime,\rm SgrA}^{\rm obs} - m_{K^\prime}^0) + \eta \text{ with } m_{K^\prime}^0 \equiv 15.8\\
\notag
 & \\
 \alpha_{H-K^\prime}^{\rm est} &=\frac{-0.4[(H-K^\prime)_{\rm est} - \langle E(H-K^\prime) \rangle] + \log_{10}(\frac{f_{0,H}}{f_{0, K^\prime}})}{\log_{10}(\lambda_{K^\prime}/\lambda_H)}  \equiv \mathcal{F}  ( \varepsilon_H; \varepsilon_{K^\prime}, m_{K^\prime}^{\text{est}},  \boldsymbol{\theta_s})
 \tag{\ref{eq:SpecIndex}}
\end {align} 
 
\begin{equation}
\begin{split}
\text{where }K^\prime_{\rm est} &=  m_{K^\prime}^{\text{est}} \\
H_{\rm est} &= -2.5 \log_{10} \left( \frac{F_{H, \rm SgrA}^{\rm obs} +  F_{H, \rm bck}^{\rm obs} + \varepsilon_H}{f_{0,H}} \right) \\
F_{H, \rm SgrA}^{\rm obs} &=  F_{K^\prime, \rm SgrA}^{\rm obs} (\lambda_H/\lambda_{K^\prime})^{\alpha_s} 10^{-0.4 E_{\rm SgrA}} \\
&= \left[f_{0, K^\prime} (10^{-0.4 m_{K^\prime}^{\text{est}}} -  10^{-0.4 m_{K^\prime, \rm bck}} )-  \varepsilon_{K^\prime} \right] \left( \frac{\lambda_H}{\lambda_{K^\prime}} \right)^{\alpha_s} 10^{-0.4 E_{\rm SgrA}}\\
F_{H, \rm bck}^{\rm obs} &= F_{K^\prime, \rm bck}^{\rm obs} (\lambda_H/\lambda_{K^\prime})^{\alpha_{\rm bck}} 10^{-0.4 E_{\rm bck}} \\
&= f_{0, K^\prime} 10^{-0.4( m_{K^\prime, \rm bck}+ E_{\rm bck})} \left( \frac{\lambda_H}{\lambda_{K^\prime}} \right)^{\alpha_{\rm bck}} \\
m_{K^\prime,\rm SgrA}^{\rm obs} &= - 2.5 \log_{10} \left(\frac{F_{K^\prime, \rm SgrA}^{\rm obs}}{f_{0, K^\prime}}\right) = - 2.5 \log_{10} \left(\frac{F_{K^\prime}^{\text{est}} -  F_{K^\prime, \rm bck}^{\rm obs} - \varepsilon_{K^\prime}}{f_{0, K^\prime}}\right)
\\ & = m_{K^\prime}^{\text{est}} - 2.5 \log_{10} \left[ 1- 
 \frac{\varepsilon_{K^\prime} }{f_{0, K^\prime}} 10^{0.4m_{K^\prime}^{\text{est}}} - 10^{0.4(m_{K^\prime}^{\text{est}}- m_{K^\prime, \rm bck})} \right] \\
        E_{\rm SgrA} &= E(\alpha_s) \\
        E_{\rm bck} &= E(\alpha_{\rm bck}) \\
        E (\alpha) &= E(H-K^\prime)_* + \kappa (\alpha -2).
\end{split}
\label{eq: subfunctions_alpha_formula}
\end{equation}

\caption{Flowchart representing the empirical model described in section~\ref{sec: model}, relating the measured spectral index ($\alpha_{H-K^\prime}^{\rm est}$) to the measured magnitude in $K^\prime$ band ($m_{K^\prime}^{\rm est}$). The free model parameters are highlighted in red. In green, we show the random variables that are sampled to account for noise. The equations constituting the model are gathered below the flowchart. Subscript/superscript ‘‘est’’  denotes the measured quantities (with noise), ‘‘obs’’ the quantities at the observer location (ie. without noise, but not corrected for extinction), ‘‘SgrA’’ denotes the quantities for Sgr~A*-NIR alone, and ‘‘bck’’ for the background alone. The following values are fixed: $\langle E(H-K^\prime) \rangle = 2.07$ for the mean color excess of Sgr~A* (see section~\ref{subsec:extinction}), and $\kappa = 0.0077 $, $E(H-K^\prime)_* = 2.09 $ for the color excess law (see Appendix~\ref{Appendix: Extinction}).}
\label{fig: model_graph}
\end{figure*}

In section~\ref{subsec: instrinsic alpha}, we discussed Sgr~A*-NIR's \ intrinsic spectral index $\alpha_s$, but the value $\alpha_{H-K^\prime}^{\rm est}$ obtained with equation~(\ref{eq:SpecIndex}) on actual measurements can differ, due to a potential background or noise-induced scatter. We now present an empirical model expressing this measured spectral index as a function of the measured $K^\prime$-band magnitude. Figure~\ref{fig: model_graph} illustrates the components of this model as a flowchart, and gathers all the corresponding equations.

The empirical model employs the following scheme (with four parameters, highlighted in red in Figure~\ref{fig: model_graph} and summarized in Table~\ref{ParamsPriors}) to compute the measured spectral index from measured magnitudes in $K^\prime$-band. Starting from a measured magnitude $m_{K^\prime}^{\rm est}$, we sample a random noise value and substract it from the flux, then also remove the flux contribution from the background (see below). This results in the $K^\prime$ flux for Sgr~A*-NIR alone, which we can then translate to $H$-band using the intrinsic spectral index from equation~(\ref{eq:linear param}), correcting for extinction via the color excess. We get a measured $H$ magnitude by adding the background contribution and random noise to the Sgr~A*-NIR $H$-band flux. Finally, the measured magnitudes in both bands are plugged in equation~(\ref{eq:SpecIndex}), yielding a measured spectral index $\alpha_{H-K^\prime}^{\rm est}$. 

\begin{deluxetable}{CCC}[hbtp]
    \tablehead{\colhead{Name} & \colhead{Definition} & \colhead{Prior} }
    \startdata
        \xi &  \text{Slope of the linear relation}  &  \mathcal{U}(-1, 2.5)\\
         &  \alpha_s = f(m_{K^\prime,\rm SgrA}^{\rm obs}) &  \\
         \eta &  \text{Spectral index value at } m_{K^\prime}^0 = 15.8 &  \mathcal{N}(-0.5, 1) \\
          m_{K^\prime}^{bck} &  \text{Magnitude of the background in } K^\prime &  \mathcal{U}(17, 22)  \\
          \alpha_{\rm bck} & \text{Spectral index of the background} & \mathcal{U}(-5, 5)
    \enddata
    \caption{Summary of the model parameters with their related priors. $\mathcal{N}(\mu,\sigma)$ denotes a Gaussian distribution with mean $\mu$ and variance $\sigma^2$, and $\mathcal{U}(a,b)$  a uniform prior between $a$ and $b$.}
    \label{ParamsPriors}
    \vspace*{-2\baselineskip}
\end{deluxetable}
\vspace*{-\baselineskip}

First, the model considers that the flux coming from Sgr~A*'s location can receive some contribution from sources other than the accretion flow - unresolved and unknown stars, for instance. This results in a transition, when Sgr~A*-NIR gets faint, from the spectral index $\alpha_s$ characterizing the accretion flow to some other value $\alpha_{\rm bck}$ describing the background emission.

Such contamination was invoked by \cite{Hornstein2007} to interpret the bluer spectral indices observed at low flux densities for 2005 July 31, using the same data (but different processing methods) from Keck as in this work. With our updated GC observations, we can now explain most of this blue background with confusion with S0-104 ($m_{K^\prime} \approx 16.75, m_H \approx 18.85$, see Table~\ref{ConfMags}). The existence of this star was only recently recognized \citep{Meyer2012}. In principle, there could be even fainter unidentified sources affecting the photometry of Sgr~A*-NIR, though they would have a smaller impact than S0-104. 

The added benefit of implementing a background into our model is that the formalism can be repurposed to describe the two-state model (see section~\ref{subsubsec:TwoState}). If the two states are powered by unrelated mechanisms, each would contribute to the flux separately (and a priori, with different spectral indices) - similar to a star and Sgr~A*-NIR being combined into a single, unresolved source. We note that this implies additive contributions from the two states, an assumption that may be inexact: \cite{Dodds-Edden2011_2states} mention, for instance, that the process responsible for flares could be triggered by the one producing the low-level variability. However, a simple model like this can serve as a first description of a transition between two spectral indices as the combined source gets faint. 

In order to keep the notations simple, we henceforth refer to both interpretations (unresolved stellar sources, or a quiescent state) as the \textit{background} ; and in the second case, all quantities pertaining to Sgr~A*-NIR in the text (e.g. $\alpha_s$) will coincide with the flaring state.

We assumed constant flux contributions $F_{K^\prime, \rm bck}^{\rm obs}, F_{H, \rm bck}^{\rm obs}$ from the background in the two bands ; and we described them using an equivalent magnitude in $K^\prime$-band ($m_{K^\prime}^{bck}$) and a spectral index ($\alpha_{\rm bck}$). A stellar background would yield $\alpha_{\rm bck} \sim 2$ ; whereas a quiescent state redder than the flaring state would correspond to $\alpha_{\rm bck} \sim -3$ \citep{Dodds-Edden2011_2states, Gravity2020_twostate}.

Second, we have to account for the uncertainties in our measurements, which can lead to significant scatter in spectral index values - in particular for faint magnitudes, when the photometry becomes less accurate. We assumed Gaussian white noise on flux (ie. uncorrelated, between data points and between bands). The variance $\sigma_{F_{\rm band}}^2$ (with $\text{band}=K^\prime, H$) was determined for each point, either from equation~(\ref{Eq:flux_err}), or propagated through confusion correction and interpolation (see section~\ref{sec:SpecIndexMeas}).

Including both the background contribution and the noise, the measured flux values can be written:
\begin{equation}
\begin{split}
    F_{K^\prime}^{\rm est} &= F_{K^\prime, \rm SgrA}^{\rm obs} +  F_{K^\prime, \rm bck}^{\rm obs} + \varepsilon_{K^\prime} \\
    F_H^{\rm est} &= F_{H, \rm SgrA}^{\rm obs} +  F_{H, \rm bck}^{\rm obs} + \varepsilon_H
\end{split}
\end{equation}

where $F^{\rm obs}$ denotes flux values at the observer location (not corrected for extinction) and $\varepsilon_{\rm band} \sim \mathcal{N}(0,\sigma_{F_{\rm band}})$.

\begin{figure*}[p]
\centering
\includegraphics[width=\linewidth]{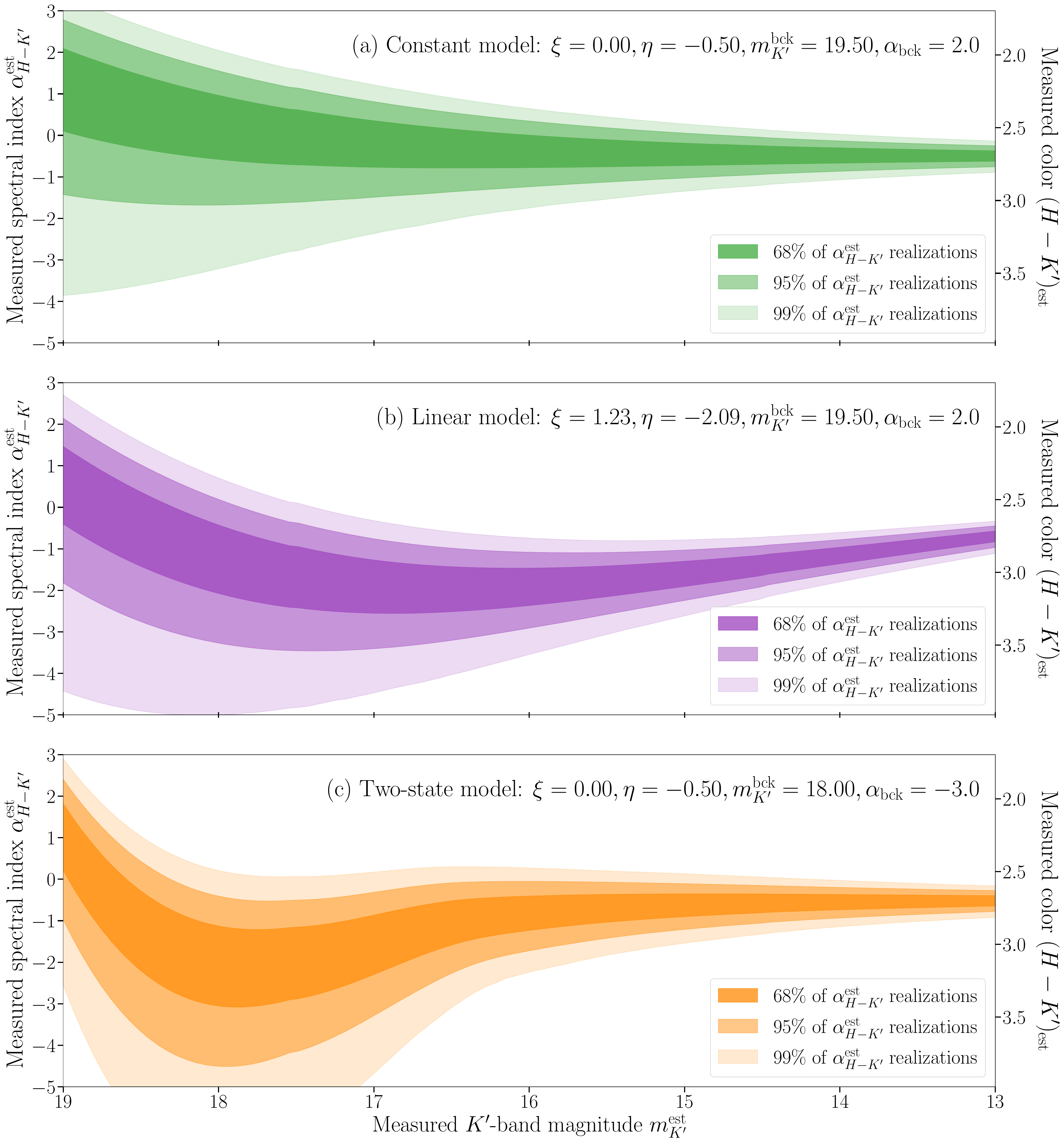} 
\caption{Predicted distribution of $H-K^\prime$ spectral index measurements at each measured magnitude in $K^\prime$, for a fixed set of model parameters $\boldsymbol{\theta_s}$. The contours are obtained by assuming some noise properties (here, equation~(\ref{Eq:flux_err}) with the mean values for $\log_{10}(C_{\rm band})$ and $\beta_{\rm band}$ displayed in Table~\ref{Noisefitvalues}), and sampling $\sim 10^5$ noise realizations for each measured magnitude $m_{K^\prime}^{\rm est}$. Three examples are presented, illustrating the various physical pictures discussed in section~\ref{sec: model}: (a) a constant spectral index for Sgr~A*-NIR, with a faint stellar background, (b) a spectral index for Sgr~A*-NIR varying linearly (with a slope $\xi =1.23$) as a function of magnitude, on top of a faint stellar background, (c) a two-state model of Sgr~A*-NIR, with a constant spectral index for the flaring state, transitioning into a redder value for the quiescent state.
We warn that the contours do not represent a joint 2D distribution on $(m_{K^\prime}^{\rm est}, \alpha_{H-K^\prime}^{\rm est})$, but rather a normalized conditional distribution on $\alpha_{H-K^\prime}^{\rm est}$ given a value $m_{K^\prime}^{\rm est}$.} 
\label{fig:alpha_contours}
\end{figure*}

Since photometric uncertainties are already estimated for each spectral index measurement (see section~\ref{sec:SpecIndexMeas}), the noise levels $\sigma_{\rm band}$ do not need to be part of the model for the fit to the data (see section~\ref{subsec:Bayesian_framework}). To generate plots showing the expected distribution of $\alpha_{H-K^\prime}^{\rm est} = f(m_{K^\prime}^{\rm est})$, however, we have to assume a noise law - ie. pick some values $C_{\rm band}, \beta_{\rm band}$ for equation~(\ref{Eq:flux_err}). For Figure~\ref{fig:alpha_contours}, the averages of $\log_{10}(C_{K^\prime})$, $\beta_{K^\prime}$, $\log_{10}(C_H)$, and $\beta_{H} $ over all seven epochs are used (see Table~\ref{Noisefitvalues}); and we display the confidence intervals for three examples of model parameters $(\xi, \eta, m_{K^\prime}^{\rm bck}, \alpha_{\rm bck})$, each representing one of the physical pictures discussed in section~\ref{subsec: instrinsic alpha}. 

We note that if Sgr~A*-NIR becomes faint enough for its $H$-band flux to be dominated by noise, then the measured spectral index will be completely anti-correlated with the $K^\prime$ magnitude by virtue of equation~(\ref{eq:SpecIndex}). This translates into a spurious trend at the faint end of the predicted $\alpha_{H-K^\prime}^{\rm est} = f(m_{K^\prime}^{\rm est})$ plot - for instance, for $m_{K^\prime}^{\rm est}\gtrsim 18$ in the bottom panel of Figure~\ref{fig:alpha_contours}.

\subsection{Bayesian inference framework}
\label{subsec:Bayesian_framework}

Our goal is to infer the posterior distribution 

\begin{equation}
\begin{split}
    p \Big( \boldsymbol{\theta_s} \Big| & \{ \alpha_{H-K^\prime}^{\text{est}, (i)}, m_{K^\prime}^{\text{est}, (i)} , \sigma_{F_{K^\prime}}^{(i)}, \sigma_{F_H}^{(i)} \} \Big) \\ 
    & \propto \pi(\boldsymbol{\theta_s}) \times \mathcal{L} \left( \{ \alpha_{H-K^\prime}^{\text{est}, (i)}\} \Big| \{m_{K^\prime}^{\text{est}, (i)} , \sigma_{F_{K^\prime}}^{(i)}, \sigma_{F_H}^{(i)} \} ; \boldsymbol{\theta_s} \right)
\end{split}
\end{equation}

where $\boldsymbol{\theta_s} = (\xi, \eta, m_{K^\prime}^{\rm bck}, \alpha_{\rm bck})$ is the set of model parameters described in sections \ref{subsec: instrinsic alpha} and \ref{subsec: alpha model}, $\pi(\boldsymbol{\theta_s})$ is the prior on this parameter space and $\mathcal{L}$ is the likelihood.

We used independent, relatively broad priors on all the parameters. For $\xi, \eta$ and $\alpha_{\rm bck}$, we adopted uniform priors on a wide range of reasonable values. We imposed a Gaussian prior on $\eta$, centered on $\alpha = -0.5$ to roughly match our observations at $m_{K^\prime} = 15.8$ (see Figure~\ref{fig:alpha_vs_mag}), but with a large standard deviation ($\sigma =1$). This choice helps to make the sampling more efficient with very little impact on the resulting posterior distributions. These choices are summarized in Table~\ref{ParamsPriors}. 

For the likelihood $\mathcal{L}$, we used the formalism presented in section~\ref{subsec: alpha model}: measurements $\{ (\alpha_{H-K^\prime}^{\text{est},(i)}, m_{K^\prime}^{\text{est},(i)}) \} $ are assumed to have uncorrelated white Gaussian noise in the two bands ($ \varepsilon_{\rm band} \sim \mathcal{N}(0,\sigma_{F_{\rm band}}^{(i)}) $), with noise levels $\sigma_{F_{K^\prime}}^{(i)}, \sigma_{F_H}^{(i)}$ computed for each data point. The likelihood function can then be written:

\begin{multline}
    \mathcal{L}(\theta_s) \equiv \mathcal{L} \left( \{ \alpha_{H-K^\prime}^{\text{est},(i)}\} \Big| \{m_{K^\prime}^{\text{est},(i)} , \sigma_{F_{K^\prime}}^{(i)}, \sigma_{F_H}^{(i)} \} ; \boldsymbol{\theta_s} \right) \\
    = \prod_{i} \mathcal{L}_1 \left( \alpha_{H-K^\prime}^{\text{est},(i)}\Big| m_{K^\prime}^{\text{est},(i)} , \sigma_{F_{K^\prime}}^{(i)}, \sigma_{F_H}^{(i)} ; \boldsymbol{\theta_s} \right)
\end{multline}

with an analytic expression for the likelihood $\mathcal{L}_1$ of an individual point, derived in detail in Appendix~\ref{Appendix: likelihood}.

From there on, a nested sampling algorithm \citep[implemented using the Python package \texttt{ultranest},][]{ultranest} allowed us to evaluate the posterior distribution on parameters, as well as the evidence:
\begin{equation}
    Z \equiv p \left( \{ \alpha_{H-K^\prime}^{\text{est},(i)}\} \Big| \{m_{K^\prime}^{\text{est},(i)} , \sigma_{F_{K^\prime}}^{(i)}, \sigma_{F_H}^{(i)} \}  \right)
\end{equation}
The posterior can be used to describe the parameter constraints imposed by the data. We use the evidence for model comparison via the Bayes factor

\begin{equation}
    B_{12} = \frac{Z_1}{Z_2} 
\end{equation}
(assuming that models 1 and 2 have equal prior probability). We will also consider the Akaike and Bayesian Information Criteria for model comparison, given by
\begin{align}
 \mathrm{AIC} & = 2k - 2 \sup_{\theta_s} \left[ \ln \mathcal{L}(\theta_s) \right] \\
 \mathrm{BIC} & = k \ln(N) - 2 \sup_{\theta_s} \left[ \ln \mathcal{L}(\theta_s) \right]
\end{align}

where $k$ is the number of free parameters in the model and $N$ is the number of measurements in the sample.

\section{Results}
\label{sec:results}

\subsection{Bright subset}
\label{subsec: bright end results}

\begin{figure}[ht!]
\vspace*{\baselineskip}
\centering
\includegraphics[width=0.95\linewidth]{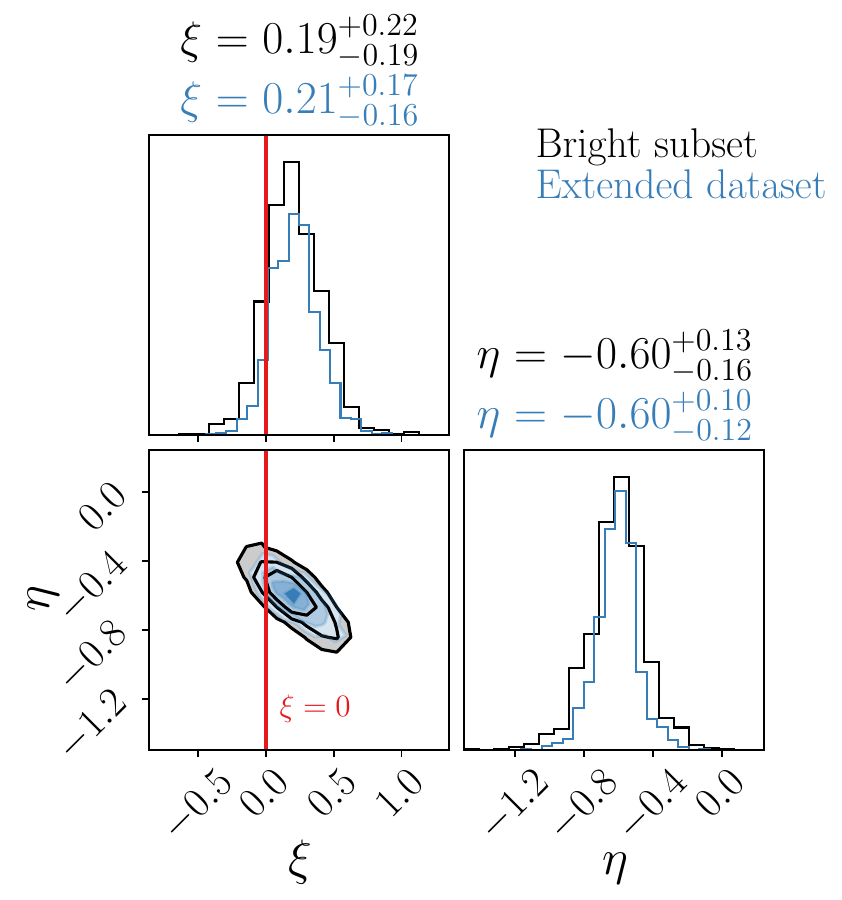}
\caption{Joint posterior distribution of $\xi$ and $\eta$, inferred from the bright subset in black, from the extended dataset in blue (see Figure \ref{fig:alpha_vs_mag}), and marginalized over the background parameters ($ m_{K^\prime}^{\rm bck}, \alpha_{\rm bck}$). The contours have been smoothed slightly for visualization purposes. The red line shows $\xi=0$.} 
\label{fig:corner_plot_xi}
\end{figure}

First, we present the results for the bright subset, ie. points with $m_{K^\prime}^{\rm est} < 16.5$ (see section~\ref{subsec: mag cuts}).
Our spectral index measurements appear to be consistent between the different nights, despite being taken over 17 years (see Figure~\ref{fig:alpha_vs_mag}). Furthermore, they seem compatible with a constant intrinsic spectral index $\alpha_s \sim -0.5$, over a $\sim 3.5$  magnitude range in $K^\prime$ (i.e. a factor $\sim 25$ in flux); and, as expected, there is larger scatter when Sgr~A*-NIR is fainter. More rigorously, we used the Bayesian framework introduced in section~\ref{subsec:Bayesian_framework} to get a posterior distribution on the model parameters $\boldsymbol{\theta_s}=(\xi, \eta, m_{K^\prime}^{\rm bck}, \alpha_{\rm bck})$. To begin with, we focus on the parameters $\xi$ and $\eta$ which describe the intrinsic spectral index (see section~\ref{subsec: instrinsic alpha}), so the result displayed in black in Figure~\ref{fig:corner_plot_xi} is marginalized over the background parameters ($ m_{K^\prime}^{\rm bck}, \alpha_{\rm bck}$).

The inferred constraint on the slope of the $\alpha_s - m_{K^\prime}$ relation is $\xi=-0.19^{+0.22}_{-0.19}$. This value shows a very slight variation in the spectral index, but it is not statistically significantly different than zero. The results are therefore consistent with the constant spectral index model ($\xi=0$, see section \ref{subsubsec:ConstAlpha}). In contrast, the prediction for the exponential cutoff model ($\xi=1.23$, see section~\ref{subsubsec:Expcutoff}) can be ruled out confidently (at $\sim 5\sigma$). Furthermore, we can use model selection criteria to compare the model $M_1$ where $\xi$ is left as a free parameter, and the ‘‘null hypothesis’’ model $M_0$ where $\xi=0$. The AIC, BIC and log-evidence are more robust comparison tools than the reduced $\chi^2$, but can be interpreted in a similar fashion: they have to be improved sufficiently to justify the addition of a free parameter. In our case, allowing non-zero values for $\xi$ actually degrades the different metrics: the Bayes factor is $B_{10} \approx 0.2 <1$, and we have  $\mathrm{AIC}_1 - \mathrm{AIC}_0 \approx 3 >0 $, $\mathrm{BIC}_1 - \mathrm{BIC}_0 \approx 5 >0$.

\begin{figure*}[ht!]
\centering
\includegraphics[width=0.7\linewidth]{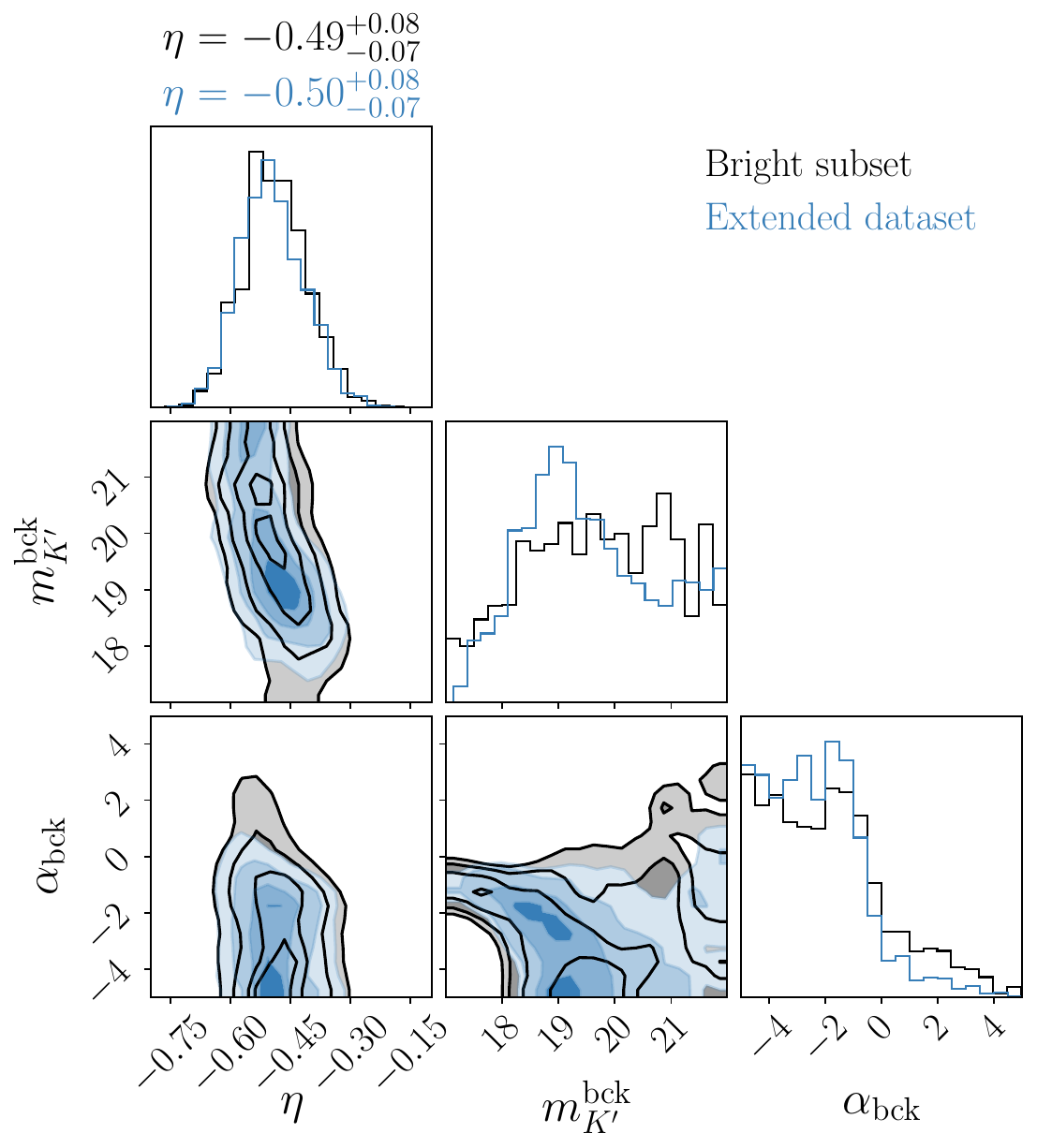} 
\caption{Joint posterior distribution of the constant spectral index $\eta$ and the background parameters ($ m_{K^\prime}^{\rm bck}, \alpha_{\rm bck}$), assuming a model with fixed slope $\xi=0$, for the bright subset (in black) and the extended dataset (in blue). The magnitude of the background is completely unconstrained when $\alpha_{\rm bck} \approx \hat{\eta} = -0.50$, since in that case, background and main emission would be impossible to distinguish with two-band photometry. The contours have been smoothed slightly for visualization purposes.} 
\label{fig:corner_plot_bck}
\end{figure*}

Therefore, $\xi=0$ is preferred, and this value can be fixed to determine the posterior distribution on the other model parameters (see Figure~\ref{fig:corner_plot_bck}). This provides a tighter constraint for the value of the constant spectral index: $\eta = - 0.49 \pm 0.08 \ [\pm 0.17]$ where the second error term reflects the uncertainty on the mean color excess. This is consistent with the spectral index $\alpha_s \sim -0.6$ usually quoted in the literature for bright phases of Sgr~A*-NIR (see section~\ref{sec: discussion} for a more advanced discussion).

Even though we don't expect a background as bright as the magnitude threshold for this subset ($m_{K^\prime} = 16.5$), we can still constrain its presence. Indeed, if a relatively bright background with $\alpha_{\rm bck}$ very different from $\alpha_s$ was present, we should still be able to see the start of a transition. Assuming a stellar background ($\alpha_{\rm bck}\sim 2$) and calculating the marginalized posterior distribution for $m_{K^\prime}^{\rm bck}$, we obtain that $m_{K^\prime}^{\rm bck}\geq 19.6$ (at 95 \% confidence level). Similarly, we can constrain the brightness of some very red underlying emission (ie. the quiescent state of the two-state model, see section~\ref{subsubsec:TwoState}): if $\alpha_{\rm bck}\leq -2$, we find that the quiescent Sgr~A*-NIR brightness must be fainter than $m_{K^\prime}^{\rm bck}\geq 18.1$ (95 \% confidence level), or equivalently, $F_{K^\prime}^{\rm bck}\leq 0.04$~mJy (not de-reddened).

\subsection{Extended dataset}
\label{subsec: extended dataset result}

We now consider the extended dataset, obtained with a less stringent magnitude cut at $m_{K^\prime}^{\rm est} < 17.2$ (see section~\ref{subsec: mag cuts} and Figure~\ref{fig:alpha_vs_mag}).  

The additional spectral index measurements seem to agree with the bright subset : no obvious change in spectral index, or discrepancy between nights (for most points, $-1.5\leq\alpha_{H-K^\prime}^{\rm est}\leq0$). We note that these points have increased scatter and larger errorbars, as expected for faint samples.
Using the results from section~\ref{subsec: bright end results}, we assessed the importance of selection effects for these additional points. Drawing many model parameters from the posterior previously inferred with $\xi=0$ on the bright subset (see Figure~\ref{fig:corner_plot_bck}, in black); and random noise values like in section~\ref{subsec: alpha model} (ie. taking mean values for $\log_{10}(C_{K^\prime})$, $\beta_{K^\prime}$, $\log_{10}(C_H)$, and $\beta_{H} $), we found that, at $m_{K^\prime}^{\rm est} = 17.2$, $ \sim 95 \%$ of the simulated points were above the $H$-band 50\% completeness limit ($m_H^{\rm est} \lesssim 20.3$, see section~\ref{subsec: mag cuts}). This confirms that, even at its faint end, the extended dataset would be sensitive to changes in Sgr~A*-NIR's spectral index.

We used again the framework presented in section~\ref{subsec:Bayesian_framework}, running a nested sampling algorithm to evaluate the posterior for various models - ie. fixing or constraining some of the model parameters. We display the evidence, $\mathrm{AIC}$ and $\mathrm{BIC}$ for each run in Table~\ref{Model Comparison criteria}.

\begin{deluxetable*}{CCCCCCC}[hbtp]
\tablehead{\colhead{$\xi$} & \colhead{$\eta$} & \colhead{$m_{K^\prime}^{\rm bck}$} & \colhead{$\alpha_{\rm bck}$}  & \colhead{$\ln Z$} & \colhead{AIC} & \colhead{BIC} }
\startdata 
=0 & \text{free} & \text{free} & \text{free}  & -385.1 & 767.5 & 776.4 \\
\text{free} & \text{free} & \text{free} & \text{free} & -386.1  & 768.5 & 780.4 \\ 
=1.23 & \text{free} & \text{free} & \text{free}  & -398.0 & 789.9 & 798.8 \\
\hline 
=0 & \text{free} & \text{free} & = -0.5  & -385.3 & 768.2 & 774.1 \\
=0 & \text{free} & \text{free} & = -3  & -386.7 & 769.6 & 775.5 \\
=0 & \text{free} & \text{free} & \leq -2 & - & 770.4 & 779.3 \\
=0 & \text{free} & \text{free} & =2  & -389.9 & 774.9 & 780.8 \\
\enddata
\caption{Values used in model comparison criteria (log-evidence, $\mathrm{AIC}$, $\mathrm{BIC}$), obtained by running the nested sampling algorithm (see section~\ref{subsec:Bayesian_framework}) on the full dataset for various choices of model parameters. NB: for the second-to-last model, the evidence is not reported since the constraint $\alpha_{\rm bck}\leq -2$ effectively changes the prior on one of the free parameters (AIC/BIC comparison is still possible in that case, however).}
\vspace*{-3\baselineskip}
\label{Model Comparison criteria}
\end{deluxetable*}
\vspace*{-2\baselineskip}

First, if we examine the updated constraints on the slope $\xi$, we obtain results in agreement with the bright subset. Leaving $\xi$ as a free parameter, the marginalized posterior is, again, consistent at $\sim 1.3 \sigma$ with $\xi=0$ (see Figure~\ref{fig:corner_plot_xi}, in blue). Furthermore, we find comparable evidence, AIC and BIC between the model where $\xi$ is free, and the model where we fix $\xi=0$ (see Table~\ref{Model Comparison criteria}). On the other hand, the exponential cutoff model is ruled out confidently (at $\sim 6 \sigma$ for the inference with free $\xi$, and by model comparison criteria if we fix $\xi = 1.23$).

Since there is still no evidence for a non-zero slope in this extended dataset, we assume again $\xi=0$ and calculate the posterior on the other model parameters under this assumption (displayed in blue in Figure~\ref{fig:corner_plot_bck}). We can then directly infer the value for the constant spectral index: $\eta = - 0.49 \pm 0.08 \ [\pm 0.17]$ (where, again, the second error term comes from the uncertainty on the mean color excess), which is very close to the one deduced from the bright subset (see section~\ref{subsec: bright end results}). This can be easily understood: since faint measurements are much noisier, most of the constraining power here comes from the bright points.

\begin{figure*}[ht!]
\centering
\includegraphics[width=0.83\linewidth]{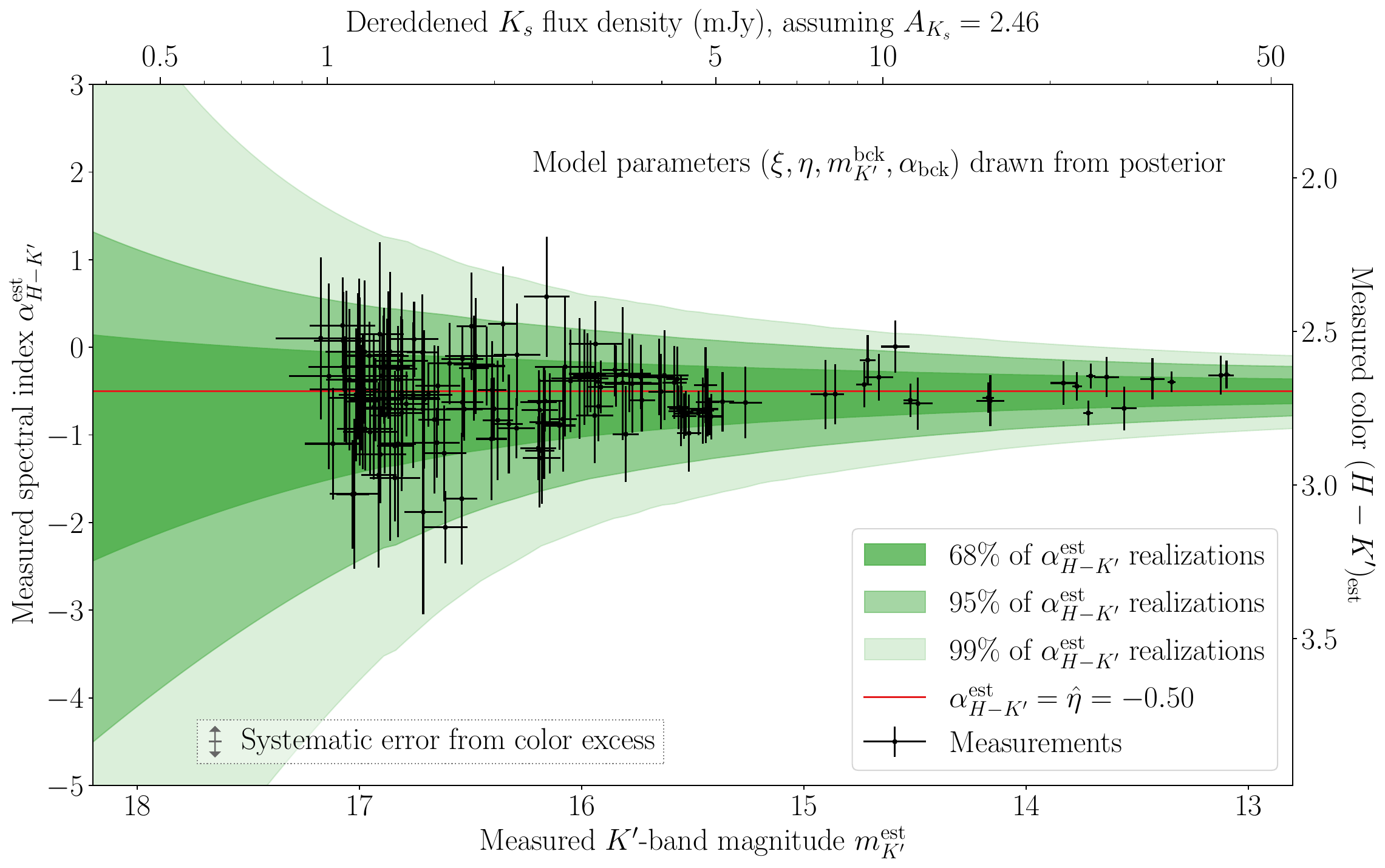} \caption{Comparison between the actual $H-K^\prime$ spectral index measurements from the full dataset (in black) and the predicted distribution of $\alpha_{H-K^\prime}^{\rm est}$ at each measured magnitude in $K^\prime$ (green). The contours are obtained similarly to Figure~\ref{fig:alpha_contours}, ie. by assuming some noise properties (here, equation~(\ref{Eq:flux_err}) with the mean values for $\log_{10}(C_{\rm band})$ and $\beta_{\rm band}$ displayed in Table~\ref{Noisefitvalues}) and sampling $\sim 10^5$ noise realizations for each measured magnitude $m_{K^\prime}^{\rm est}$. This time, however, the model parameters $\boldsymbol{\theta_s}$ are not fixed (except for $\xi=0$), but sampled simultaneously from the posterior shown in blue in Figure~\ref{fig:corner_plot_bck}. The red line indicates the best estimate for the constant spectral index ($\alpha = \hat{\eta}=-0.50$, see Figure~\ref{fig:corner_plot_bck}). The systematic error from the color excess is shown separately, since changing the value of $\langle E(H-K^\prime)\rangle$ would simply shift every spectral index by the same amount.}
\label{fig:full_set_posterior_comparison}
\end{figure*}

The model with $\xi=0$ can convincingly reproduce the observations, as illustrated in Figure~\ref{fig:full_set_posterior_comparison}. This compares the measurements from the extended dataset to the expectation from the model under parameters given by the posterior distribution (in blue, Figure~\ref{fig:corner_plot_bck}). Contours for the measured spectral index were computed by sampling ($10^4$ times for each magnitude) model parameters from the posterior, as well as uncorrelated Gaussian white noise ($\varepsilon_{K^\prime}, \varepsilon_{H}$) in the two bands (mean values are assumed for $\log_{10}(C_{K^\prime})$, $\beta_{K^\prime}$, $\log_{10}(C_H)$, and $\beta_{H} $ like in section~\ref{subsec: alpha model}). These contours indicate how the measured spectral indices should be distributed, which is partially determined by the noise properties of our experiment. When comparing with other works, only the \textit{noise-free} constraints on the spectral index of Sgr~A*-NIR should be considered. These constraints for our work are shown in Figures~\ref{fig:literature_comparison_phot} and \ref{fig:literature_comparison_spec}, and we discuss their consistency with the literature in section~\ref{sec: discussion}.

The extended dataset can also be help constrain the presence of background emission. Still keeping $\xi=0$, we tested different prescriptions for the spectral index of the background. More specifically, we examined a model with $\alpha_{\rm bck}=2$ (confusion from unknown stellar sources), one with $\alpha_{\rm bck}=-3$ (redder, quiescent state, see section~\ref{subsubsec:TwoState}), and one with $\alpha_{\rm bck}= -0.5$ (same as the main emission, which amounts to no background). Between the three, the first model is the only one that is disfavored (see Table~\ref{Model Comparison criteria}, e.g. $\Delta \text{AIC} > 6, \Delta \text{BIC} > 4$ compared to the model with free $\alpha_{\rm bck}$). The other two have evidence and $\mathrm{AIC}/\mathrm{BIC}$ that are comparable to the model with $\alpha_{\rm bck}$ left as a free parameter. In addition, we tested a more permissive case for the red, quiescent background; imposing $\alpha_{\rm bck} \leq -2$ instead of choosing $\alpha_{\rm bck} = -3$. Since this changes the prior on $\alpha_{\rm bck}$ while keeping it as a free parameter, comparing the evidence is not possible, but we find no reason to favor this model when looking at the AIC and BIC.

We conclude that there is no evidence for the presence of a background causing deviations in the spectral index, or at least, one that is bright enough to have a discernible influence on our dataset. In fact, using the same method as in section~\ref{subsec: bright end results}, we can update the constraints on the equivalent magnitude of a potential background at Sgr~A*'s location. If we consider a stellar background ($\alpha_{\rm bck}\sim 2$), we must have $m_{K^\prime}^{\rm bck}\geq 20.3$ (at 95 \% confidence level), or equivalently, that the flux contribution must be $F_{K^\prime}^{\rm bck}\leq 0.005$ mJy (not de-reddened). Similarly, assuming a red quiescent state ($\alpha_{\rm bck}\leq -2$), we find $m_{K^\prime}^{\rm bck}\geq 18.3$ (at 95 \% C.L.), or equivalently, $F_{K^\prime}^{\rm bck}\leq 0.03$~mJy (not de-reddened).

\section{Discussion}
\label{sec: discussion}

\subsection{Comparison to other works}

The best model explaining our measurements of the $H-K^\prime$ color of Sgr~A*-NIR involves a constant value for the intrinsic spectral index, independently of the level of emission. This is quite remarkable given the wide range of flux densities probed by our experiment ($\approx 0.1-4$~mJy observed in $K^\prime$, or $\approx 1-42$~mJy in $K_s$, dereddened with $A_{K_s}=2.46$) - the largest to date by a factor of $\sim 3$.  

In particular, we are able to confidently rule out the exponential cutoff model proposed by \cite{Witzel2018}, with a slope $\xi =1.23$ for the linear dependance of $\alpha$ on magnitude. This does not imply that there are no shifts in the synchrotron cutoff frequency over time, rather that the NIR variability cannot be exclusively explained by such a mechanism. 

\begin{figure*}[ht!]
\centering

\includegraphics[width=0.92\textwidth]{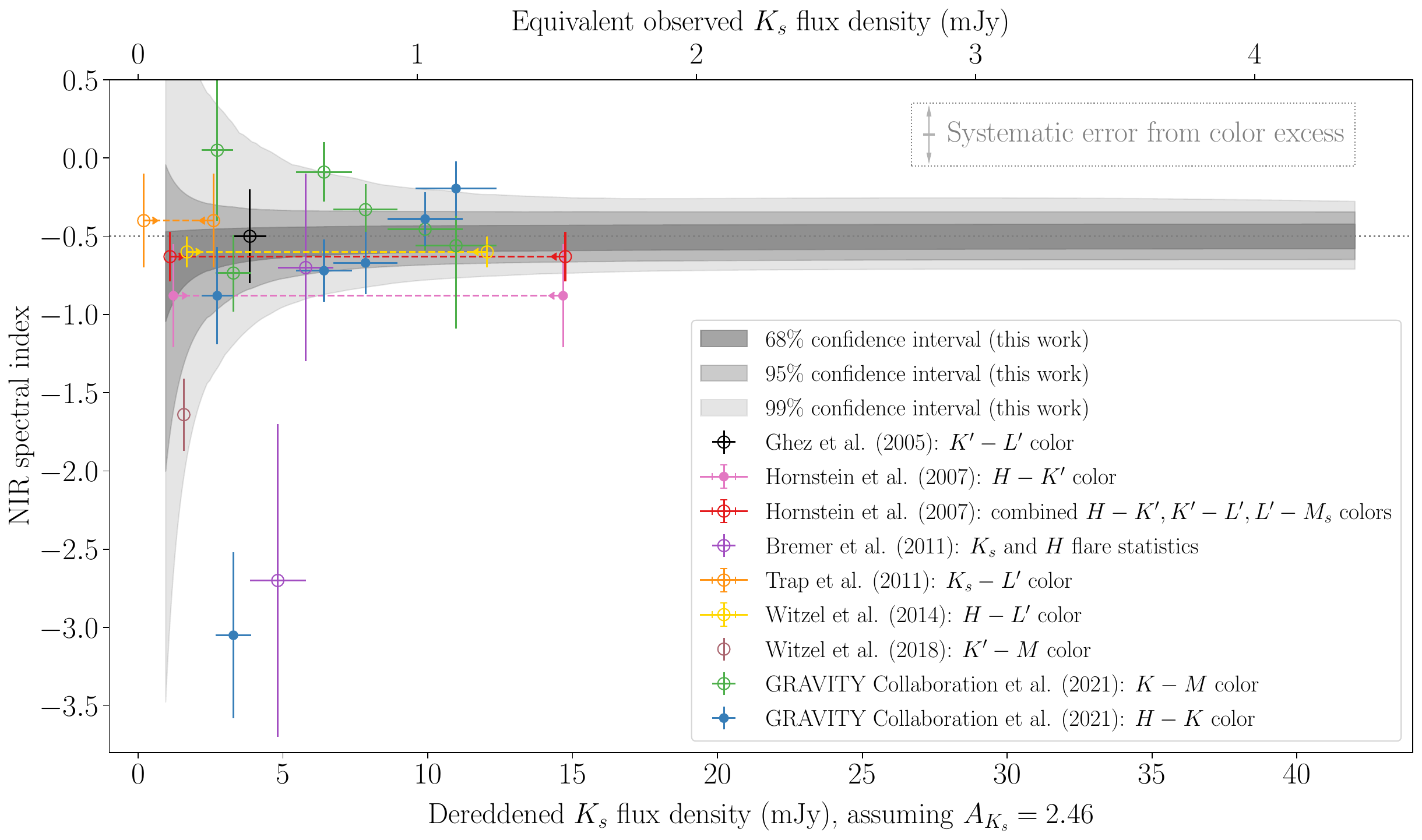}
\caption{Comparison between the spectral index constraints obtained in this work, and values found in other photometric studies in the NIR. The confidence intervals were obtained by sampling model parameters from the posterior plotted in blue in Figure~\ref{fig:corner_plot_bck} (ie. from the extended dataset, with $\xi=0$), then computing the spectral index with zero noise. Points indicate individual measurements, and dotted lines the best-fit values over some range, when measurements were found to be consistent with a constant spectral index. Filled circles specify when color measurements between $H$ and $K^\prime$ (or $K$) bands are used to determine the spectral index, open circles when other methods or NIR bands are considered. We show the systematic error from extinction correction separately: changing the color excess would shift our entire posterior distribution up/down.} 
\label{fig:literature_comparison_phot}

\includegraphics[width=0.92\textwidth]{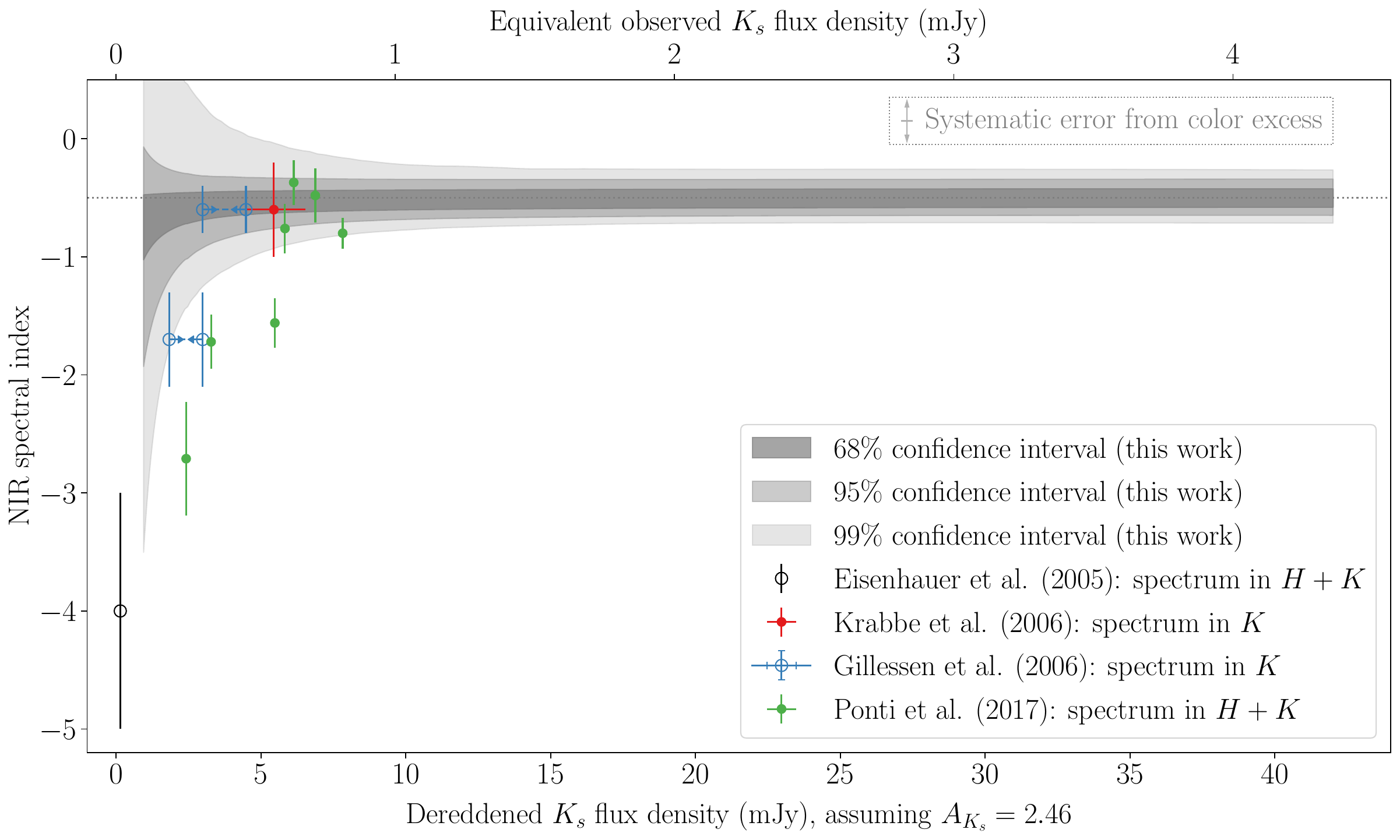}
\caption{Comparison between the spectral index constraints obtained in this work, and values found in spectroscopic studies in the NIR. Open (resp. filled) circles denote measurements with (resp. without) precursor/off-state substraction, a method that can bias spectral index measurements by removing flux that should be attributed to Sgr~A*-NIR.}
\label{fig:literature_comparison_spec}
\end{figure*}

Assuming that the spectral index is indeed constant, we find $\alpha = \hat{\eta} = -0.50 \pm 0.08 _{\rm stat}  \pm 0.17_{\rm sys}$, with the first error term corresponding to statistical uncertainty, and the second to the systematic uncertainty on the mean color excess. This is in good agreement with the value commonly quoted in the literature for bright states of Sgr~A*-NIR  \citep[$\alpha\approx -0.6$,][]{Ghez2005_color, Gillessen2006, Krabbe2006, Hornstein2007, Bremer2011, Trap2011, Witzel2014, Ponti2017, GRAVITY2021_spectral_index}. We extend the validity of this value up to flux densities $\approx 42$~mJy in $K_s$ (dereddened with $A_{K_s}=2.46$), which is $\approx 2.8$ brighter than any previous work. In addition, thanks to the large number of points in our dataset, we improve the errorbars on this best-fitting constant spectral index compared to previous works (e.g., $\sim 2$ times smaller than the $H-K^\prime$ value in \cite{Hornstein2007}, which uses a similar approach). In more detail, Figure~\ref{fig:literature_comparison_phot} compares our posterior for the $H-K^\prime$ spectral index to the result of other NIR photometric studies, as a function of the equivalent flux in $K_s$. The same method as in section~\ref{subsubsec:TwoState} is used to convert between $K,K^\prime$ and $K_s$ fluxes, and we present dereddened values by adopting a common extinction coefficient $A_{K_s}=2.46$ \citep{Schodel2010_extinction}.  For the sake of clarity, and because they have different systematics, spectroscopic studies are plotted separately in Figure~\ref{fig:literature_comparison_spec}.

We warn that in some cases, a direct comparison can be ambiguous. In early investigations of Sgr~A*-NIR, it was common practice to account for local background contamination by subtracting the ‘‘precursor’’ or ‘‘off-state’’, i.e. the lowest flux observed at the location of Sgr~A*-NIR's \citep[e.g.][]{Eisenhauer2005,Gillessen2006, Krabbe2006}. More recent studies prove that Sgr~A*-NIR never actually turns off, displaying stochastic emission even at flux densities $ \sim 0.1 $ mJy \citep{Weldon2023}. This means that the off-state subtraction removes flux that should be attributed to Sgr~A* \citep[sometimes a substantial amount, e.g. $3.5$ mJy in $K$-band for][]{Krabbe2006}, which can bias the spectral index measurement \citep[e.g., $\Delta \alpha \approx -2$ for ][]{Krabbe2006}, especially at the faint end. In section \ref{subsec: alpha model}, we have presented a more robust method to deal with potential background contamination: including it in the model parameters instead of subtracting it directly from the data. For \cite{Krabbe2006}, we plot the spectral index measurement without precursor subtraction. This is unfortunately not reported in \cite{Eisenhauer2005} and \cite{Gillessen2006}, so the apparent discrepancy with our results at  faint flux levels (see Figure~\ref{fig:literature_comparison_spec}) is likely explained by this difference in the background handling method. For \cite{Hornstein2007}, off-state subtraction was used only for the $H-K^\prime$ color, biasing those measurements towards a slightly redder spectral index $\alpha_{H-K^\prime} = -0.88 \pm 0.33$. However, the final value ($\alpha_{\rm NIR} = -0.63 \pm 0.16$) is dominated by the $K^\prime-L^\prime$ color measurements ($\alpha_{K^\prime-L^\prime} = -0.55 \pm 0.18$) which do not use this method.

We find no evidence for the dramatic reddening of Sgr~A*-NIR at faint flux levels claimed by some other studies \citep{Eisenhauer2005,Gillessen2006, Bremer2011, Ponti2017, GRAVITY2021_spectral_index}. The recent measurements from \cite{GRAVITY2021_spectral_index} are actually consistent with our results, except for a single point with $\alpha \sim -3$ (see Figure \ref{fig:literature_comparison_phot}). It is conceivable that this is simply an outlier, especially since three other points in the study are reported at the same flux densities with $\alpha \sim -0.5$, which is consistent with our preferred model.

In fact, our model also provides constraining upper limits on the brightness of a possible quiescent state of Sgr~A*, assuming a red spectral index ($\alpha \lesssim -2$). \cite{Dodds-Edden2011_2states} propose a median flux for the quiescent state $\approx 1.1$ mJy (in $K_s$, dereddened with $A_{K_s}=2.46$), corresponding to $m_{K^\prime}^{\rm bck} \sim 17$ (see section \ref{subsubsec:TwoState}). \cite{Gravity2020_twostate} update this prediction to a typical $K_s$ quiescent flux $= 0.65^{+0.17}_{-0.14}$ mJy (dereddened with $A_{K_s}=2.46$) - ie. $m_{K^\prime}^{\rm bck} = 17.65 \pm 0.25$ (see section \ref{subsubsec:TwoState}). With our bright subset, we find that $m_{K^\prime}^{\rm bck} \geq 18.1$ at 95\% confidence ($F_{K_s}^{\rm bck}\leq 0.38$~mJy, dereddened with $A_{K_s}=2.46$); and with the extended dataset, we have $m_{K^\prime}^{\rm bck} \geq 18.3$ at 95\% confidence ($F_{K_s}^{\rm bck}\leq 0.32$~mJy, dereddened with $A_{K_s}=2.46$). These results are in slight tension with the predictions.

We remark that our measurements do not rule out the existence of a quiescent state altogether. Rather, they challenge one of the arguments in favor of the two-state model (ie., the steepening of the spectral slope at faint fluxes). The two states might be powered by mechanisms with similar spectral indices (with values close enough to not be detectable with our measurements). It is also possible that a redder quiescent state contributes at flux levels where our sample is no longer sensitive ($\lesssim 0.3$ mJy in $K_s$, dereddened). Finally, we have not considered a case where the reddening of Sgr~A*-NIR at faint flux levels is exactly cancelled out by a blue stellar background, which can probably be tuned to be degenerate with the constant spectral index description.

\subsection{Physical interpretation of a constant spectral index model}

In this section, we discuss physical pictures that can lead to a constant, flux-independent value for the NIR spectral index. In general, we can write the energy distribution of the underlying population of electrons $\frac{dN}{d\gamma} = N_0 F(\gamma)$ (where $\gamma$ is the Lorentz factor) in the range relevant for NIR emission. Our results imply that the mechanism behind the variability does not alter the shape $F(\gamma)$, but only the total number of electrons in that range (ie., the normalization $N_0$).

For instance, the NIR emission could be dominated by the synchrotron contribution from a population of non-thermal electrons with $\frac{dN}{d\gamma} \propto \gamma^{-p}$ \citep[e.g.,][]{Dibi2014, Connors2017, GRAVITY2021_spectral_index, Chatterjee2021}, where $p$ is the power-law index of the electron distribution. These non-thermal electrons can form as they are accelerated out of the thermal population, via processes such as diffusive shocks or magnetic reconnection events  \citep[e.g.,][]{Markoff2001,Liu2004,Dodds-Eden2010}. If this acceleration process injects electrons at a variable rate, but with the same power-law distribution, it could explain the flux variations at a constant spectral index. In that picture, there is a direct relation between $p$ and the spectral index of the source ($p=1-2\alpha$, yielding $p\approx 2$ in our case).

The stability of the measured spectral index also has interesting implications for energy loss (cooling) in the electron population, suggesting that there is no significant redistribution of electrons as flares decay. This is most apparent in the 2019 May 13 observations, which include a drop of $\sim 3 $ magnitudes in $K^\prime$ at the end of an extremely bright flare (see Figure~\ref{fig:lightcurves_appendix}) without any detectable color change. 

A complete description is out of the scope of this work - in fact, there exists a vast literature discussing semi-analytic models or simulations of Sgr~A*'s spectral energy distribution and variability \citep[e.g.,][and references therein]{Markoff2001,Yuan2003, Dodds-Eden2009,Trap2011, Dibi2014,Ponti2017, Dexter2020, Witzel2021,Chatterjee2021,GRAVITY2021_spectral_index, Boyce2022}. As one example of added complexity, other radiative processes like thermal synchtrotron or synchrotron-self-Compton (SSC) might contribute in the NIR. At that point, monitoring other wavelengths (in particular in the sub-mm and X-ray), and how they correlate to the NIR variability, becomes essential to get a clear picture. In this work, we provide the best observational constraint to date on the NIR spectral index, which is one piece that can help discriminate between the various models.

\subsection{Current limitations and future improvements}

The spectral index of Sgr~A*-NIR is constrained quite precisely for bright flux densities (see Figure~\ref{fig:literature_comparison_phot}), so this work will likely be refined mostly in the low-flux regime. For this purpose, improving the spatial resolution and sensitivity will be critical. This would not only reduce photometric uncertainties and facilitate fainter detections of Sgr~A*-NIR, but also mitigate the impact of stellar confusion - a significant source of uncertainty for faint fluxes. Improvements of the sensitivity will be particularly impactful in $H$-band, since this is currently limiting for the study of color variations. We note that refining color measurements requires improving precision in (at least) \textit{two} bands, meaning that better spatial resolution and/or sensitivity in only one band (e.g., by using interferometry in $K$-band with GRAVITY) will not necessarily result in a more accurate spectral index.

More multi-wavelength observations, even with the current data quality, would also help in several ways. First, further monitoring of GC stars close to the SMBH would allow a better determination of their orbits and magnitudes, reducing the systematics associated with confusion correction. For example, in this work, we had to make an educated guess for the intrinsic color of our confusing sources. This would be unnecessary if a larger pool of $H$-band observations was available (since their magnitudes could be measured directly). Second, even if the GC is a crowded field, there some periods during which no known source is confused with Sgr~A*-NIR \citep[ie., within $60$ mas for our data quality,][]{Weldon2023}. We had access to only one epoch with both $H$ and $K^\prime$ data that happened to be unconfused (2019 May 13, which is also a very unique night due to the historically intense flare). The number of such nights will increase as observations are carried out over many more years, which will lessen the need for confusion correction.

Another improvement would be to take synchronous observations with multiple instruments, which has been done before \citep[e.g.,][]{Witzel2014, GRAVITY2021_spectral_index} and reduces the uncertainties introduced by interpolation. Though the MOGP regression method presented here is well-motivated and fruitful, shorter temporal separations would make interpolation more accurate.

Finally, at the level of precision attained in this work, analyzing data with the same methodology in more than two NIR filters might be interesting. Comparing the measured spectral index values between different pairs of bands could give additional hints on the nature of Sgr~A*'s NIR emission: for instance, a wavelength-dependent spectral index would suggest a non-power-law behavior for the NIR spectral energy distribution.

\section{Conclusion}
\label{sec: summary}

Using a dataset formed with 7 epochs of broad-band photometry in the near-infrared $H$ and $K^\prime$ bands, we have examined potential changes in spectral index $\alpha$ ($F_\nu \propto \nu^\alpha$) for the highly variable source Sgr~A*-NIR. We have presented a rigorous procedure to correct for confusion with known stellar sources, as well as a new method to efficiently interpolate between observations that are interleaved in the two bands, using a Multi-Output Gaussian Process. 

We have introduced a flexible empirical model for the spectral index measurements: Sgr~A*-NIR's intrinsic spectral index is allowed to vary linearly with magnitude, and is contaminated by background emission along with uncorrelated white noise in the two bands. The slope and intercept of the linear relation, as well as the magnitude and spectral index of the background, are parameters that can be constrained by the observations. This allowed us to test three different physical pictures: the first with a constant spectral index, the second with a linear dependence on magnitude and a predicted slope, the third with additive contributions from a flaring state and a redder, quiescent state.

We considered two datasets for our final analysis: (1) a ‘‘bright’’ subset which uses a conservative magnitude cut at $m_{K^\prime}^{\rm est}=16.5$, expected to be more robust since the various systematics (uncertainties in the extraction of photometry, and confusion, mainly) are smaller at high flux densities ; (2) an ‘‘extended’’ dataset where we increase the number of points with a fainter magnitude threshold at $m_{K^\prime}^{\rm est}=17.2$.

The two datasets give consistent results : our data prefers a spectral index $\alpha = -0.50 \pm 0.08 _{\rm stat}  \pm 0.17_{\rm sys}$ that remains constant over a large range of fluxes ($\approx 0.1-4$ mJy observed in $K^\prime$, ie. a factor $\sim 40$ in brightness). The model with a strong linear dependence on magnitude is confidently ruled out by our measurements. This result is in good agreement with the literature concerning the spectral index for bright Sgr~A*-NIR states. We find no evidence, however, for a transition towards redder spectral indices when Sgr~A*-NIR gets faint. In fact, we get constraining upper limits on the flux contribution of a red ($\alpha_{\rm bck}\leq -2$) background: $\leq 0.03$~mJy (resp. $\leq 0.04$ mJy) observed $K^\prime$ flux at 95\% confidence level, or equivalently $\leq 0.3$~mJy (resp. $\leq 0.4$ mJy) dereddened $K_s$ flux, using the extended dataset (resp. only the bright subset). This is in minor tension with proposed flux values for the transition to a red quiescent state in the NIR. 

One possible interpretation of this flux-independent spectral index is that the NIR emission is produced by a part of the electron energy distribution which retains the same shape as a variable number of electrons are injected. Combined with observations at other wavelengths, this result may help to better understand the physical mechanisms powering the variability of Sgr~A*.

\begin{acknowledgments}
Support for this work was provided by the Gordon and Betty Moore Foundation under award No. 11458, and the National Science Foundation under grant No. 1909554. HP acknowledges partial support from the Monahan Foundation. We thank Mark Morris for helpful discussion and comments, and the astronomers and staff at the Keck Observatory for their help in taking the observations. The W. M. Keck Observatory is operated as a scientific partnership among the California Institute of Technology, the University of California, and the National Aeronautics and Space Administration. The Observatory was made possible by the generous financial support of the W. M. Keck Foundation. The authors wish to recognize that the summit of Maunakea has always held a very significant cultural role for the indigenous Hawaiian community. We are most fortunate to have the opportunity to observe from this mountain.
\end{acknowledgments}

\vspace{5mm}
\facilities{Keck}


\software{\texttt{KAI} \citep{KAI}, \texttt{StarFinder} \citep{Starfinder_code}, \texttt{AIROPA} \citep{AIROPA}, \texttt{NStarOrbits} \citep{NSOtalk}, \texttt{GPy} \citep{gpy2014}, \texttt{ultranest} \citep{ultranest}, \texttt{corner} \citep{corner}, \texttt{astropy} \citep{astropy_1,astropy_2, astropy_3}}

\vspace{5mm}
\newpage

\appendix
\vspace{-2mm}

\section{Reduction method for Sgr~A*-NIR photometry}

\subsection{\texttt{StarFinder} ‘‘force’’ mode}
\label{Appendix: SF_force}

Identification and characterization of point sources in the images was accomplished using the PSF-fitting code \texttt{AIROPA} \citep{AIROPA}, based on the IDL package \texttt{StarFinder} \citep{Starfinder_code}. More specifically, we employed a adaptation of \texttt{StarFinder} called ‘‘force’’ mode \citep{Hornstein2007, Weldon2023} on the individual frames.

During routine \texttt{StarFinder} runs (‘‘non-force’’ mode), a PSF is constructed for each frame using some reference bright stars, then sources are found by sweeping across that frame to find locations where the image correlates with the PSF above some threshold value (in our case, with a correlation value $\geq 0.8$). This process is repeated several times to improve the PSF estimate and allow for more reliable measurements. This typical method is not optimized for the study of Sgr~A*-NIR, though: in the individual frames, it tends to not find sources when they are too faint. \texttt{StarFinder} can be helped if a list of expected positions is passed as an input in order to drive detections. Thus, if we can predict Sgr~A*'s location accurately enough, this ‘‘force’’ mode can improve the number of frames where its NIR counterpart is detected \citep{Weldon2023}.

On that account, the following procedure was adopted: we first reduced the data in non-force mode (for both individual and composite images, see section~\ref{subsec:data_reduction}). Since composite images have better quality, sources are more reliably detected in those, so we translated these detections into a list of expected source positions for the individual frames. To this list, we added the expected position of Sgr~A*, determined thanks to an IR astrometric reference frame - constructed using SiO masers with very precise radio positions \citep{Sakai2019_refframe}. \texttt{StarFinder} was then run a second time on individual frames with the following prior information: the pre-constructed PSF and background from the non-force run, and the list (including Sgr~A*) of expected source positions.

\subsection{Comparison of \texttt{AIROPA} versions}
\label{Appendix: Single_vs_legacy} 

In this work, we utilized \texttt{AIROPA} with a version of \texttt{StarFinder} called legacy - the same as the one employed in \cite{Gautam2019} and \cite{Do2019flare}. There exists a more recent version called single-PSF \citep{AIROPA}, employed in \cite{Weldon2023} and \cite{Gautam2024}, and originally introduced to improve point-source detection in images of the GC. However, we find that for our purposes, legacy mode has better performance.

In their Appendix A, \cite{Gautam2024} state that single-PSF mode finds more stars overall (and especially faint ones), but detects fewer artifact sources near the edge of the field of view. They also find that the two modes have comparable photometric precision. Single-PSF mode was thus more apt for their science objective (detecting binaries in  $10\arcsec$ wide images). Our study focuses on a much smaller region - only sources within a few $0.1\arcsec$ of Sgr~A* will matter for the photometry of Sgr~A*-NIR. Considerations on artifact sources have no relevance in this context since Sgr~A* is near the center of our images. On the other hand, we find that single-PSF mode sometimes misses detections very close to Sgr~A* compared to legacy mode. This can lead to significant changes in the detected magnitude, and in turn impact the spectral index measurements. The most extreme example is 2019 May 13, where the bright star S0-2 is detected ($\sim 80$ mas away from Sgr~A*) with legacy but not single-PSF in the composite frames, leading to a roughly constant flux offset for the individual frames: $\Delta F \equiv F_{\rm single} - F_{\rm legacy} = - 0.09 \pm 0.03$ mJy in $K^\prime$. Since points with magnitudes $m_{K^\prime}\sim 17$ have comparable flux densities, faint detections of Sgr~A*-NIR (and the corresponding spectral index measurements) are heavily biased by this missed detection. Other epochs in our dataset show less dramatic shifts, but legacy mode consistently detects more sources in the $\sim 0.2 \arcsec$ around Sgr~A*. Therefore, legacy \texttt{StarFinder} is the better choice for our science goal, since flux will be attributed to Sgr~A*-NIR more reliably when sources nearby are detected more often.

We note that \cite{Weldon2023} studied the $K^\prime$ flux distribution of Sgr~A*-NIR using single-PSF \texttt{StarFinder}, but their scientific conclusions are unlikely to differ significantly with legacy mode. Flux offsets between the two versions vary from epoch to epoch with alternating signs, so these shifts should average out when considering distributions over many nights (which is the case for their pre-2019 and post-2019 distributions). For instance, grouping the 2022 epochs of our dataset, the offsets between single-PSF and legacy have a mean and standard deviation $\Delta F = 0.01 \pm 0.03$ mJy. The 2019-only flux distribution might be slightly impacted because of the particularly high difference in 2019 May 13 - but this shift is towards higher fluxes, so it would only increase the median and reinforce the claim of elevated flux levels in 2019.

\section{Strehl ratio cuts}
\label{Appendix: Strehlcut}

To avoid large photometric errors due to bad seeing conditions, we removed observations below some thresholds in Strehl ratio, taken as a measure of AO performance: $S=0.2$ in $K^\prime$-band, $S=0.175$ in $H$-band. To determine these values, we looked at some stars close to Sgr~A*, with magnitudes in the band considered comparable to Sgr~A* (mean magnitudes $\langle m_{K^\prime}\rangle =16.7$ and $\langle m_{H}\rangle =18.7$), and detected in a large number of epochs, and which are either non-variable or variable on long timescales ($t\gtrsim 1 $ yr) \citep{Gautam2019, Gautam2024}. More specifically, we considered S1-31 ($\langle m_{K^\prime}\rangle = 15.7$, non variable), S0-17 ($\langle m_{K^\prime}\rangle = 16.11$, variable on long timescales), and S0-55 ($\langle m_{K^\prime}\rangle = 16.53$, variable on long timescales ) in $K^\prime$-band; S0-16 ($\langle m_{H}\rangle = 17.75$, variable on long timescales) and S0-17 ($\langle m_{K^\prime}\rangle = 18.29$, not variable) in $H$-band.

Since these stars are expected to be photometrically stable within a night, 
the median flux value of the night $\mu_{1/2}^{{\rm night}(i)}$ can be used as an estimator for the ‘‘true’’ flux value during that night. Low Strehl ratios (i.e. poor AO performance) should result, on average, in statistically significant deviations of the measurements $F_i \pm \sigma_i$ compared to that value. This motivates the use of the following quantitative criterion: splitting the dataset into bins of Strehl ratio, we compute the weighted standardized mean square error (wSMSE) in each bin 

\begin{equation}
\begin{split}
    \text{wSMSE} = \frac{\langle \big( F-\mu_{1/2}^{\rm night} \big)^2 \rangle}{\langle \sigma^2 \rangle} 
    & = \frac{\sum\limits_{S_i \in \text{bin}} w_i \big( F_i-\mu_{1/2}^{{\rm night}(i)} \big)^2 } { \sum\limits_{S_i \in \text{bin}} w_i \sigma_i^2} \text{ with } w_i = 1/\sigma_i ^2\\
    & = \frac{1}{n_{\rm bin}} \sum_{S_i \in \text{bin}}  \Bigg( \frac{ F_i-\mu_{1/2}^{{\rm night}(i)}}{\sigma_i} \Bigg)^2
\end{split}
\end{equation}
where $n_{\rm bin}$ is the number of data points in the bin. If the image quality is good enough in a bin for measurements to be photometrically consistent with the rest of their respective nights, we expect $\text{wSMSE} \lesssim 1 $. On the other hand, $\text{wSMSE} \gg 1 $ indicates that photometric measurements with images in the corresponding Strehl range can yield large systematic errors and thus should not be considered in the dataset.

\begin{figure*}[ht!]
\centering
\includegraphics[width=0.78\linewidth]{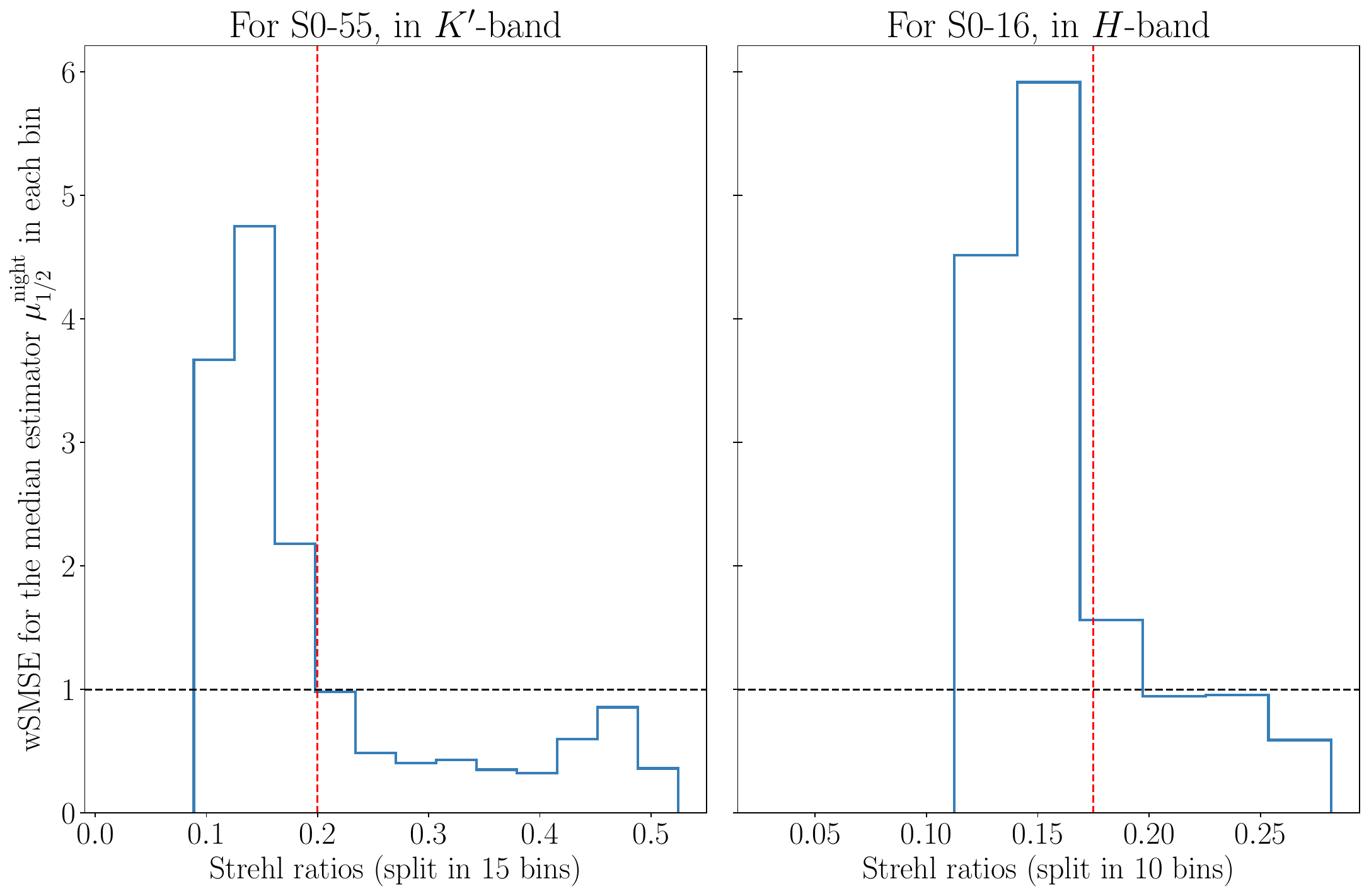}
\caption{Weighted standardized mean square error for the median estimator, as a function of Strehl ratio, for a reference star in each band. The red dashed lines indicate the chosen threshold values for the Strehl ratio cuts. } 
\label{fig:Strehl_hist}
\end{figure*}

For the chosen stars, in $K^\prime$-band, there is a transition  between the two regimes for Strehl values $S \approx 0.2$ (see left panel of Figure~\ref{fig:Strehl_hist}), matching the Strehl ratio cut chosen in \cite{Do2009_variability} and \cite{Weldon2023}.  $H$-band displays a similar turning point around $S \approx 0.175$ (see right panel of Figure~\ref{fig:Strehl_hist}). Different binning choices were explored to ensure the robustness of these values. We note that they are consistent with the fact that Strehl ratio typically increases with wavelength \citep{Strehl_paper}: thus, the Strehl ratios corresponding to a ‘‘good’’ AO performance should be be lower in $H$ than in $K^\prime$.

All measurements with a Strehl value below these empirical thresholds were discarded for the spectral index analysis presented above. Table~\ref{NbObs} reports the number of data points before and after this cut.

\section{Extinction correction}
\label{Appendix: Extinction}

Extinction at the Galactic center is so large that observations of the region are not possible at visible wavelengths and have to be performed in the infrared - so a precise determination of the extinction properties in the NIR is critical for the study of objects in the GC. For color measurements, two approaches exist to correct for reddening: either (1) adopt a model for the NIR extinction curve, or (2) compare with nearby sources of known intrinsic spectral properties.

We first consider method (1), which requires a precise determination of the wavelength dependence of the NIR extinction in order to accurately recover the properties of reddened sources. The extinction coefficient (in magnitudes) is generally approximated as a power law \citep[e.g.,][]{Nishiyama2006_extinction, Schodel2010_extinction, Fritz2011_extinction}:

\begin{equation}
    A_\lambda = A_0 (\lambda/\lambda_0)^{-\beta}
    \label{eq:extinction_curve}
\end{equation}
where $\beta$ is the extinction index. Recent studies point towards a more complicated behaviour over large wavelength ranges \citep[e.g.,][]{Hosek2018_extinction, Nogueras-Lara2019_extinction, Nogueras-Lara2020_extinction}, but since we only consider $H$ and $K^\prime$ bands in this work, the power-law approximation is adequate. Considering a source with an intrinsic spectral index $\alpha_{\rm int}$ (such that $F_\nu \propto \nu^{\alpha_{\rm int}}$ or equivalently, $F_\lambda \propto \lambda^{-2-\alpha_{\rm int}}$), the spectral index that would be measured using equation~(\ref{eq:SpecIndex}) for a given color excess $E(H-K^\prime)$ is \citep{Zero_pt_flux_densities}:

\begin{equation}
    \alpha_{H-K^\prime}^{\rm meas} = \log_{10}\left(\frac{\lambda_{K^\prime}}{\lambda_H}\right)^{-1} \left[ 0.4 E(H-K^\prime) + \log_{10} \left( \frac{\int_H \lambda^{-1-\alpha_{\rm int}}S(\lambda)10^{-0.4A_\lambda}d\lambda}{\int_{K^\prime}\lambda^{-1-\alpha_{\rm int}}S(\lambda)10^{-0.4A_\lambda}d\lambda} \right)\right]
    \label{eq:filter_integrals}
\end{equation}

Here, the intrinsic flux density of the source is $F_\lambda$, so $\lambda F_\lambda$ is proportional to the number of photoelectrons detected per second (which is the measured quantity). We assume that the total system response $S(\lambda)=T(\lambda) R(\lambda)$ is determined by the filter response function $R(\lambda)$ and the atmospheric transmission $T(\lambda)$. In the following calculations, we used the filter transmission from the NIRC2 website\footnote{\url{https://www2.keck.hawaii.edu/inst/nirc2/filters.html}}, and the atmospheric transmission above Mauna Kea from the Gemini website\footnote{\url{https://www.gemini.edu/observing/telescopes-and-sites/sites}} with an airmass of 1.5 and a water vapor column of 1.6 mm.
Equation~(\ref{eq:filter_integrals}) implies that, for a given extinction curve ie. for a given choice of $A_0, \beta$ in equation~(\ref{eq:extinction_curve}), there is a value $\hat{E}(H-K^\prime)$ for the color excess  such that $\alpha_{H-K^\prime}^{\rm meas} = \alpha_{\rm int}$. This value depends on $\alpha_{\rm int}$ and on the extinction law $A_\lambda$ - explicitely, inverting equation~(\ref{eq:filter_integrals}):
\begin{equation}
    \hat{E}(H-K^\prime) = 2.5 \log_{10} \left( \frac{\lambda_{K^\prime}}{\lambda_H} \right) \alpha_{\rm int} +  2.5 \log_{10} \left( \frac{\int_H \lambda^{-1-\alpha_{\rm int}}S(\lambda)10^{-0.4A_\lambda}d\lambda}{\int_{K^\prime}\lambda^{-1-\alpha_{\rm int}}S(\lambda)10^{-0.4A_\lambda}d\lambda} \right)
    \label{eq:optimal_color_excess}
\end{equation}

\begin{figure*}[hbtp]
\centering
\includegraphics[width=0.78\linewidth]{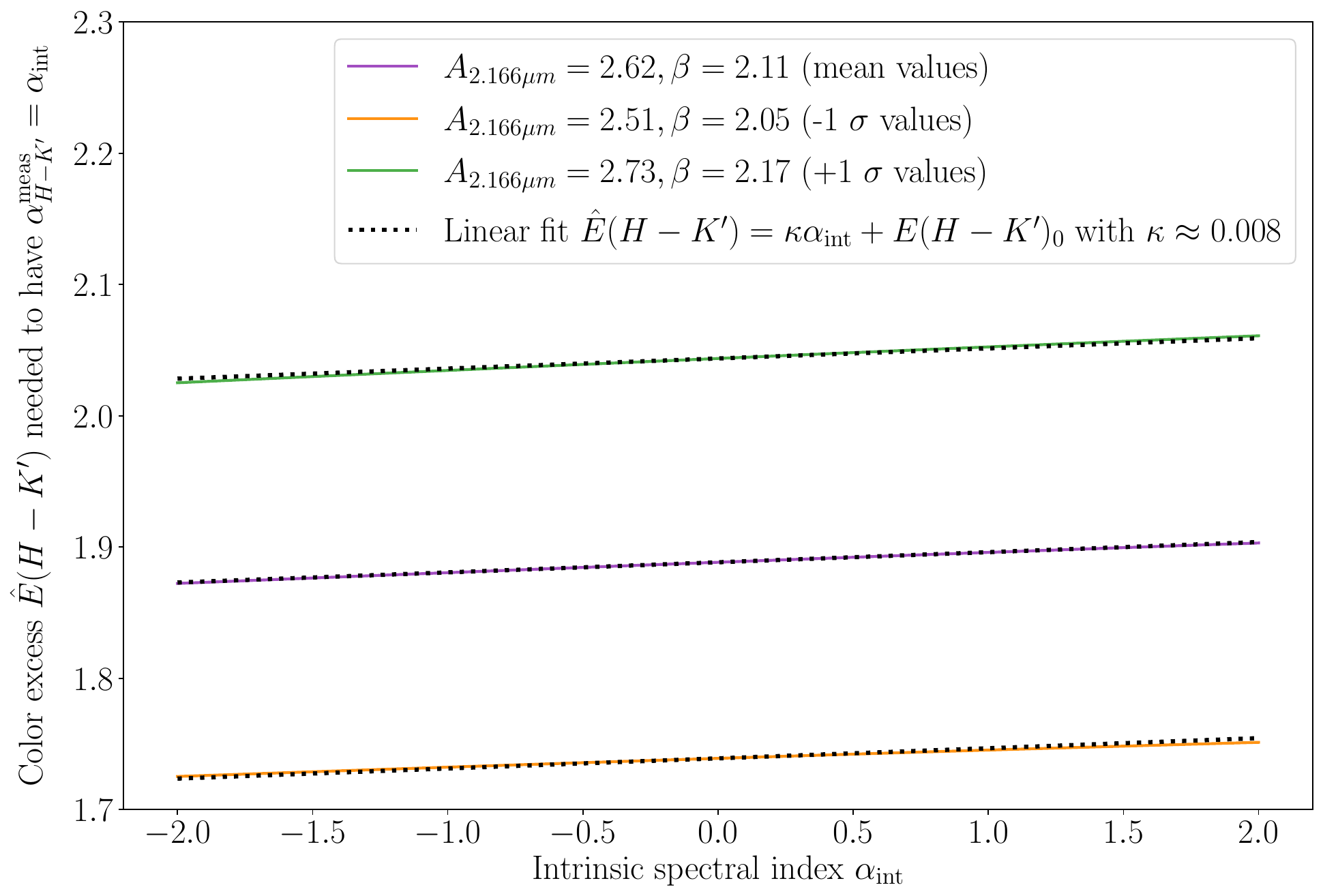}
\caption{Color excess $ \hat{E}(H-K^\prime)$ from equation~(\ref{eq:optimal_color_excess}), obtained by setting $\alpha_{H-K^\prime}^{\rm meas} = \alpha_{\rm int}$ in equation~(\ref{eq:filter_integrals}), as a function of the intrinsic spectral index $\alpha_{\rm int}$ of the source, for different extinction curves. We show three examples consistent with the values from \cite{Fritz2011_extinction}: $A_0 = A_{2.166 \mu m} = 2.62 \pm 0.11$, $\beta = 2.11 \pm 0.06$, and compare them to a linear relation with a fitted slope $\kappa \approx0.008$.} 
\label{fig:extinction_correction}
\end{figure*}

Unfortunately, the NIR extinction curve is currently not determined reliably enough to obtain very precise color measurements with method (1). For instance, \cite{Fritz2011_extinction} relied on measurements of the Brackett-$\gamma$ emission line ($\lambda_0 = 2.166 \mu$m) to infer the following values: $A_0 = A_{2.166 \mu m} = 2.62 \pm 0.11$, $\beta = 2.11 \pm 0.06$. Figure~\ref{fig:extinction_correction} shows that even if these quantities seem to have relatively small errorbars, choosing the $\pm 1 \sigma$ values leads to a substantial difference ($\approx \pm 0.15$) in the color excess $\hat{E}(H-K^\prime)$ required to accurately correct for reddening. This would add an uncertainty $\delta \alpha_{H-K^\prime} \approx 0.5$ to each spectral index measurement.

Uncertainties are set to increase even further when considering other studies on the GC NIR extinction curve: regarding the extinction index between $H$ and $K_s$ for instance, \cite{Nishiyama2008_extinction} find $\beta = 1.99 \pm 0.02$, \cite{Schodel2010_extinction} obtain $\beta = 2.21 \pm 0.24$, \cite{Hosek2018_extinction, Hosek2019_extinction} quote $\beta=2.14$, \cite{Nogueras-Lara2018_extinction} find $\beta = 2.24 \pm 0.11$, ... We also note that extinction towards the GC varies significantly on arcsecond scales, and that the values quoted above are given as averaged over many sightlines that are not necessarily relevant to Sgr~A*-NIR.  Some more specific extinction values (0.5$\arcsec$ around Sgr~A*) are reported by \cite{Fritz2011_extinction} ($A_H=4.21\pm 0.10$, $A_{K_s}=2.42\pm 0.10$) and \cite{Schodel2010_extinction} ($A_H=4.35 \pm 0.12$, $A_{K_s}=2.46\pm 0.03$), but they generate comparable uncertainties on the spectral index ($\delta \alpha_{H-K^\prime} \gtrsim 0.4$) which would limit our study.

These considerations motivate the use of method (2): instead of making specific assumptions about the extinction curve, we can  determine the color excess around Sgr~A* by measuring the colors of nearby stars (in this work, S0-2 which has a known intrinsic color). Indeed, Figure~\ref{fig:extinction_correction} shows that the color excess $ \hat{E}(H-K^\prime)$ from equation~(\ref{eq:optimal_color_excess}) depends very weakly on the intrinsic spectral index of the source $\alpha_{\rm int}$. In fact, for a reasonable range of $A_0$ and $\beta$, the color excess is well described by:
\begin{equation}
\begin{split}
        \hat{E}(H-K^\prime) & = \kappa \alpha_{\rm int} + E(H-K^\prime)_0 \\
                & = \kappa (\alpha_{\rm int}-2) + E(H-K^\prime)_*
\end{split}
\end{equation}
where we fit $\kappa \approx 0.008 \ll 1$. We adopt $E(H-K^\prime)_* = 2.09 \pm 0.05$ for the color excess on stars ($\alpha_{\rm int}\sim 2$), based on the mean detected color of S0-2 (see section~\ref{subsec:extinction}), ie. without assuming a specific extinction curve. The difference in spectral type between Sgr~A*-NIR ($\alpha_{H-K^\prime} \lesssim -0.5$) and S0-2 ($\alpha_{\rm int}\sim 2$) does not matter much, since we have $\Delta E \equiv E(H-K^\prime)_* - E(H-K^\prime)_{\rm SgrA} \approx 0.02$, well within the errorbars of the mean color excess.

\section{Star-planting for confusion correction}
\label{Appendix: conf_corr_starplanting}

\subsection{Star-planting simulations}
\label{Appendix: starplanting_sim}

In order to correct our measurements for flux contamination by stars confused with Sgr~A*, we used star-planting simulations, creating synthetic images where Sgr~A*-NIR was injected at a known magnitude, then running them through our photometry extraction pipeline (that we call \texttt{StarFinder} in this section for simplicity, see more details in section~\ref{subsec:data_reduction} and Appendix~\ref{Appendix: SF_force}).

\begin{figure*}[ht!]
\centering
\includegraphics[width=\linewidth]{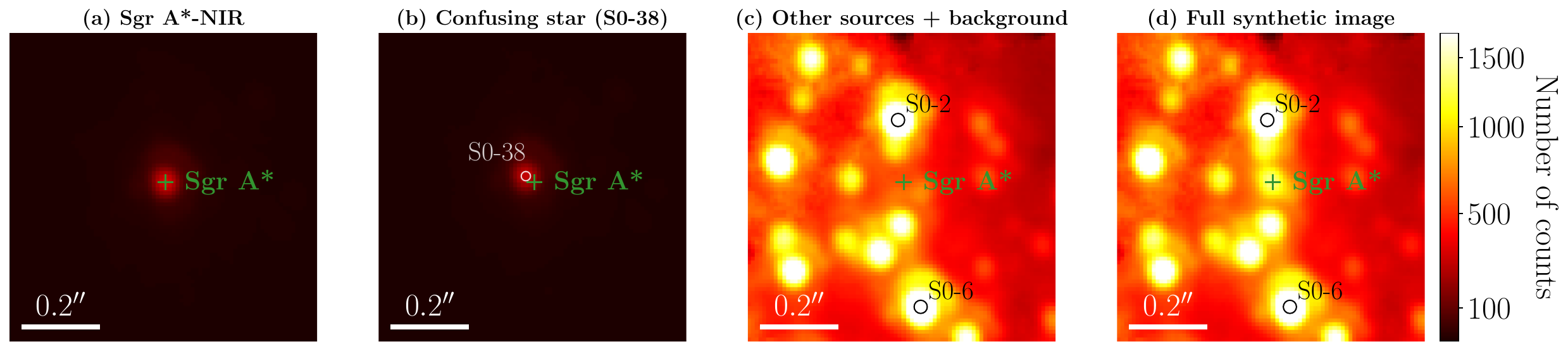}
\caption{Example of synthetic frame construction in star-planting simulations. Here, the observed frame used as a reference is the first $K^\prime$ frame of 2022 May 21. From left to right, the panels show $0.8\arcsec \times 0.8\arcsec $ cutouts around Sgr~A* of: (a) the simulated source representing Sgr~A*-NIR (planted with $m_{\rm plant}=16.8$), (b) one realization of the simulated source representing S0-38, the confused star (sampling the magnitude and position from known distributions), (c) the background and other sources found by \texttt{StarFinder} in the observed image (d) the final synthetic frame (summing the previous contributions and adding white noise). } 
\label{fig:starplanting_process}
\end{figure*}

First, each observed frame was analyzed using \texttt{StarFinder}, yielding a reconstructed PSF, a fitted background, and a list of detections (with their positions and magnitudes). As illustrated in Figure~\ref{fig:starplanting_process}, synthetic images were then constructed by summing contributions from:
\begin{enumerate}
    \item a simulated source representing Sgr~A*-NIR, planted using the reconstructed PSF at a chosen magnitude $m_{\rm plant}$ and at the known location of Sgr~A*,
    \item a simulated source representing the star confused with Sgr~A*-NIR (S0-104 for 2005 July 31, S0-38 for the 2022 epochs), planted using the reconstructed PSF at a randomly sampled magnitude $m_{\rm conf}$ and a randomly sampled location relative to Sgr~A* $(\Delta x_{\rm conf}, \Delta y_{\rm conf})$,
    \item every other detected source (ie. removing the one at Sgr~A*'s location), planted using the reconstructed PSF at its fitted magnitude and position,
    \item the background that was fitted on the observed frame,
    \item white noise, at a level of $\sim 20$ counts/pixel estimated from sky frames.
\end{enumerate}
Each synthetic image was in turn analyzed with \texttt{StarFinder}, recovering (most of the time) a detection at the location of Sgr~A*. The magnitude $m_{\rm rec}$ of this detection combines the flux contributions from Sgr~A*-NIR and the star it is confused with, so we should have $m_{\rm rec} \leq m_{\rm plant}$.

Because they are not perfectly known, we sampled the properties of the confusing sources taking their uncertainties into account. The magnitude $m_{\rm conf}$ was sampled from a Gaussian distribution, with the mean and standard deviation determined by measurements at other unconfused epochs (see Table~\ref{ConfMags}). The location relative to Sgr~A* $(\Delta x_{\rm conf}, \Delta y_{\rm conf})$ was sampled from posteriors on the position of the stars at the relevant epochs (see an example in the inset of Figure~\ref{fig:corrected_lightcurve}). We used the \texttt{NStarOrbits} software \citep{NSOtalk} to compute this posterior, from an orbital fit to long-term, unconfused astrometric data from the Galactic Center Orbits Initiative (O'Neil et al., in prep.).

For a given frame and choice of $m_{\rm plant}$ (planted Sgr~A*-NIR magnitude), 18 random samples of $(m_{\rm conf}, \Delta x_{\rm conf}, \Delta y_{\rm conf})$ were taken, yielding 18 values of $m_{\rm rec}$ after the star-planting simulations. 
We used the median value and the half-width of the 68\% central confidence interval as the estimate $\langle m_{\rm rec} \rangle$ and the uncertainty $\sigma_{\rm conf}$, respectively. We chose these estimators because they are more robust to outliers than the mean and standard deviation - and we could only take 18 samples due to computational limitations.

\begin{figure*}[ht!]
\centering
\includegraphics[width=\linewidth]{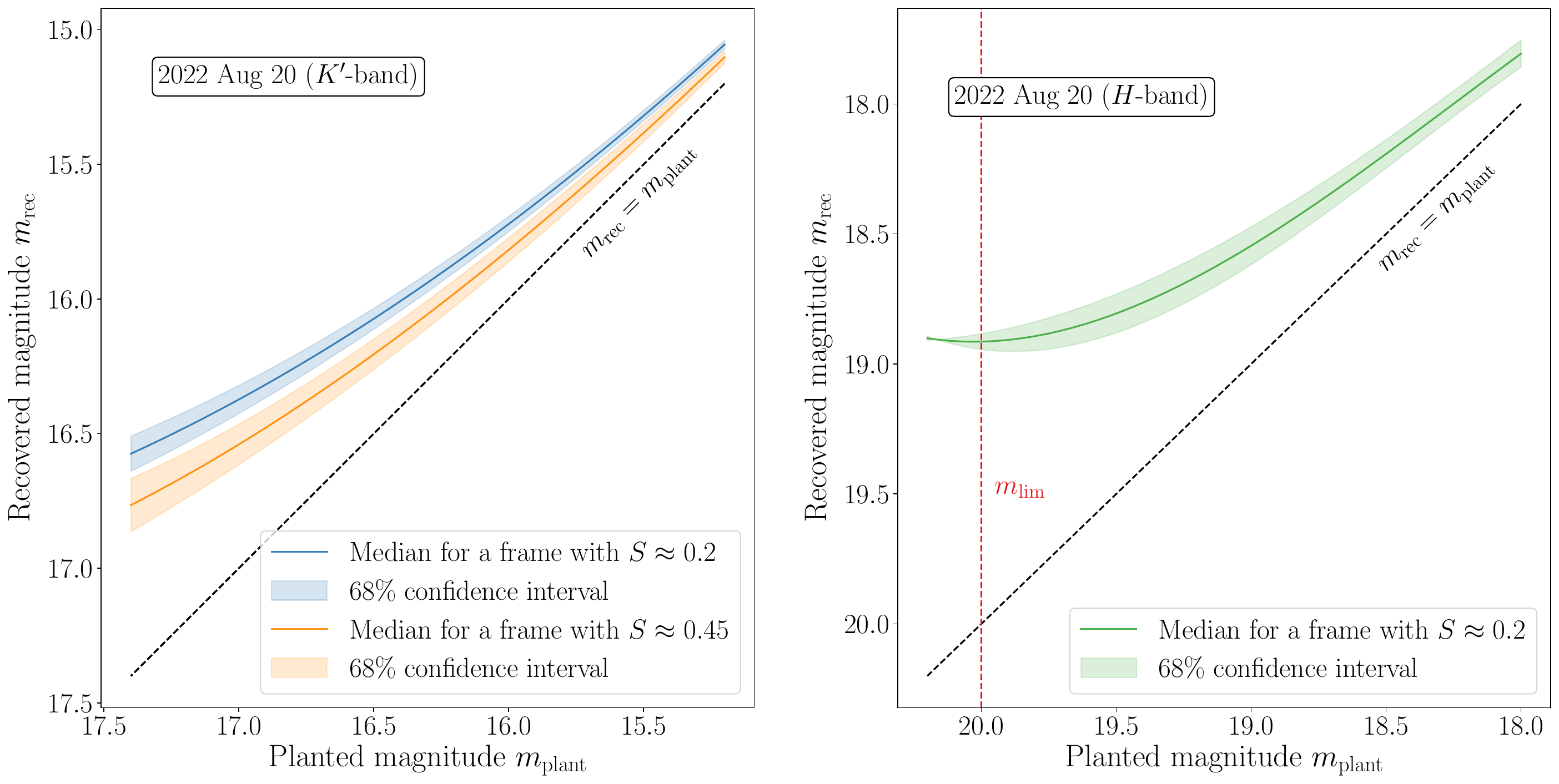}
\caption{Relation between the recovered and planted magnitude of Sgr~A*-NIR obtained with star-planting simulations for three example frames in 2022 Aug 20: two in $K^\prime$-band (left panel), one in $H$-band (right panel). Sgr~A* is confused with S0-38 for this epoch. The median and confidence interval are determined by randomly sampling the magnitude and location of S0-38, from a distribution informed by unconfused observations. Since the confusing source adds to the detected flux, we have $m_{\rm rec} \leq m_{\rm plant}$. This flux contribution depends on the PSF: in $K^\prime$, the frame with a low Strehl ratio ($S\approx 0.2$ ie. close to the cut, see Appendix~\ref{Appendix: Strehlcut}) shows larger deviations of $m_{\rm rec}$ from $m_{\rm plant}$ than the frame with a relatively high Strehl ratio ($S\approx 0.45$). Fainter than some magnitude $m_{\rm lim}$ (shown in red for the $H$-band frame), confusion correction becomes ambiguous since a recovered magnitude $m_{\rm rec}$ would not correspond to a unique $m_{\rm plant}$ value. We note that the curves have been slightly smoothed for the sake of visualization.} 
\label{fig:confusion_correction}
\end{figure*}

For each observed frame, we repeated these star-planted simulations for many values of $m_{\rm plant}$, obtaining values that depend on the planted magnitude for the median recovered magnitude and uncertainty : $\langle m_{\rm rec} \rangle (m_{\rm plant})\pm \sigma_{\rm conf}(m_{\rm plant})$. Figure~\ref{fig:confusion_correction} shows such an output for two example frames: we essentially obtain a relation between recovered and planted magnitude, with an uncertainty on this relation. Because the star confused with Sgr~A*-NIR is slightly offset from Sgr~A*'s location, its flux contribution varies with the PSF: if the PSF is more spread out (ie., for lower values of the Strehl ratio), the flux contribution from the confusing star is more important, so the difference between recovered and planted magnitude is larger (see Figure~\ref{fig:confusion_correction}).

\subsection{Confusion correction and uncertainty estimation}
\label{Appendix: conf_corr+err}

For a given frame, we have: (a) a detected magnitude $m_{\rm det, confused}$ at the location of Sgr~A* , which combines the flux contributions from Sgr~A*-NIR and the star it is confused with; and (b) a median confusion correction law $ \langle m_{\rm rec} \rangle = f(m_{\rm plant})$ with an uncertainty $\sigma_{\rm conf}(m_{\rm plant})$ from star-planting simulations (see Appendix~\ref{Appendix: starplanting_sim}). We can simply invert this relation to correct the magnitude of Sgr~A*-NIR in that frame:
\begin{equation}
    m_{\rm est} = f^{-1}(m_{\rm det, confused})
\end{equation}

The existence of the inverse function $f^{-1}$ is guaranteed as long as $f$ is monotonous. In theory, this condition is verified, because decreasing the flux of Sgr~A*-NIR while keeping the confusion contribution fixed should decrease the total flux in the combined detection (in other words, $m_{\rm rec}$ should always increase with $m_{\rm plant}$, see Figure~\ref{fig:confusion_correction}).

However, in our actual star-planting simulations, when Sgr~A*-NIR becomes fainter than some value $m_{\rm lim}$, $f^{-1}$ is not always well-defined (see Figure~\ref{fig:confusion_correction}): either because not enough samples have detections at Sgr~A*'s location, or because scatter in $m_{\rm rec}$ produces non-monotonous variations with $m_{\rm plant}$. The limiting magnitude $m_{\rm lim}$ depends on the data quality so is different for each frame, but we can report a median value of $ \tilde{m}_{\rm lim} \approx 20.3$ mag for $H$-band frames and a median upper limit $ \tilde{m}_{\rm lim} \gtrsim 18.6$ mag for $K^\prime$-band frames. If $m_{\rm det, confused} \geq m_{\rm lim}$, confusion correction becomes ambiguous, so we remove the frame from our sample. This occurs mostly in $H$-band, since Sgr~A*-NIR is redder than stars, ie. fainter relative to the confusing source than in $K^\prime$-band: $\lesssim 30\%$ of the total frames are removed in $H$, $\lesssim 5\%$ in $K^\prime$ (see Table~\ref{NbObs} for the exact count per epoch).

The uncertainty on the confusion corrected magnitude $m_{\rm est}$ comes from two sources. First, the uncorrected magnitude itself has some uncertainty $\delta m_{\rm det, confused}$
(determined with equation~(\ref{Eq:flux_err}), see section~\ref{subsec:data_reduction}). Second, the relation between $m_{\rm rec}$ and $m_{\rm plant}$ is only known up to some precision $\sigma_{\rm conf}(m_{\rm plant})$, due to the imperfect knowledge of the properties of the stars confused with Sgr~A* (see section~\ref{Appendix: starplanting_sim}). Therefore, we estimate the final uncertainty as:

\begin{equation}
    \delta m_{\rm est} \approx \sqrt{ \left( \frac{\partial f^{-1}}{\partial m_{\rm rec}} \Big|_{m_{\rm det, confused}} \delta m_{\rm det, confused}\right)^2 + \sigma_{\rm conf}^2 (m_{\rm det, corrected})}
\end{equation}

\section{Multi-Output Gaussian Process interpolation}

\subsection{Mathematical framework for Gaussian Processes}
\label{Appendix: GP math}

This appendix is meant as a quick introduction to the formalism behind Gaussian Process (GP) regression - see \cite{GP_book} for a more complete mathematical presentation ; or \cite{GP_astro_review} for a review of applications to astronomical time series. The broad idea of Gaussian Processes is to generalize the approach of parametric Bayesian inference to distributions over functions. 

We start with a few formal definitions. A random function (also called a random process, or a stochastic process) is a collection of random variables indexed by some variable $t$ (often interpreted as a time variable). In other words, it is a function $f:t \mapsto f(t)$ where $f(t)$ is a random variable for all $t$. A Gaussian process is a random function $f$ such that, for any finite set of indices $(t_1,..., t_k)$, the random vector $(f(t_1),..., f(t_k))$ is distributed according to a multivariate Gaussian distribution: $(f(t_1),..., f(t_k)) \sim \mathcal{N}(\mu, \Sigma)$ where $\mu$ is the mean vector and $\Sigma$ is the covariance matrix.

\newcommand{\cov}{\mathrm{Cov}}
Given a Gaussian process, we can define a mean function $m:t\mapsto m(t) = \mathbb{E} [f(t)]$ and a covariance function $k:(t,t^\prime)\mapsto k(t,t^\prime) = \cov [f(t),f(t^\prime)] = \mathbb{E} [(f(t)-m(t))\cdot(f(t^\prime)-m(t^\prime))]$, where $\mathbb{E}[.]$ denotes the expectation value. Conversely, if we specify a mean function and a well-behaved (ie., positive semi-definite) covariance function, every set of random variables in the GP is determined since it follows a multivariate Gaussian distribution. Therefore the GP is completely described by $m$ and $k$, so we usually write $f \sim \mathcal{G}(m(t), k(t, t^\prime))$.

The covariance function $k$ is commonly called a kernel and plays a fundamental role: it essentially quantifies output similarity between any two data points of the input space. Various types of kernels exist, each having different properties, and their choice typically captures some assumptions about the underlying structure of the data. In the case of GP interpolation, the kernel reflects the expected appearance (smoothness, periodicity, etc.) of the target function. For this type of problem, it is common to use kernels that are stationary (i.e. depend only on $\lVert t-t^\prime \rVert$), since they encode the fact that that nearby points are more likely to be correlated.

Typically, the kernels are described by a functional form with some free parameters (called hyperparameters to avoid confusing GPs with parametric models), that are optimized during the regression. In the following, we only present kernels that are relevant to this work, and refer the reader to \cite{GP_book} for a more exhasutive list. We note that kernels can also be combined  (summed, multiplied, convolved,...) for added flexibility.

The most popular choice of kernel is the squared-exponential or gaussian RBF (Radial Basis Function) kernel:
\begin{equation}
    k_{\rm RBF}(t,t^\prime ; \sigma, l) = \sigma^2 \exp \left(- \frac{\lVert t-t^\prime \rVert^2}{2l^2}\right)
\end{equation}
with the hyperparameters $l$ and $\sigma$ specifying a characteristic length-scale and an output scale amplitude of variation, respectively. GPs using this kernel yield very smooth interpolated functions, ideal to encapsulate long-term trends.

The RBF kernel is sometimes considered to be too smooth, in which case one can use the exponential (or Ornstein-Uhlenbeck) kernel:
\begin{equation}
    k_{\rm exp}(t,t^\prime ; \sigma, l) = \sigma^2 \exp \left(- \frac{\lVert t-t^\prime \rVert}{2l}\right)
\end{equation}
where the hyperparameters $l$ and $\sigma$ have the same interpretation. This kernel corresponds to more ‘‘choppy’’ functions, and is helpful for continuous stochastic variability. We employ it to model short-term fluctuations in the lightcurves.

In practice, the mean function of the Gaussian process is often fixed at $m(t)=0$ ; instead of shifting the mean to match the data, it is ordinary practice to add a constant kernel:
\begin{equation}
    k_c(t,t^\prime ; \sigma) = \sigma^2
\end{equation}

Finally, to account for measurement uncertainties, a common choice is to estimate a global noise level $\sigma$ in the data using a white noise kernel:
\begin{equation}
    k_{\rm WN}(t,t^\prime ; \sigma) = \sigma ^2 \delta (t-t^\prime)
    \label{eq: white noise kernel}
\end{equation}
where $\delta$ denotes the Kronecker delta function.

In the following paragraphs, we focus on GP regression, ie. applying the Gaussian Process framework to interpolation. We use the example of times series (ie., $t \in \mathbb{R}$ is a time variable) ; keeping in mind that the same formalism can be applied to other types of problems.
Let us consider a set of times $\{t_i\}_{1\leq i \leq N}$ with associated measurements $\{y_i \}_{1\leq i \leq N}$. We want to interpolate a function $y=f(t)$ from those measurements, that is, in practice, to predict the outputs $\{\tilde{y}_k = f(\tilde{t}_k)\}$ for a new set of points $\{\tilde{t}_k\}_{1\leq k \leq m}$.

To do this, we choose an appropriate kernel $k_\theta$ depending on a set of hyperparameters $\theta$, and consider a GP model  $f \sim \mathcal{G}(0, k_\theta(t, t^\prime))$. Then, by definition, we have:

\begin{equation}
        \mathbf{y} = f(\mathbf{t})
\sim \mathcal{N}(\mathbf{0}_N, \mathbf{K}_\theta(\mathbf{t},\mathbf{t}))
\text{ and }
    \begin{bmatrix}
        \mathbf{y} \\
        \mathbf{\tilde{y}}
\end{bmatrix} 
\sim \mathcal{N}\left(  \mathbf{0}_{N+m} \ , 
\begin{bmatrix}
        \mathbf{K}_\theta(\mathbf{t},\mathbf{t}) & \mathbf{K}_\theta(\mathbf{t},\mathbf{\tilde{t}}) \\
        \mathbf{K}_\theta(\mathbf{\tilde{t}}, \mathbf{t}) & \mathbf{K}_\theta(\mathbf{\tilde{t}}, \mathbf{\tilde{t}})
\end{bmatrix} \right)
\label{eq:joint_obs_pred_distrib}
\end{equation}

where $\mathbf{t} = [t_1,..., t_N]^T$, $\mathbf{y} = [y_1,..., y_N]^T$, $\mathbf{\tilde{t}} = [\tilde{t}_1,..., \tilde{t}_m]^T$, $\mathbf{\tilde{y}} = [\tilde{y}_1,..., \tilde{y}_m]^T$, $\mathbf{0}_{p} \in \mathbb{R}^p$ is the null vector

and for $(\mathbf{t}^*, \mathbf{t}^\prime) \in \mathbb{R}^n\times\mathbb{R}^p$, $\mathbf{K}(\mathbf{t}^*, \mathbf{t}^\prime)$ is the covariance matrix constructed from the kernel:
\begin{equation}
\mathbf{K}_\theta(\mathbf{t}^*, \mathbf{t}^\prime)  =  \begin{bmatrix}
        k_\theta(t^*_1, t_1^\prime) & \dots & k_\theta(t_1^*, t_p^\prime) \\
        \vdots & \ddots & \vdots \\
        k_\theta(t_n^*, t_1^\prime) & \dots & k_\theta(t_n^*, t_p^\prime)
\end{bmatrix} 
\label{eq:cov_matrix}
\end{equation}

First, we can improve our GP model by fitting the set of hyperparameters. One possible strategy, for instance, is to choose the maximum marginal likelihood estimate, based on the measured data points:  
\begin{equation}
    \hat{\theta} = \text{arg}\max\limits_{\theta} \mathbb{P}[\{y_i\}| \{t_i, \sigma_i\}, \theta] = \text{arg}\max\limits_{\theta} \left[ \frac{1}{(2\pi)^{N/2} \left| \det \mathbf{K}_\theta(\mathbf{t},\mathbf{t})\right|^{1/2} } \exp\left(-\frac{1}{2}\mathbf{y}^T\mathbf{K}_\theta(\mathbf{t},\mathbf{t})^{-1}\mathbf{y} \right) \right]
\end{equation}

Then, making use of the Gaussian nature of the joint distribution, we can make get a predictive distribution on the new points by conditioning the joint distribution of equation~(\ref{eq:joint_obs_pred_distrib}) on the observations:

\begin{equation}
    \mathbf{\tilde{y}} \big| \mathbf{\tilde{t}}, \mathbf{t}, \mathbf{y} 
    \sim \mathcal{N} (\mathbf{\mu}, \mathbf{\Sigma}) \text{ where }  
    \begin{cases}
        \mathbf{\mu} = \mathbf{K}_{\hat{\theta}}(\mathbf{\tilde{t}}, \mathbf{t}) \mathbf{K}_{\hat{\theta}}(\mathbf{t}, \mathbf{t})^{-1} \mathbf{y} \\
        \mathbf{\Sigma} = \mathbf{K}_{\hat{\theta}}(\mathbf{\tilde{t}}, \mathbf{\tilde{t}}) - \mathbf{K}_{\hat{\theta}}(\mathbf{\tilde{t}}, \mathbf{t}) \mathbf{K}_{\hat{\theta}}(\mathbf{t}, \mathbf{t})^{-1} \mathbf{K}_{\hat{\theta}}(\mathbf{t}, \mathbf{\tilde{t}})
    \end{cases}
    \label{eq:prediction_GP}
\end{equation}

Since this predictive distribution is also Gaussian, one can directly obtain best-guess interpolated values ($\{\mu_k\}_{1\leq k \leq m}$) as well as uncertainties on these values:
$$ \tilde{y}_k = \mu_k \pm \sqrt{\Sigma_{k,k}} \ , 1\leq k \leq m$$

In practice, interpolation is often performed on observations $\{y_i \pm \sigma_i \}$ that have measurement uncertainties. We mentioned above that a white noise kernel can account for this. However, equation~(\ref{eq: white noise kernel}) assumes that the noise is homoscedastic, ie., that it has the same level for all random variables (in other words, it is independent of $t$). This assumption is not always valid. In our case, for example, we know from equation~(\ref{Eq:flux_err}) that uncertainties depend on the flux level, so they vary with time. This can be remedied using a heteroscedastic (= different for each measurement) white noise kernel:
\begin{equation}
    k_{\rm HWN}(t,t^\prime) =  \sum_{i=1}^N \sigma_i^2  \delta (t-t_i) \delta (t^\prime-t_i)
\end{equation}

We warn that in this case, the prediction does not include any noise from this kernel - it only has an impact during the ‘‘learning’’ phase, when the hyperparameters are optimized. We also note that the $\sigma_i$ are not hyperparameters, since they correspond to the measurement uncertainties. In fact, a more intuitive way to view this contribution is to consider that, compared to the noise free case, a diagonal matrix is added to the covariance matrix of the observations: $\mathbf{K}_\theta(\mathbf{t},\mathbf{t}) \mapsto \mathbf{K}_\theta(\mathbf{t},\mathbf{t}) + \text{diag} \left(\sigma_1,..., \sigma_N \right)$ while $\mathbf{K}_\theta(\mathbf{t},\mathbf{\tilde{t}}), \mathbf{K}_\theta(\mathbf{\tilde{t}}, \mathbf{t}), \mathbf{K}_\theta(\mathbf{\tilde{t}}, \mathbf{\tilde{t}})$ are left unchanged.

\subsection{Joint interpolation of two lightcurves using a Multi-Output Gaussian Process}
\label{Appendix: MOGP math}

The GP regression method presented in the previous section applies to functions with a single scalar output (ie. $f(t)\in \mathbb{R}$ for all $t$). In this work, however, we want to understand the relationship between multiple outputs (the lightcurves in two bands) - in which case we talk about Multi-Output Gaussian Processes (MOGPs). The most straightforward way uses independent GPs to model each output separately, but this approach ignores any correlation that may exist between the outputs (in our case, when Sgr~A*-NIR is brighter in one band, it should also be in the other one !). One solution is to extend the GP regression approach to vector-valued outputs \citep{Alvarez_MOGP}. While MOGPs can be generalized to $N$-dimensional outputs, we only aim to jointly interpolate two lightcurves, so we restrict ourselves to the 2D case in the following:

$f:t \mapsto \begin{bmatrix}
        f_1(t) \\
        f_2(t) 
\end{bmatrix}$ 
where for example, $f_1(t)$ represents the $K^\prime$-band magnitude and $f_2(t)$ the $H$-band magnitude. \\

The most common method used to account for multiple correlated outputs is known as the linear model of coregionalization (LMC). Each output is expressed as a linear combination of independent random functions $\{u^{(i)}_q(t)\}$ \citep{GP_geostats, Alvarez_MOGP_2}:   

\begin{equation}
    \begin{cases}
    f_1(t) = \sum\limits_{q=1}^{Q} \sum\limits_{i=1}^{R_q} a^{(i)}_{1,q} u^{(i)}_q(t) \\
    f_2(t) = \sum\limits_{q=1}^{Q} \sum\limits_{i=1}^{R_q} a^{(i)}_{2,q} u^{(i)}_q(t)
\end{cases}
\end{equation}

where the $\{a^{(i)}_{k,q}\}$ are real-valued coefficients, and the functions $\{u^{(i)}_q(t)\}$ are zero-mean (single-output) GPs. The later are grouped by covariance function: there are $Q$ groups of samples, and for each group $q$ of samples, $R_q$ samples are obtained independently from a GP with the same kernel $k_q(t,t^\prime)$. Other approaches employed in MOGP regression can be seen as simplified versions of the LMC: for instance, the Intrinsic Coregionalization Model \citep[ICM,][]{ICM_book} when $Q=1$, or the Semiparametric Latent Factor Model \citep[SLFM,][]{SLFM} when $R_q=1$.

In the LMC, the covariance function can then be written as a ($2\times2$ in our case) matrix-valued function \citep{Alvarez_MOGP_2}:

\begin{equation}
    \cov[f(t), f(t^\prime)] = \sum\limits_{q=1}^Q \mathbf{A}_q\mathbf{A}_q^T k_q(t,t^\prime) = \sum\limits_{q=1}^Q \mathbf{B}_q k_q(t,t^\prime) \text{ where } \mathbf{A}_q = \begin{bmatrix}
        a^{(1)}_{1,q} & ... & a^{(R_q)}_{1,q} \\
        a^{(1)}_{2,q} & ... & a^{(R_q)}_{2,q} 
        \end{bmatrix}
\end{equation}
        
The matrices $\mathbf{B}_q = \mathbf{A}_q^T\mathbf{A}_q$ are known as the coregionalization matrices and are positive semi-definite. The rank of each $\mathbf{B}_q $ is $R_q$, which has to be less than the output dimension in order for the $u^{(i)}_q$ to be independent: $R_q\leq 2$ here.

Considering a set of times $\{t_i\}_{1\leq i \leq N}$, the covariance matrix between the outputs can be easily expressed from the coregionalization matrices, using the Kronecker product $\otimes$:
\begin{equation}
    \begin{bmatrix}
        f_1(\mathbf{t})\\
        f_2(\mathbf{t})
    \end{bmatrix} \sim \mathcal{N} \left( \begin{bmatrix}
        \mathbf{0}_N \\
        \mathbf{0}_N
\end{bmatrix} , \mathbf{K}_{\rm full} = \sum\limits_{q=1}^Q \mathbf{B}_q \otimes \mathbf{K}_q(\mathbf{t},\mathbf{t}) \right)
\label{eq:MOGP_cov_matrix}
\end{equation}
where $f_j(\mathbf{t}) = [f_j(t_1),..., f_j(t_N)]^T$ for $j=1,2$ and $\mathbf{K}_q$ is given by equation~(\ref{eq:cov_matrix}).

MOGP regression does not require for both outputs to be measured at all times $t_i$ - which is convenient for our purposes, since the two lightcurves are never sampled simultaneously. We can split the times into a group $\{t_{1,i}\}$ where only the first output $f_1$ is measured, and a group $\{t_{2,i}\}$ where only the second output $f_2$ is measured. To optimize the hyperparameters, we simply need to marginalize the distribution (\ref{eq:MOGP_cov_matrix}) over the outputs that are not measured. In other terms, if we build a $N\times N$ matrix $\mathbf{K}_{\rm marg}$ by selecting in $\mathbf{K}_{\rm full}$ only the rows and columns corresponding to actual measurements $\{f_1(t_{1,i})\}$ or $\{f_2(t_{2,i})\}$, we have:
\begin{equation}
    \begin{bmatrix}
        f_1(\mathbf{t}_1)\\
        f_2(\mathbf{t}_2)
    \end{bmatrix} \sim \mathcal{N} \left( 
        \mathbf{0}_N , \mathbf{K}_{\rm marg} \right) 
\end{equation}
After this, predictions for any unknown output can be obtained similarly to equation~(\ref{eq:prediction_GP}), by considering the joint distribution then conditioning over the observations. 

In this work, the final form of the covariance function that we use is:
\begin{equation}
\begin{split}
    k_f(t,t^\prime) = \ & \mathbf{B}_{\rm RBF} \cdot k_{\rm RBF}(t,t^\prime ; \sigma =1, l_{\rm RBF}) + \mathbf{B}_{\rm exp} \cdot k_{\rm exp}(t,t^\prime ; \sigma=1, l_{\rm exp}) \\ & + \mathbf{B}_c \cdot k_c(t,t^\prime ; \sigma=1)  + \sum_{i=1}^N \begin{bmatrix}
        \epsilon_i & 0 \\
        0 & 1-\epsilon_i 
        \end{bmatrix} \sigma_i^2  \delta (t-t_i) \delta (t^\prime-t_i) 
\end{split}
\label{eq:final_kernel}
\end{equation}
where $\epsilon_i = \begin{cases}
    1 \text{ if the measurement at $t_i$ is in $K^\prime$-band}\\
    0 \text{ if the measurement at $t_i$ is in $H$-band}
\end{cases}$

and for $q \in \{ {\rm RBF, exp},c \}$:
$\mathbf{B}_q = W_q^T W_q + \begin{bmatrix}
        \kappa_q^{(1)} & 0 \\
        0 & \kappa_q^{(2)} 
\end{bmatrix}$ \text{ with }$
W_q  = [w_q^{(1)}, w_q^{(2)}]$ \text{, ensuring that $\mathbf{B}_q$ is positive definite. } \\

The coregionalization matrices for the RBF, exponential and constant kernels have four real-valued hyperparameters each: $w_q^{(1)}, w_q^{(2)}, \kappa_q^{(1)}, \kappa_q^{(2)}$. However, the constant kernel is only designed to fit the mean of the two outputs, so there is no need to include correlation there: we impose $w_c^{(1)} = w_c^{(2)} =0$. We also fix $\sigma=1$ inside the $k_q$ since the matrices $\mathbf{B}_q$ already encode the amplitude scale.

The last term in equation~(\ref{eq:final_kernel}) corresponds to a white heteroscedastic kernel adapted to the multi-output case. If we write $\{ \sigma_{i}^{(j)} \}$ the uncertainties on the measurements $\{f_j(t_{j,i})\}$ of the $j$-th output (for $j=1,2$), the inclusion of this kernel amounts (compared to the noise-free case) to adding $\text{diag} \left(\sigma_{1}^{(1)},..., \sigma_{n}^{(1)},\sigma_{1}^{(2)},..., \sigma_{p}^{(2)} \right) $ to the marginalized covariance matrix $\mathbf{K}_{\rm marg}$.

In total, we have $12$ hyperparameters that need to be optimized: two timescales $l_{\rm RBF}, l_{\rm exp}$, six auto-correlation coefficients $ \{ \kappa_q^{(1)}, \kappa_q^{(2)} \}_{q= {\rm RBF, exp}, c}$ and four cross-correlation coefficients $\{ w_q^{(1)}, w_q^{(2)} \}_{q= {\rm RBF, exp}}$. Because we want the RBF kernel to represent the long-term trends and the exponential kernel to encode the short-term variability, we impose the constraint $0 \leq l_{\rm exp} \leq 10 $ min during the optimization. 

The interpolated lightcurves obtained using this method are either presented in the bottom panel of Figure~\ref{fig:lightcurve} (for 2022 May 21), or in Figure~\ref{fig:lightcurves_appendix} (for all the other epochs in our dataset).

{ 
\nolinenumbers
\centering
\begin{longtable}{c}
\includegraphics[width=0.9\linewidth]{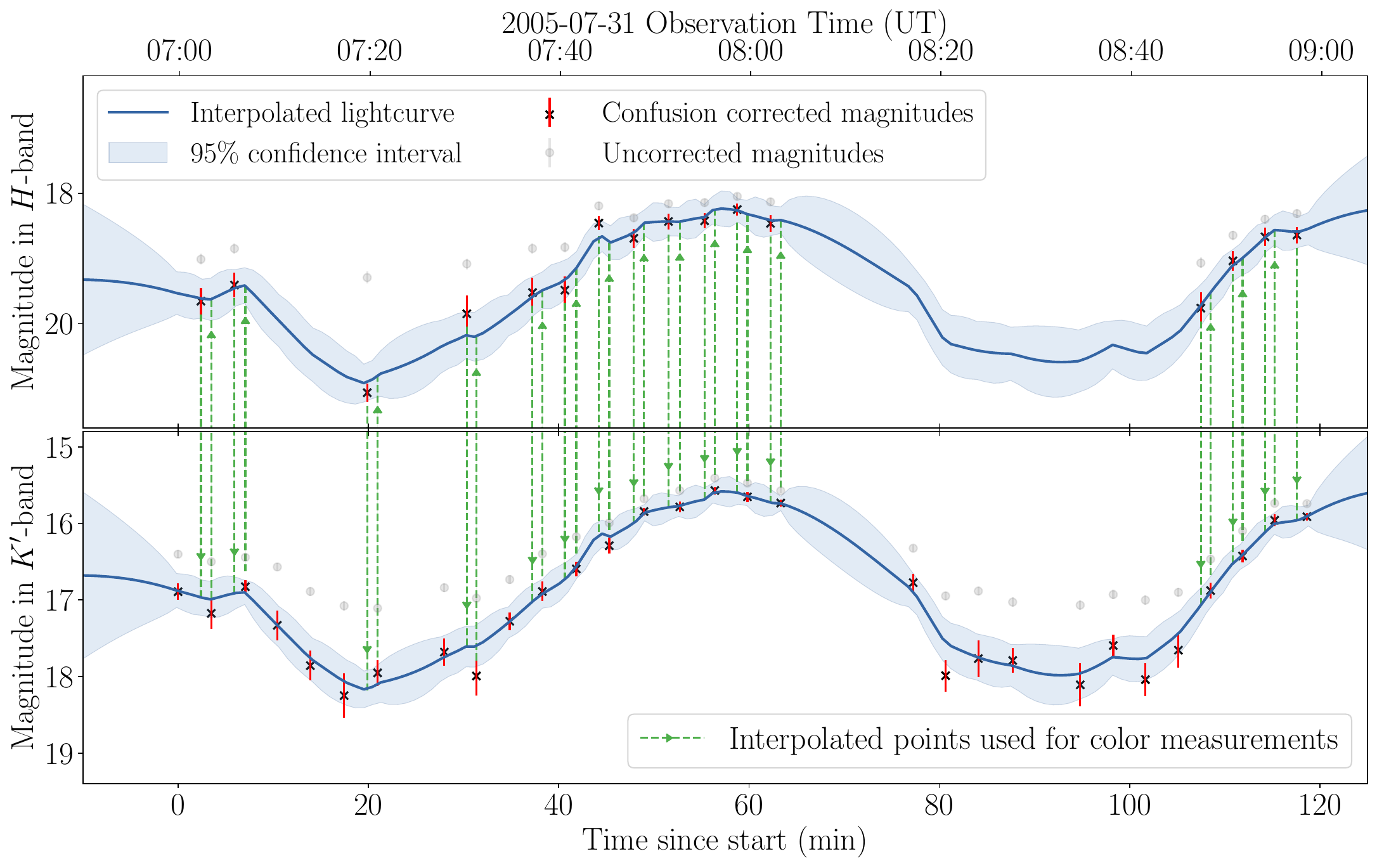}\\
\vspace{0.3cm} \\
\includegraphics[width=0.9\linewidth]{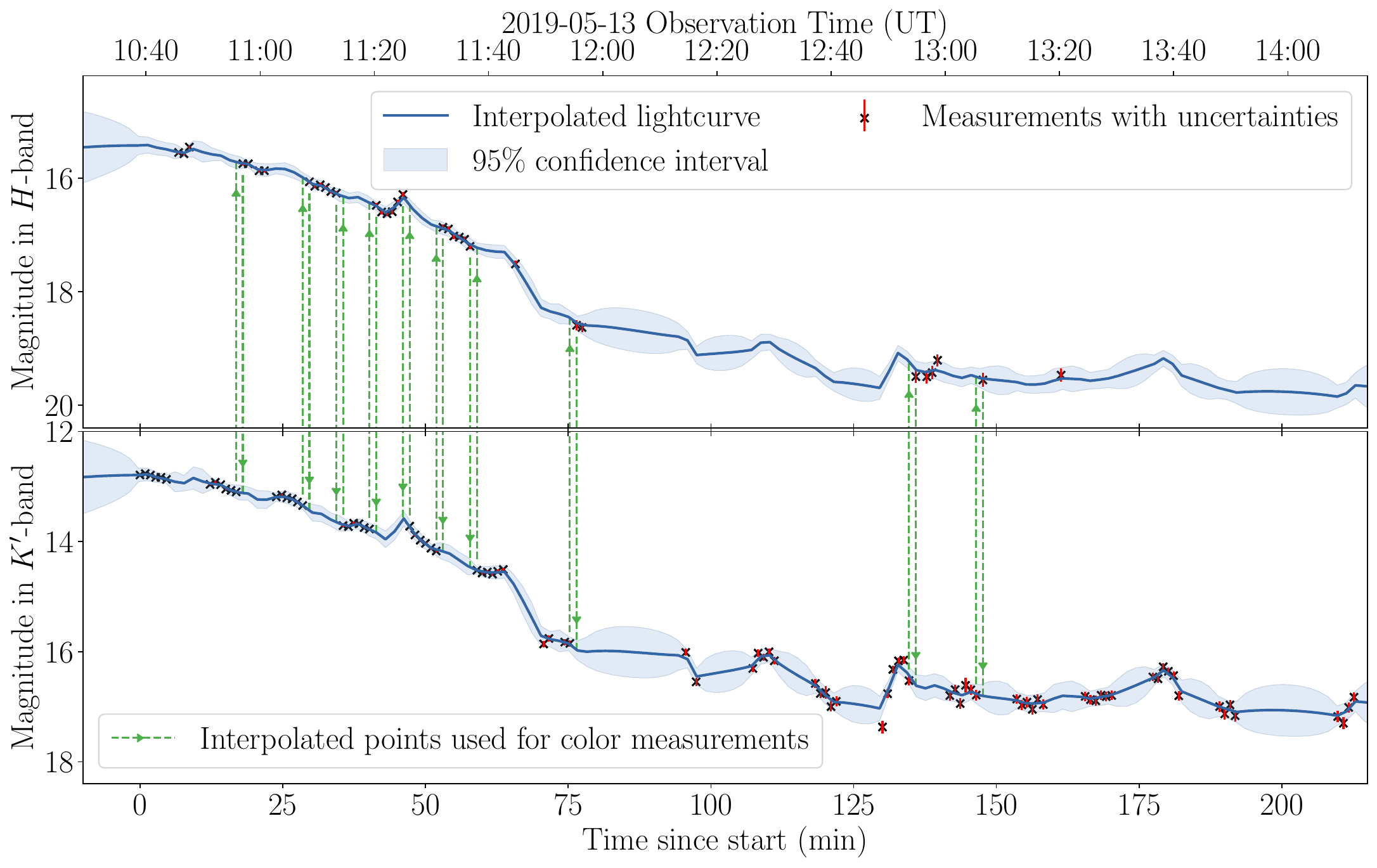}\\ 
\vspace{0.3cm} \\
\includegraphics[width=0.9\linewidth]{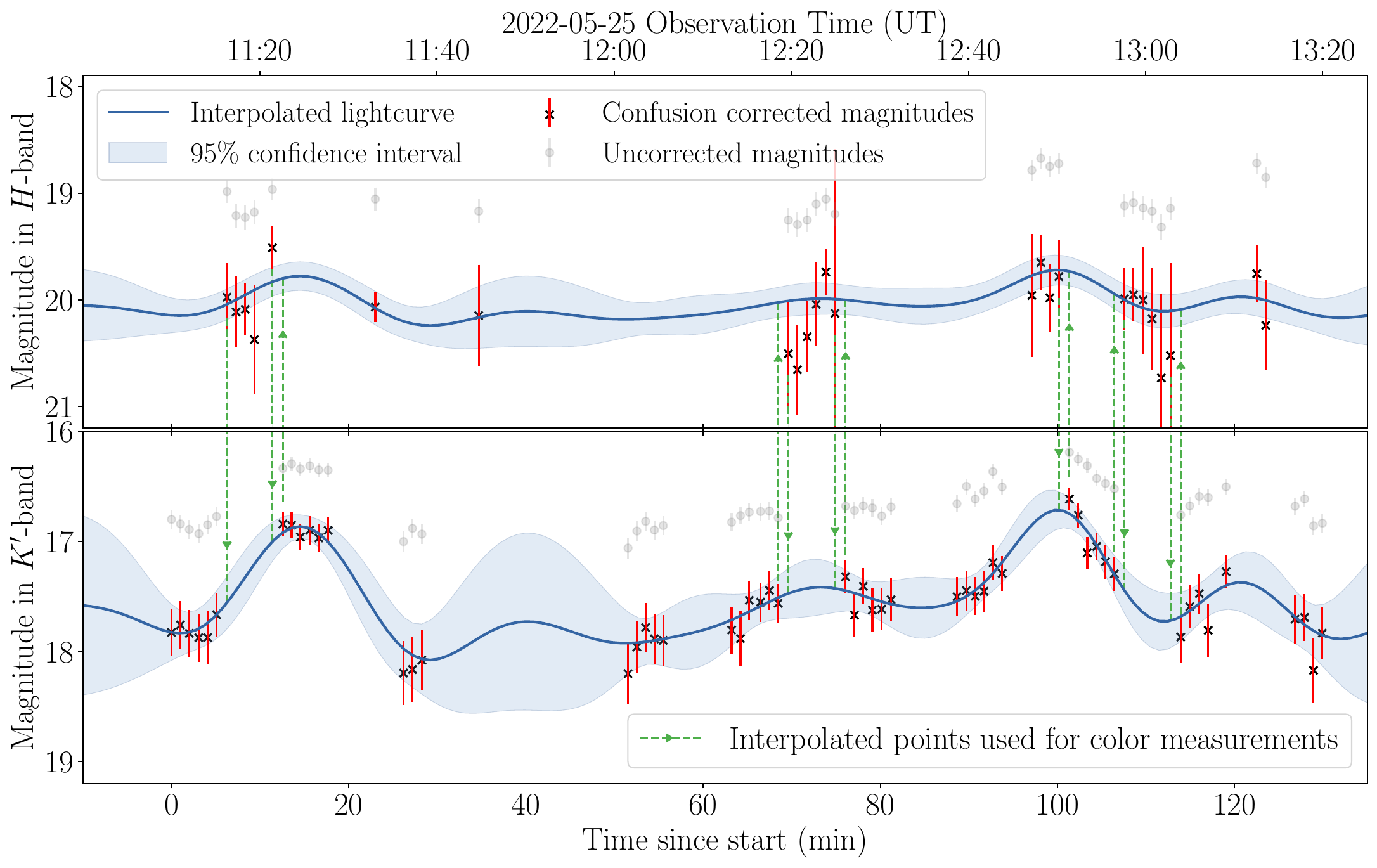}\\
\vspace{0.3cm} \\
\includegraphics[width=0.9\linewidth]{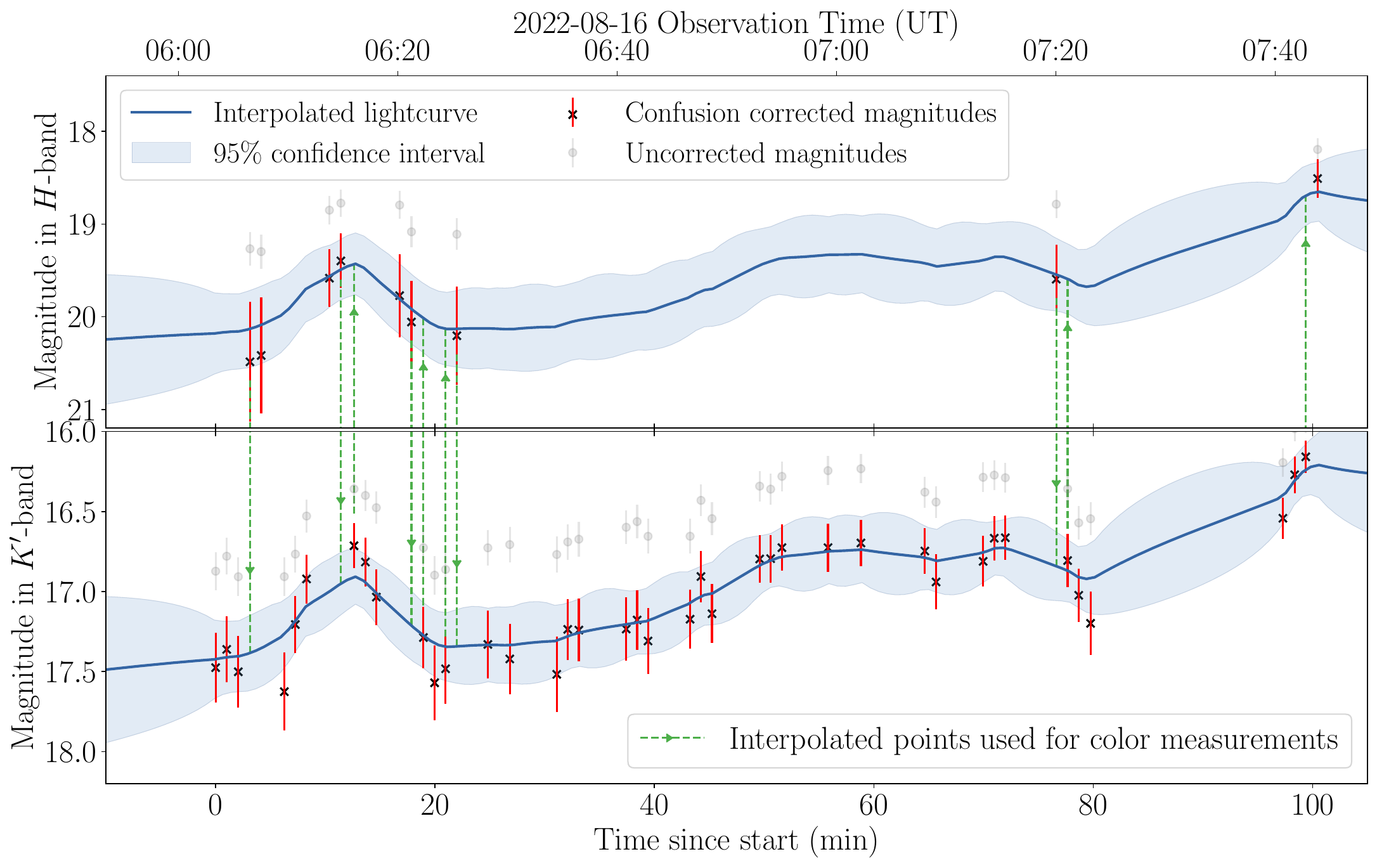}\\ 
\vspace{0.3cm} \\
\includegraphics[width=0.9\linewidth]{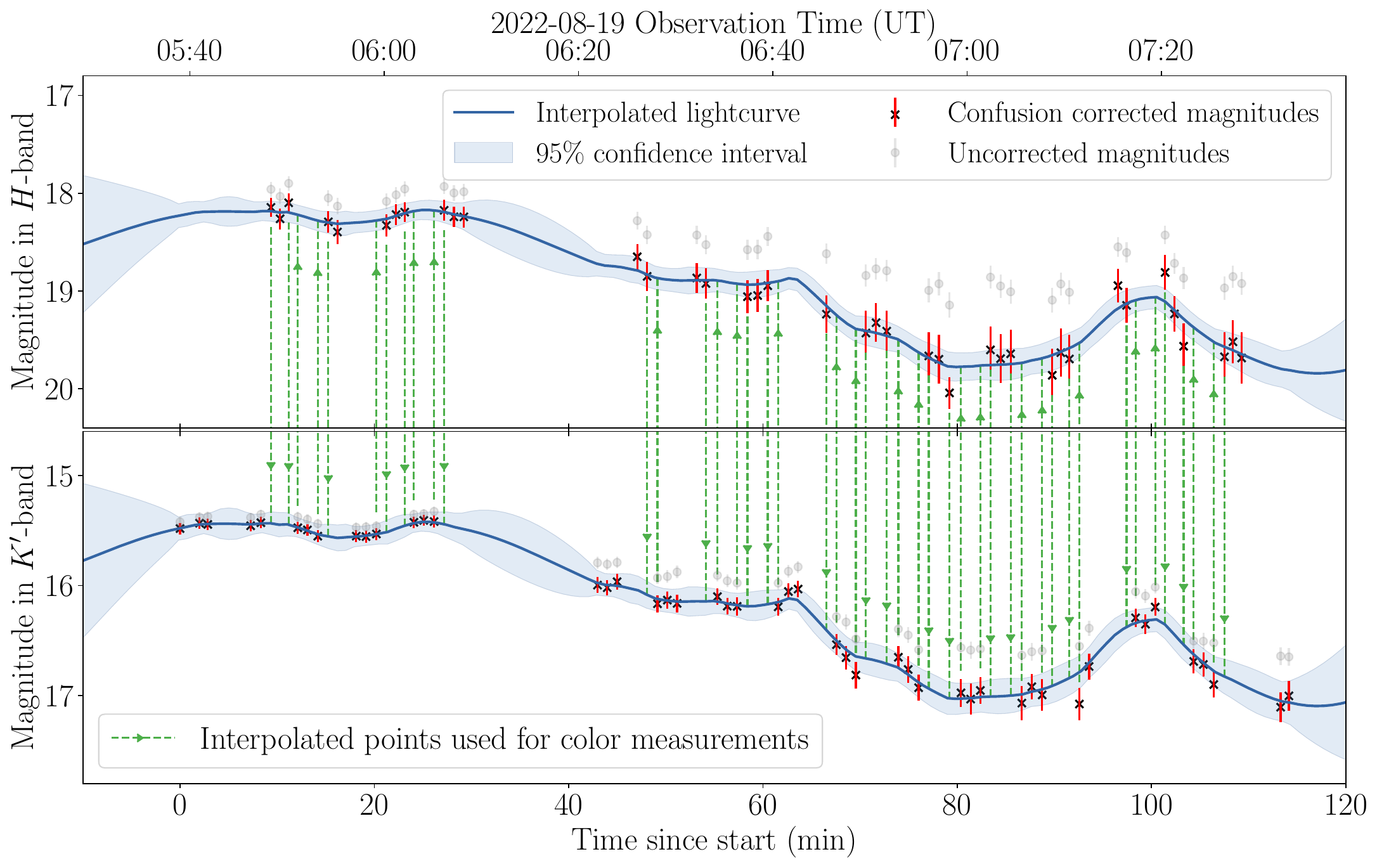}\\
\vspace{0.2cm} \\
\includegraphics[width=0.9\linewidth]{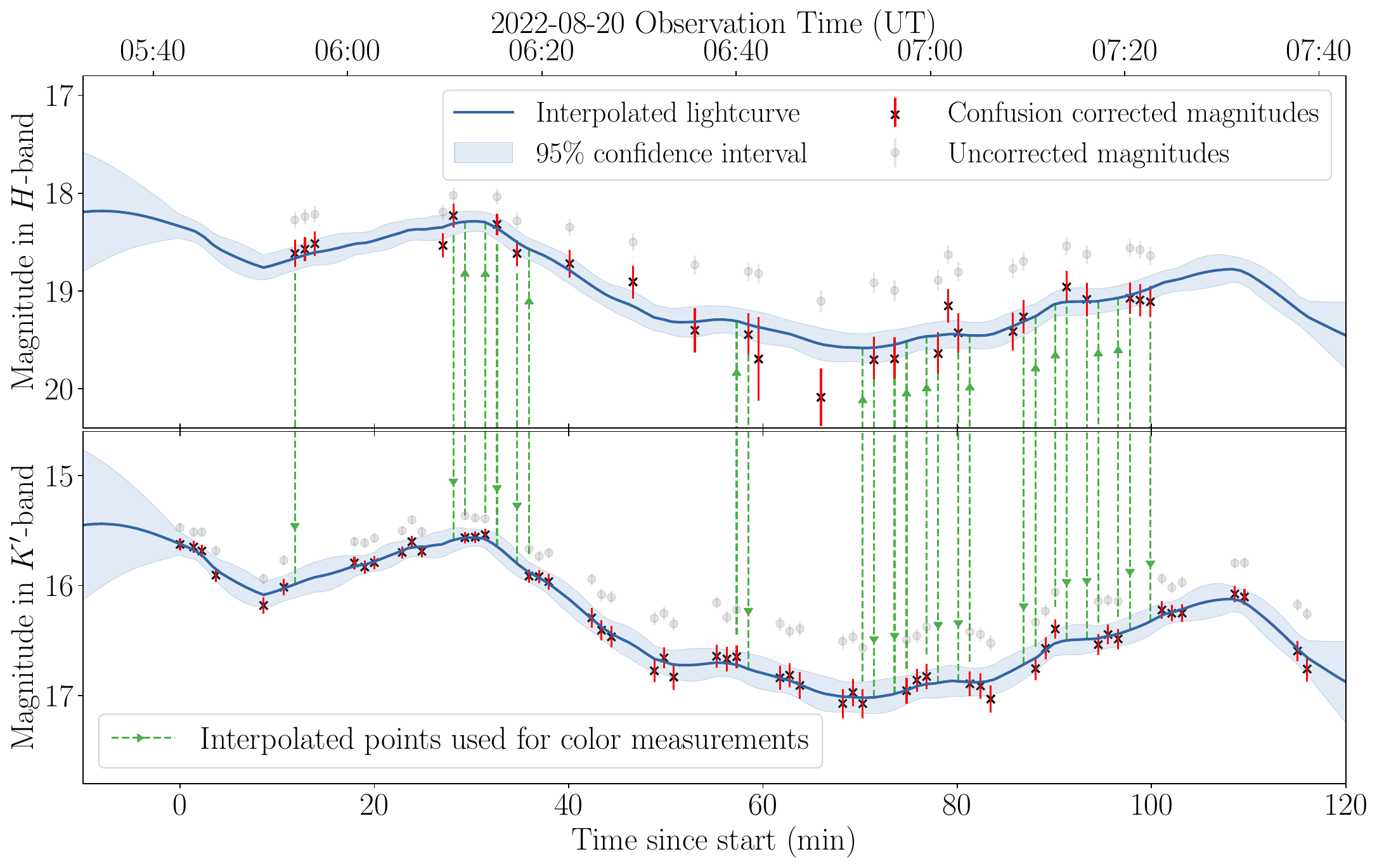}
\end{longtable}
\captionof{figure}{Same as the bottom panel of Figure~\ref{fig:lightcurve}, but for the other epochs of our dataset (from top to bottom: 2005 July 31, 2019 May 13, 2022 May 25, 2022 Aug 16, 2022 Aug 19, 2022 Aug 20). We also show the lightcurves before confusion correction (except for 2019 May 13 since Sgr~A*-NIR is not confused with any known source). The lightcurves (with flux/magnitude values with uncertainties, before/after confusion correction) are also available as a machine-readable table. }
\label{fig:lightcurves_appendix}
}
\vspace{0.3cm}

\subsection{Testing the interpolation method.}
\label{Appendix: GP tests}

In order to assess the performance of the MOGP regression method described in Appendix~\ref{Appendix: MOGP math} for the joint interpolation of lightcurves, we examined the distribution of leave-one-out normalized errors. Within a given epoch, we dropped each magnitude measurement $m_{\rm meas}\pm \sigma_{\rm meas}$ (in each band), then applied the MOGP model on the remaining measurements in order to get a prediction with uncertainty $m_{\rm pred} \pm \sigma_{\rm pred}$ at the time of the dropped point. The normalized error (or standard score) is then:

\begin{equation}
    E_n = \frac{(m_{\rm pred}-m_{\rm meas})}{\sqrt{\sigma_{\rm pred}^2+\sigma_{\rm meas}^2}}
\end{equation}

\begin{figure*}[ht!]
\centering
\includegraphics[width=0.99\linewidth]{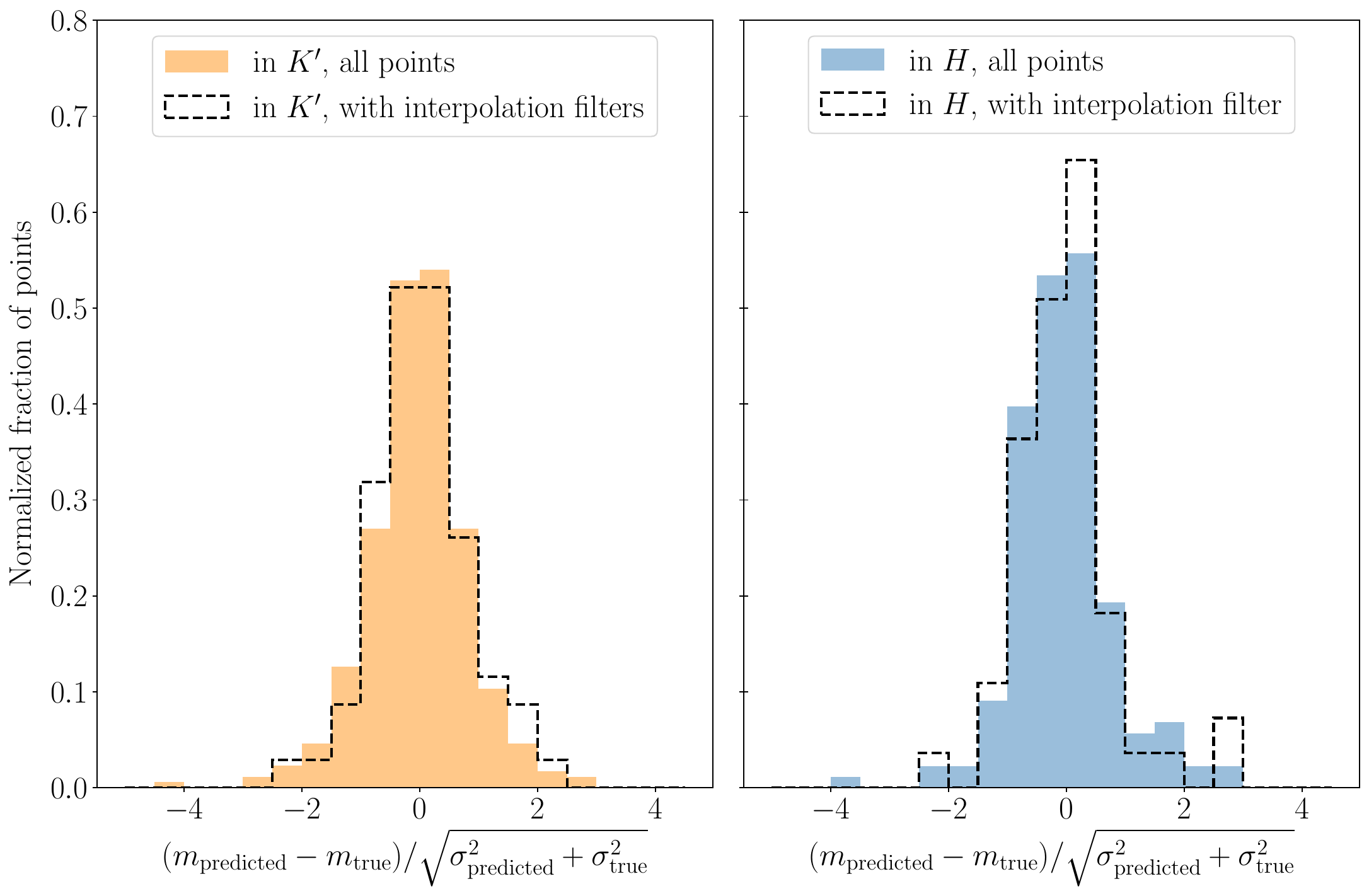}
\caption{Distribution of normalized leave-one-out errors for the MOGP interpolation (all epochs combined).} 
\label{fig:MOGP_loo_test}
\vspace{-0.5\baselineskip}
\end{figure*}

(the expected uncertainty is $\sqrt{\sigma_{\rm pred}^2+\sigma_{\rm meas}^2}$ since measurement uncertainties described by the heteroscedastic white noise kernel are not already added to the prediction, see Appendix~\ref{Appendix: GP math}). Repeating the procedure for every data point and every epoch, we can estimate the  distribution of $E_n$. Figure~\ref{fig:MOGP_loo_test} shows that this distribution is close to a unit normal distribution in the two bands, whether or not the interpolation filters (see section~\ref{subsec: GP interp}) are applied. In particular: (a) it shows no bias towards positive or negative errors, since we always have $|\langle E_n \rangle| \leq 0.04$ where $\langle . \rangle$ is the mean, (b) the errors are consistent with the uncertainties, since $\sqrt{\langle E_n^2 \rangle} \approx 0.85 \sim 1$. We note that the interpolation removes the few extreme ($\sim 4\sigma$) outliers.

\newpage
\section{Added uncertainties from confusion correction \& interpolation}
\label{Appendix: uncertainty_comparison}

In addition to the photometric uncertainties, the uncertainties on the spectral index measurements presented in this paper account for the additional uncertainties introduced by the two correction steps (extinction, see section~\ref{subsec:extinction}, and source confusion, see section~\ref{subsec:confusion correction}), as well as by the interpolation (section~\ref{subsec: GP interp}). Table~\ref{OoMcorrection} gives ranges for the uncertainties added by those steps. The uncertainty from extinction correction is accounted for as a systematic error term, since choosing a different value for the mean color excess is equivalent to shifting all the spectral index measurements by the same amount. In this Appendix, we discuss in more detail the uncertainties from confusion correction and interpolation, which vary between measurements.

The uncertainty on the spectral index added by the interpolation step is estimated using:
\begin{equation}
    \sigma_{\rm add}^{\rm interp} (\alpha) = \frac{-0.4}{\log_{10}(\lambda_{K^\prime}/\lambda_H)} \sigma_{\rm pred}(m_{\rm band})
\end{equation}
where $\sigma_{\rm pred}$ is the uncertainty predicted by the Gaussian Process (see Appendix~\ref{Appendix: GP tests}) and band=$K^\prime$ (resp. $H$) if there is a measurement in $H$ (resp. $K^\prime$) for the considered point.

Confusion correction adds uncertainty to the magnitudes (see section~\ref{Appendix: conf_corr+err}), so in order to assess the impact on the spectral index uncertainties, it is only possible to compare the errorbars after interpolation. We estimate the contribution from confusion correction to the uncertainty on $\alpha$ as:
\begin{equation}
    \sigma_{\rm add}^{\rm conf} (\alpha) = \sqrt{\sigma_{\rm corrected} (\alpha)^2-\sigma_{\rm uncorrected} (\alpha)^2}
\end{equation}
where $\sigma_{\rm corrected} (\alpha)$ (resp. $\sigma_{\rm uncorrected} (\alpha)$) is the final uncertainty on the spectral index after interpolating the confusion corrected (resp. uncorrected) lightcurve.

\begin{figure*}[ht!]
\centering
\includegraphics[width=0.55\linewidth]{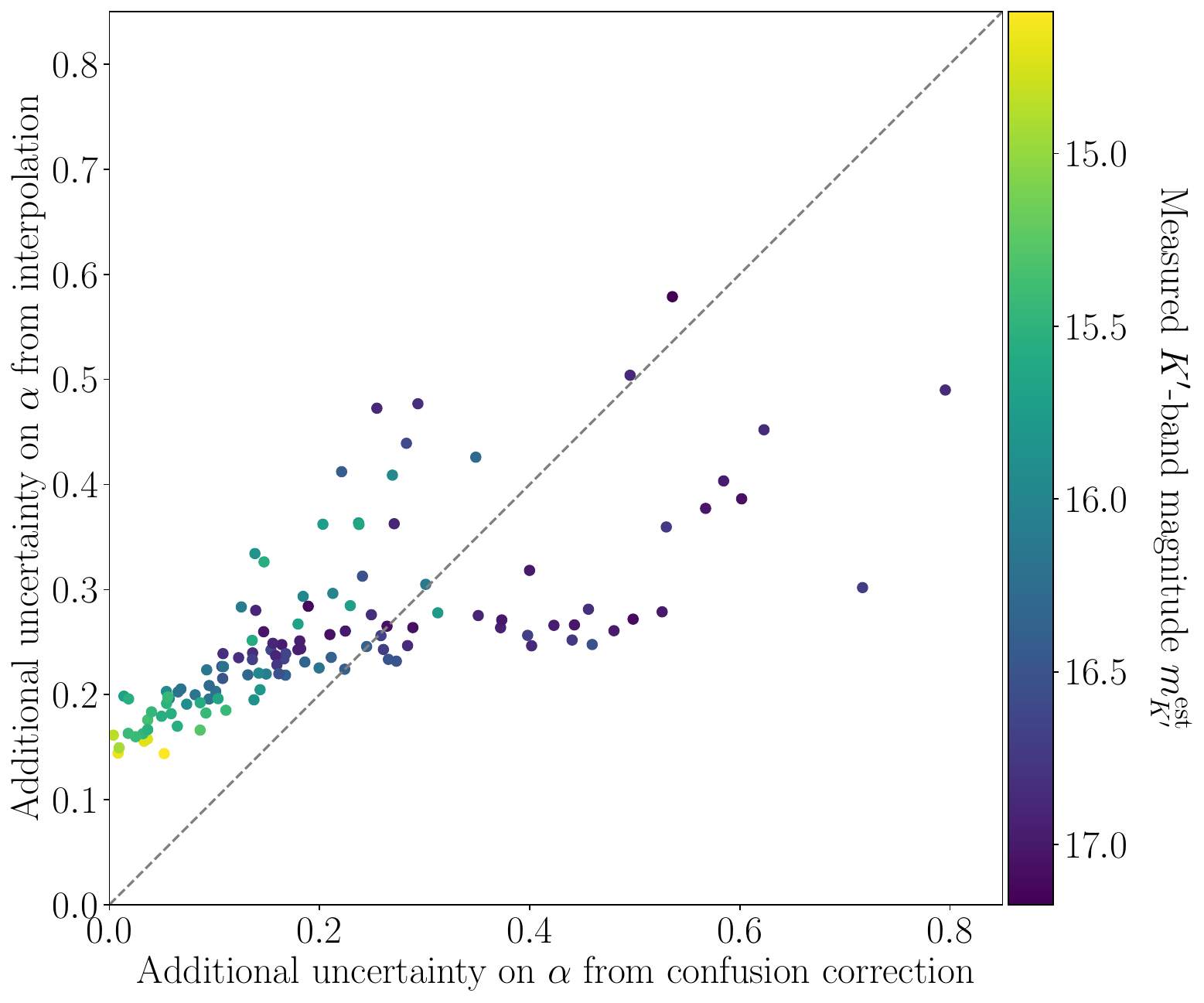}
\caption{Comparison between the uncertainties added during the confusion correction step, and the uncertainties added during the interpolation step, as a function of $K^\prime$ magnitude. Interpolation tends to adds more uncertainty for brighter points, but as Sgr~A*-NIR gets fainter, the uncertainty added by confusion correction starts to dominate.} 
\label{fig:comparison_uncertainties}
\end{figure*}

These added uncertainties are reported as a range in Table~\ref{OoMcorrection} for points belonging to our extended dataset (excluding 2019 May 13 from the range on $\sigma_{\rm add}^{\rm conf} (\alpha)$ since there is no source confusion for that epoch). We compare these two sources of uncertainty in more detail in Figure~\ref{fig:comparison_uncertainties}: we can see that, for bright states of Sgr~A*-NIR, interpolation is the dominant source of added uncertainty. For fainter states, confusion correction becomes more substantial and therefore adds comparatively more uncertainty.

\vspace{-0.5\baselineskip}

\section{Derivation of the likelihood}
\label{Appendix: likelihood}

In this section, we derive an analytic formula for the likelihood $\mathcal{L}_1 $ of a single spectral index measurement $\alpha_{H-K^\prime}^{\text{est},(i)}$ at a magnitude $m_{K^\prime}^{\text{est},(i)}$.
For a set of model parameters $\boldsymbol{\theta_s} = (\xi, \eta, m_{K^\prime}^{\rm bck}, \alpha_{\rm bck})$, the expected value $\alpha_{H-K^\prime}^{\rm est}$ for the measured spectral index at that magnitude is completely determined by the noise values in the two bands ($\varepsilon_{H}$, $\varepsilon_{K^\prime}$). Indeed, from equations (\ref{eq:SpecIndex}) and (\ref{eq: subfunctions_alpha_formula}), we have: $\alpha_{H-K^\prime}^{\rm est} = \mathcal{F}  ( \varepsilon_H; \varepsilon_{K^\prime}, m_{K^\prime}^{\text{est},(i)},  \boldsymbol{\theta_s})$. Fixing the noise value $\varepsilon_{K^\prime}$, this defines a bijective mapping $\varepsilon_H \overset{\mathcal{F}}{\longmapsto} \alpha_{H-K^\prime}^{\rm est}$. We can then consider the reverse mapping $ \alpha_{H-K^\prime}^{\rm est} \overset{\mathcal{G}}{\longmapsto} \varepsilon_H$, explicitely given by:

\begin{equation}
\begin{split}
    \mathcal{G}( \alpha_{H-K^\prime}^{\rm est} ; \varepsilon_{K^\prime}, m_{K^\prime}^{\text{est},(i)}, \boldsymbol{\theta_s}) &= f_{0,H} 10^{-0.4 H_{\rm est}}  - F_{H, \rm SgrA}^{\rm obs} -  F_{H, \rm bck}^{\rm obs} \\
    \text{with } H_{\rm est} &= K^\prime_{\rm est} + \langle E(H-K^\prime) \rangle + 2.5 \log_{10}(f_{0,H}/f_{0,K^\prime}) - 2.5\ \alpha_{H-K^\prime}^{\rm est} \log_{10}(\lambda_{K^\prime}/\lambda_H) 
\end{split}
\label{eq: reverse_mapping}
\end{equation}
where $K^\prime_{\rm est} = m_{K^\prime}^{\text{est},(i)}$ and $ F_{H, \rm SgrA}^{\rm obs},  F_{H, \rm bck}^{\rm obs}$ are given by the same expressions as in the group of equations (\ref{eq: subfunctions_alpha_formula}).

Then, assuming uncorrelated Gaussian noise values in the two bands ($\varepsilon_{\rm band} \sim \mathcal{N}(0,\sigma_{F_{\rm band}}^{(i)}) $), we can write the likelihood for that measurement as

\begin{equation}
\begin{split} 
     \mathcal{L}_1 & \left( \alpha_{H-K^\prime}^{\text{est},(i)}  \Big| m_{K^\prime}^{\text{est},(i)} , \sigma_{F_{K^\prime}}^{(i)}, \sigma_{F_H}^{(i)}  ; \boldsymbol{\theta_s} \right) 
     = \int d\varepsilon_H \ d\varepsilon_{K^\prime} \ p\left( \alpha_{H-K^\prime}^{\text{est},(i)} \Big| m_{K^\prime}^{\text{est},(i)} , \sigma_{F_{K^\prime}}^{(i)}, \sigma_{F_H}^{(i)}, \varepsilon_H, \varepsilon_{K^\prime} ; \boldsymbol{\theta_s} \right) p \left( \varepsilon_H \Big| \sigma_{F_H}^{(i)} \right) p \left( \varepsilon_{K^\prime} \Big| \sigma_{F_{K^\prime}}^{(i)} \right) \\
     &= \frac{1}{2\pi \sigma_{F_H}^{(i)} \sigma_{F_{K^\prime}}^{(i)}} \int d\varepsilon_H \ d\varepsilon_{K^\prime} \ \delta \left[ \alpha_{H-K^\prime}^{\text{est},(i)} - \mathcal{F} ( \varepsilon_H ; \varepsilon_{K^\prime}, m_{K^\prime}^{\text{est},(i)}, \boldsymbol{\theta_s}) \right] \exp \left( -\frac{\varepsilon_H^2}{2 \sigma_{F_H}^{(i) \ 2} } - \frac{\varepsilon_{K^\prime}^2}{2 \sigma_{F_{K^\prime}}^{(i) \ 2} } \right) \scalebox{1.25}{$\mathbbm{1}$}(F_{K^\prime, \rm SgrA}^{\rm obs}) \\
     &=  \frac{1}{2\pi \sigma_{F_H}^{(i)} \sigma_{F_{K^\prime}}^{(i)}} \int  d\varepsilon_{K^\prime} \ \left| \frac{\partial \mathcal{F}}{\partial \varepsilon_H} \right|^{-1} \exp \left( -\frac{\mathcal{G} ( \alpha_{H-K^\prime}^{\text{est},(i)} ; \varepsilon_{K^\prime}, m_{K^\prime}^{\text{est},(i)}, \boldsymbol{\theta_s})^2}{2 \sigma_{F_H}^{(i) \ 2} } - \frac{\varepsilon_{K^\prime}^2}{2 \sigma_{F_{K^\prime}}^{(i) \ 2} } \right) \scalebox{1.25}{$\mathbbm{1}$}(F_{K^\prime, \rm SgrA}^{\rm obs}) \\
\end{split}
\end{equation}

where $\delta$ denotes the Kronecker delta function, and \scalebox{1.25}{$\mathbbm{1}$} the Heaviside step function, introduced to avoid the $(\varepsilon_H, \varepsilon_{K^\prime})$ space corresponding to non-physical solutions where $10^{-0.4 m_{K^\prime}^{\text{est},(i)}} - \varepsilon_{K^\prime} = F_{K^\prime, \rm SgrA}^{\rm obs} < 0$. The last equation is obtained using the change of variables $\varepsilon_H \rightarrow \alpha_{H-K^\prime}^{\rm est} = \mathcal{F} (\varepsilon_H ; \varepsilon_{K^\prime}, m_{K^\prime}^{\text{est},(i)},  \boldsymbol{\theta_s}) $ for which we can compute the Jacobian explicitely:
\begin{equation}
     \left( \frac{\partial \mathcal{F}}{\partial \varepsilon_H} \right) = 
     \frac{-0.4}{\log_{10}(\lambda_{K^\prime}/\lambda_H)} \frac{\partial H_{\rm est}}{\partial \varepsilon_H} = \frac{10^{-0.4 H_{\rm est}}}{f_{0,H} \ln(\lambda_{K^\prime}/\lambda_H)} 
\end{equation}
where $H_{\rm est}$ is computed in the same way as equation~(\ref{eq: reverse_mapping}), with $ \alpha_{H-K^\prime}^{\rm est} = \alpha_{H-K^\prime}^{\text{est},(i)}$.

\bibliography{biblio}{}
\bibliographystyle{aasjournal}

\end{document}